\pdfoutput=1
\documentclass[preprint]{aastex}

\usepackage{psfrag}
\usepackage{textcomp}
\usepackage{arydshln}
\usepackage{graphicx}
\usepackage{natbib}
\usepackage[caption = false]{subfig}
\usepackage{float}
\usepackage{threeparttable}
\usepackage{multirow}

\slugcomment{Accepted for publication in ApJS}

\begin{document}

\title{THE \textit{SPITZER} INFRARED SPECTROGRAPH SURVEY of PROTOPLANETARY DISKS IN ORION A: I. DISK PROPERTIES}

\author{K. H. Kim\altaffilmark{1, 2},
Dan M. Watson\altaffilmark{2},
P. Manoj\altaffilmark{2, 13},
W. J. Forrest\altaffilmark{2},
Elise Furlan\altaffilmark{3},
Joan Najita\altaffilmark{4},
Benjamin Sargent\altaffilmark{5},
Jes\'{u}s Hern\'{a}ndez\altaffilmark{6},
Nuria Calvet\altaffilmark{7},
Luc\'{i}a Adame\altaffilmark{8},
Catherine Espaillat\altaffilmark{9}, 
S. T. Megeath\altaffilmark{10},
James Muzerolle\altaffilmark{11},
M. K. McClure\altaffilmark{12}
}

\altaffiltext{1}{Korea Astronomy and Space Science Institute, Korea Astronomy and Space Science Institute (KASI), 776, Daedeokdae-ro, Yuseong-gu, Daejeon 305-348, Republic of Korea; {\sf quarkosmos@kasi.re.kr}}
\altaffiltext{2}{Department of Physics and Astronomy, University of Rochester, Rochester, NY 14627, USA}
\altaffiltext{3}{Infrared Processing and Analysis Center, Caltech, 770 S. Wilson Ave., Pasadena, CA, 91125, USA}
\altaffiltext{4}{National Optical Astronomy Observatory, 950 North Cherry Avenue, Tucson, AZ 85719, USA}
\altaffiltext{5}{Center for Imaging Science and Laboratory for Multiwavelength Astrophysics, Rochester Institute of Technology, 54 Lomb Memorial Drive, Rochester, NY 14623, USA}
\altaffiltext{6}{Centro de Investigaciones de Astronom\'{i}a, Apdo. Postal 264, M\'{e}rida 5101-A, Venezuela}
\altaffiltext{7}{Department of Astronomy, University of Michigan, 830 Dennison Building, 500 Church St, Ann Arbor, MI 48109, USA}
\altaffiltext{8}{Facultad de Ciencias F\'{i}sico-Matem\'{a}ticas, Universidad Aut\'{o}noma de Nuevo Le\'{o}n, Av. Universidad S/N, San Nicol\'{a}s de los Garza, Nuevo Le\'{o}n, C.P. 66451, M\'{e}xico}
\altaffiltext{9}{Department of Astronomy, Boston University, 725 Commonwealth Avenue, Boston, MA 02215, USA}
\altaffiltext{10}{Ritter Astrophysical Research Center, Department of Physics and Astronomy, University of Toledo, 2801 W. Bancroft St., Toledo, OH 43606, USA}
\altaffiltext{11}{Space Telescope Science Institute, 3700 San Martin Dr., Baltimore, MD 21218, USA}
\altaffiltext{12}{ESO, Karl-Schwarzschild-Str. 2, D-85748, Garching bei M\''{u}nchen, Germany}
\altaffiltext{13}{Department of Astronomy and Astrophysics, Tata Institute of Fundamental Research, Homi Bhabha Road, Mumbai 400 005, India}

\begin{abstract}

We present our investigation of 319 Class II objects in Orion A observed by \textit{Spitzer}/IRS. We also present the follow-up observation of 120 of these Class II objects in Orion A from IRTF/SpeX.
We measure continuum spectral indices, equivalent widths, and integrated fluxes that pertain to disk structure and dust composition from IRS spectra of Class II objects in Orion A. We estimate mass accretion rates using hydrogen recombination lines in the SpeX spectra of our targets. Utilizing these properties, we compare the distributions of the disk and dust properties of Orion A disks to those of Taurus disks with respect to position within Orion A (ONC and L1641) and to the sub-groups by the inferred radial structures, such as transitional disks vs. radially continuous full disks. Our main findings are as follows. (1) Inner disks evolve faster than the outer disks.  (2) Mass accretion rate of transitional disks and that of radially continuous full disks are statistically significantly displaced from each other. The median mass accretion rate of radially continuous disks in ONC and L1641 is not very different from that in Taurus. (3) Less grain processing has occurred in the disks in ONC compared to those in Taurus, based on analysis of the shape index of the 10 $\micron$ silicate feature ($F_{11.3}/F_{9.8}$). (4) The 20--31 $\mu$m continuum spectral index tracks the projected distance from the most luminous Trapezium star, $\theta^{1}$ Ori C. A possible explanation is the effect of UV ablation of the outer part of the disks. 

\end{abstract}


\section{INTRODUCTION}
In the process of star formation, young stars have circumstellar disks called ``protoplanetary disks". Protoplanetary disks evolve from optically thick flared disks toward optically thin, flat, tenuous disks, and these disks are known as the birth-places of planets. 
The infrared-based classification of the spectral energy distributions (SEDs) of young stellar objects (YSOs) by \citet{Lada1987-YSO-classes} divides this evolution into three stages, Classes I-II-III. These classifications are based on the spectral slope ($\alpha$) of the infrared SEDs between near- ($\sim$ 2 $\micron$) and mid- ($\sim$ 25 $\micron$) infrared. Class I objects ($\alpha$ $>$ 0) have SEDs with increasing infrared emission dominated by the envelope, as expected from protostars. Class II objects (-2 $<$ $\alpha$ $<$ 0) have relatively flat or negative SED slopes, corresponding to excess emission over infrared wavelength ranges produced by a dusty and optically thick disk around a pre-main sequence star. Class III objects ($\alpha$ $<$ -2) have SEDs with very little or no disk emission at infrared wavelengths indicating that a disk is much evolved and settled toward to the midplane with large sized grains or has largely or completely dissipated.

\citet{Andre_1993_class0} extended the classification to an earlier Class 0 based on millimeter wavelength emission. \citet{AdamsLadaShu_1988ApJ...326..865A_FS} modeled T Tauri stars with flat infrared spectrum and \citet{Greene1994_flat} classified spectra with spectral indices between -0.3 and 0.3 as flat spectrum (FS) sources since their spectral slopes are between Class I and II. The SED classes are closely, though not perfectly, matched to evolutionary stages \citep[cf.][]{Robitaille+2006ApJS}. Classes 0, I and FS correspond to the progress of the accretion and dispersal of the protostellar envelope, and the reduction of the central star's average accretion rate through the range $10^{-4}$--$10^{-7}$ $M_\odot yr^{-1}$. In the Class II phase, the disk's gaseous and small-grain components evolve and dissipate, while the central star's accretion rate decreases from $10^{-7}$ to $10^{-10}$ $M_\odot yr^{-1}$. Effects of orientation can make an object from one \emph{evolutionary} stage appear to be in a different SED class; for example, a protoplanetary disk without an envelope, viewed edge on, may belong to the Class I SED type as in case of DG Tau B \citep{Watson_2004ApJS..154..391W_ClassI-IRS-DGTauB}.

Throughout a disk's life, its material is accreted toward the central star. A classification based on accretion indicators (for instance, $H_\alpha$ equivalent width being greater/less than 10 {\AA}) divides T Tauri stars into classical T Tauri stars (CTTS) and weak-lined T Tauri stars (WTTS), which generally correspond closely to Class II and Class III, respectively.

The transitional disks (TDs) have the appearance of a Class III YSOs at shorter infrared wavelengths and Class II YSOs at longer wavelengths \citep[e.g.][]{Espaillat_obsTDs_2014arXiv1402.7103E}. 
Therefore, TDs are young systems which have AU-scale radial gaps or central clearing in their dust distribution while they still have optically thick and gas-rich outer disks. There are several mechanisms, such as photoevaporation, dust coagulation and grain growth, MRI instability, and giant planet formation, to explain the origin of gaps and holes in TDs. An interesting finding from the previous studies is that recent observations \citep[e.g.][]{andrews2011_sma_CTDs, Olofsson+2011_TCha_VLTI, Casassus_2013Natur.493..191C} of TDs and the statistical study of the properties of TDs (Kim et al. 2009, 2013) support the idea that giant planet formation produces the gaps of TDs. \citet{khkim2013-OriATD} presented observationally and statistically significant trends in mass accretion rates ($\dot{M}$) showing that mass accretion rates of TDs are reduced by about one order of magnitude compared to the typical $\dot{M}$ of CTTS \citep{najita07, Espaillat_TDclass_2012}. This strongly suggests gap opening and disk clearing by newly-formed planets in disks \citep{Lubow_Dangelo_2006}.

The direction of evolution of protoplanetary disks does not proceed in one universal way. Some disks with weak mid-IR excess suggest that their disk material depleted in a global and homologous manner~\citep[e.g.][]{Hernandez_2007ApJ...671.1784H_OriOB1, Currie_2009_homodep}. 
YSOs are also known to evolve along different tracks depending on their masses. A young star called a Herbig Ae/Be star has spectral type of A or B, with masses between 2 and 10 $M_\odot$. T Tauri star has spectral type between F and M, with masses less than 2 $M_\odot$. The more massive Herbig Ae/Be stars and their disks evolve much faster than T Tauri stars. T Tauri stars will end up as stars similar to the Sun.

In this paper, we focus on Class II YSOs including TDs, associated with T Tauri stars, because the seeds of planets are thought to be formed in the dusty and massive disks present during this phase of evolution. Especially we will explore the characteristics of Class II YSOs in the Orion A star-forming region, and we will compare their propoerties with those of YSOs in other nearby star-forming regoins, such as NGC 1333, Chamaleon I and Taurus.

The entire Orion A molecular cloud appears to be a filament that is roughly perpendicular to the line of sight at a distance of 414 pc \citep{Menten_2007_Oriondistance}. Fits of stellar-evolution models to the color-magnitude diagrams of their member stars give ages of $<$1 Myr in the Orion Nebular Cluster (ONC) \citep{Hillenbrand97_ONCpopulation_age} and $\sim$1 Myr for the southern part of Orion A LDN 1641 (L1641) \citep{galfalk_olofsson_L1641N}. YSO association age is a difficult subject. We will assume that the ages of the young clusters addressed in this paper lie in the range 0.5--5 Myr, and that the color-magnitude diagrams correctly rank the typical ages, with ONC, L1641 and NGC 1333 being the youngest regions, Chamaeleon I being the on the oldest, and Taurus star-forming region in between. 
Orion A differs from other nearby ($\leq$ 500 pc) YSO associations studied with \textit{Spitzer}, in that it is the only one that harbors an OB association. Yet it also includes regions in the L1641 molecular cloud with stellar density and UV intensity as low as in the fiducial low-mass star-formation region, Taurus. 

To reveal the evolutionary characteristics of protoplanetary disks in  Orion A, we use Spitzer-IRS observations of 319 protoplanetary disks in Orion A and follow-up observations of 120 objects with IRTF/SpeX. To explore the effects of environment on disk evolution and planet formation, we compare the structural and dust properties of Orion A disks to those of the much sparser Taurus association (Furlan et al., 2006, 2009, 2011).

This paper is organized as the following ways. In Section 2, we describe the observations and data reduction process.
We report the basic stellar properties and disk properties derived from IRS spectra in Section 3 and Section 4. Then, we compare the observed and derived stellar and disk properties of Orion A young members to those of Taurus members in Section 5. The main purpose of this paper is to make available measurements of the properties we measure from IRS spectra and conduct basic analysis to find how Orion A disks differ from disks in other regions. We will discuss many detailed correlations between properties as key clues of disk evolution in  a forthcoming paper \citep[][in preparation]{kim2015c}. In Section 6, we discuss how the evolution of protoplanetary disks is affected by the ultraviolet radiation field, by comparing disk properties by the distance from $\theta^1$ Ori C, the most luminous O star in ONC. Finally, we summarize our results in Section 7.

\section{OBSERVATIONS AND DATA REDUCTION}
We present 319 objects observed with \textit{Spitzer}/IRS in the Orion A star-forming region in this paper. Some of the objects classified as TDs in section 4.3.2 of this paper were presented in \citet{khkim2013-OriATD}. We described the \textit{Spitzer}/IRS and IRTF/SpeX observations and data reduction process in \citet{khkim2013-OriATD}. Therefore, here, we will describe our observations and data reduction process more briefly. 

\subsection{Spitzer/IRS}
\begin{figure}[t]
\epsscale{0.8}
\plotone{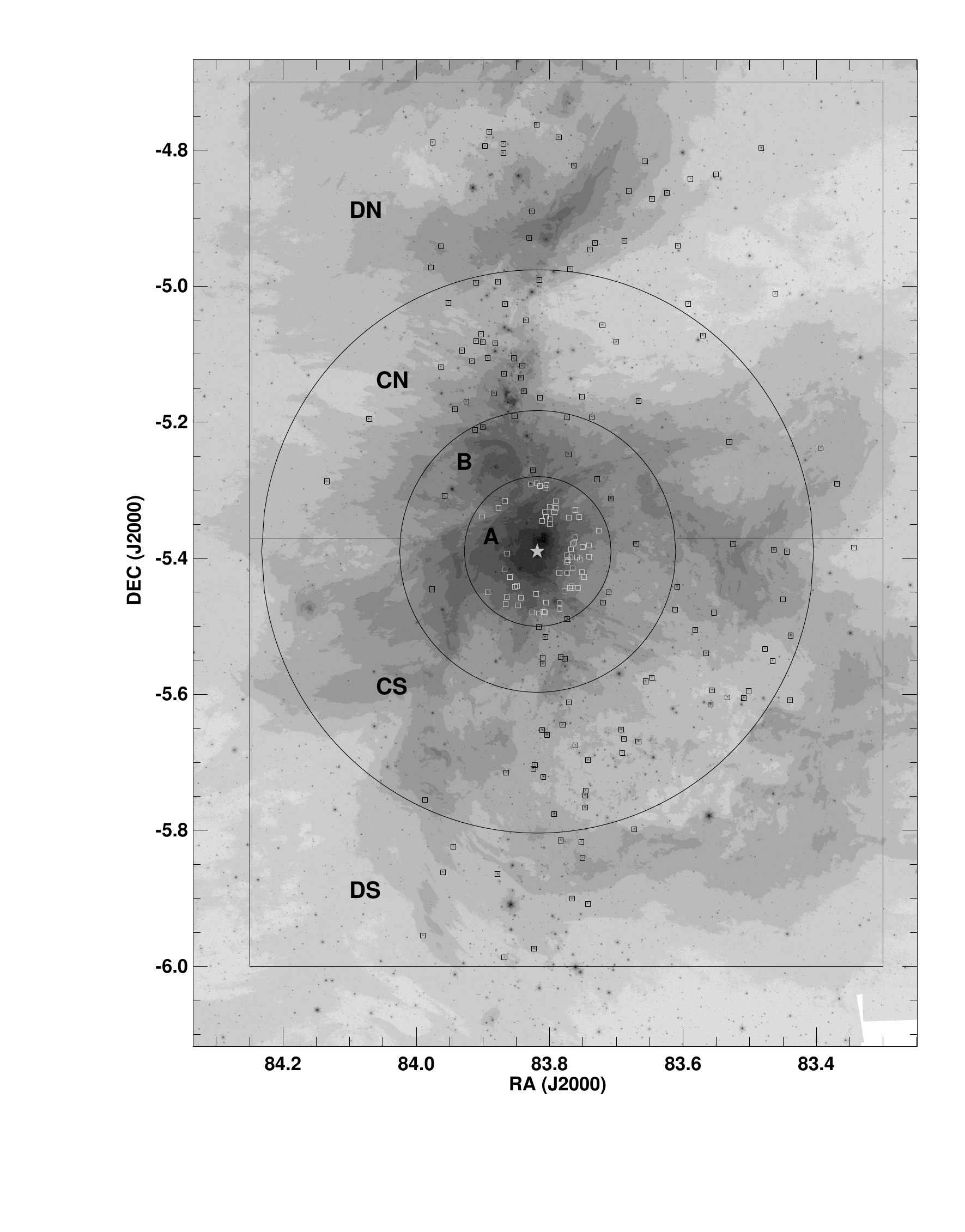}
\caption{IRS targets in ONC plotted over the IRAC/Spitzer ch2 (4.5 $\micron$) image. The squares indicate the Class II objects observed with IRS. The Black squares are for objects observed in the SLLL mode. The gray squares are for objects observed in the SL mode only. Sub-sections of ONC separated by the projected distance ($d$) from $\theta^1$ Ori C (a five point yellow star) are indicated. The criteria of sub-sections: A: $d$ $\leq$ 0.7 pc; B: 0.7 pc $<$ $d$ $\leq$ 1.5 pc; C: 1.5 pc $<$ $d$ $\leq$ 2.5 pc; D: $d$ $>$ 2.5 pc and DEC $>$ -6$^o$. Sub-section B, C, and D are also separated by N and S from the DEC of $\theta^1$ Ori C.   \label{fig-ONC-big}}
\end{figure}

\begin{figure}[t]
\epsscale{0.8}
\plotone{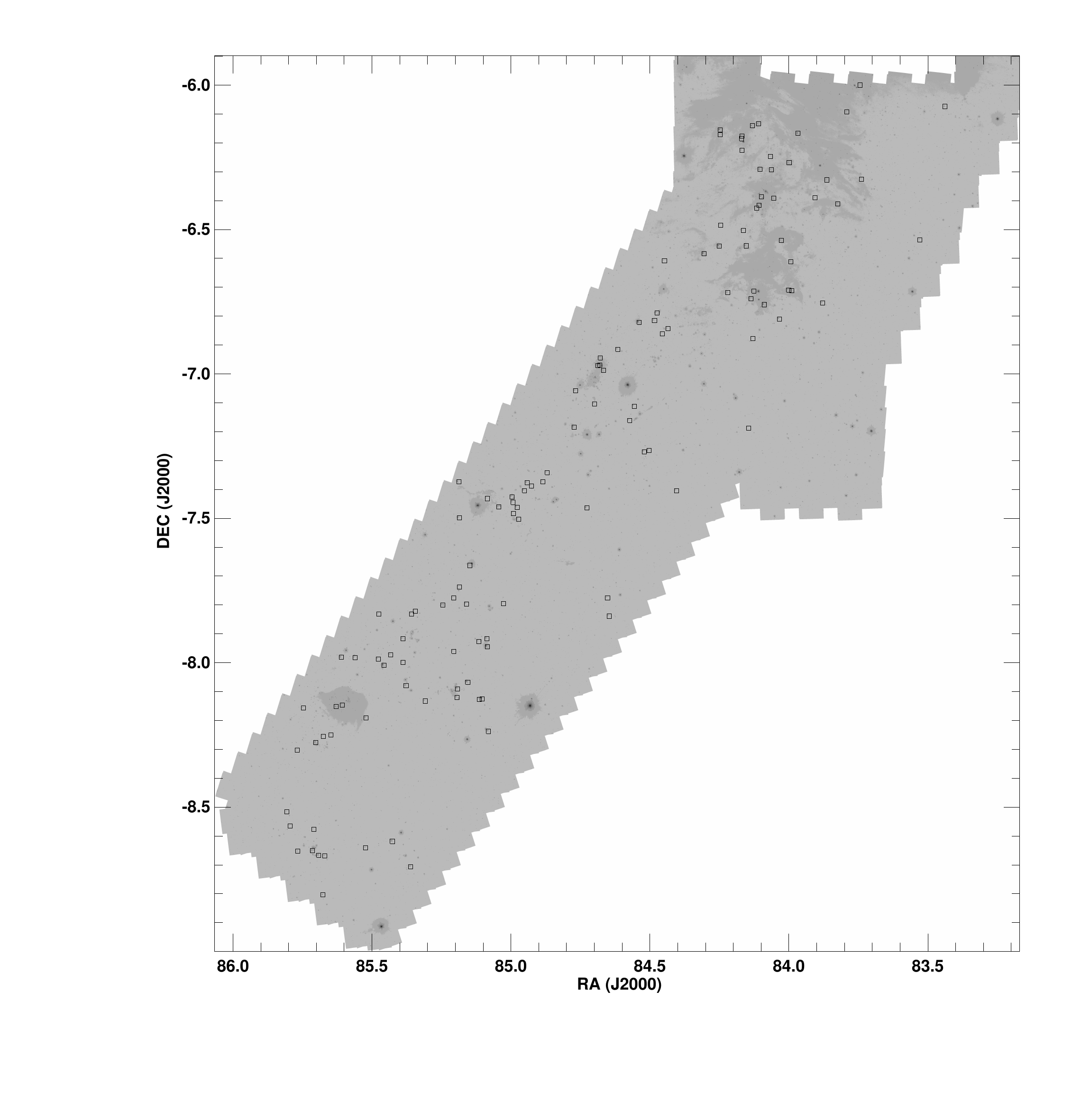}
\caption{IRS targets in L1641 plotted over the IRAC/Spitzer ch2 (4.5 $\micron$) image. The squares indicate the Class II objects observed with IRS. 
\label{fig-L1641-big}}
\end{figure}

\begin{figure}[t]
\epsscale{0.8}
\plotone{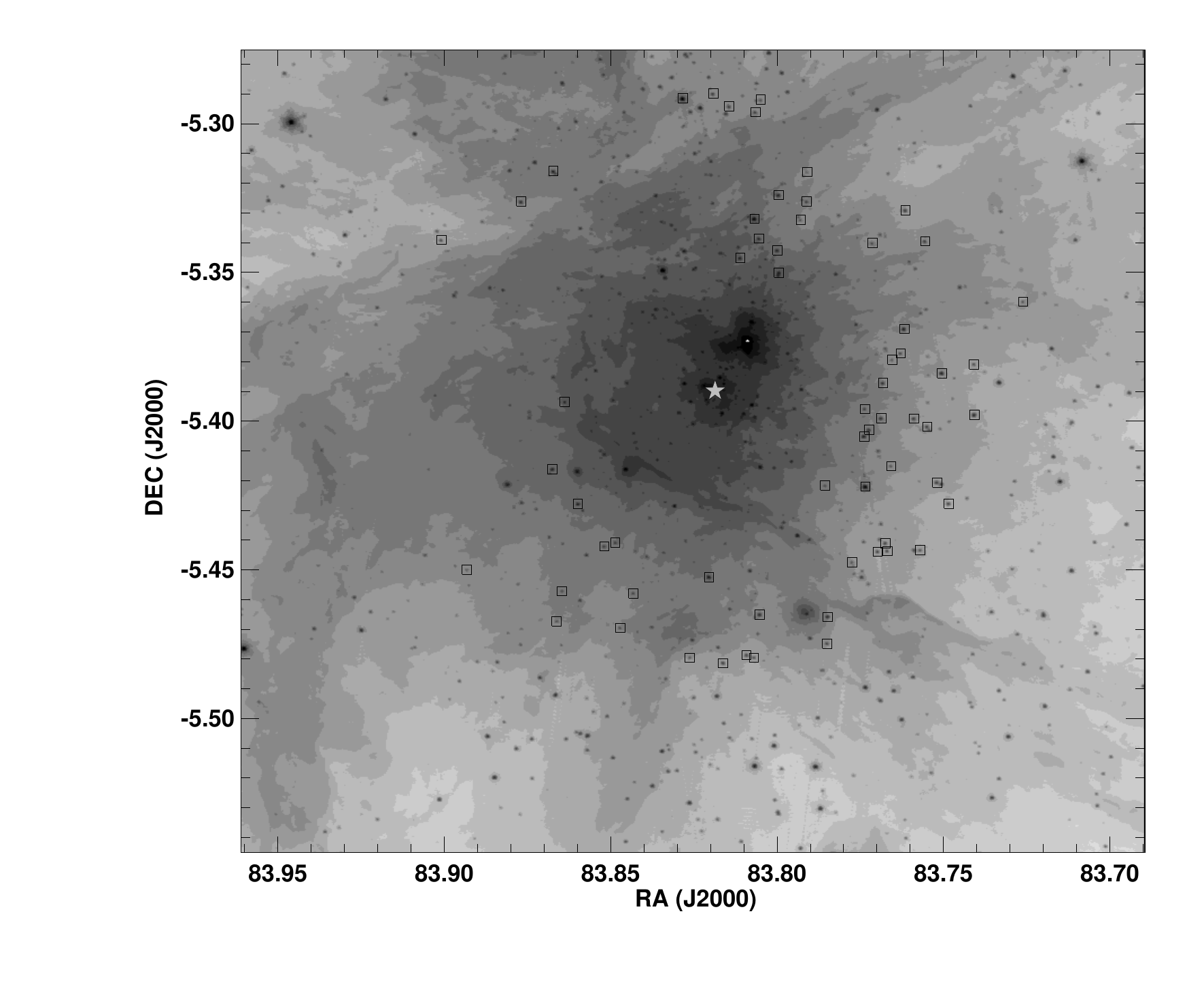}
\caption{Close up of sub-section A (Trapezium region) in Fig~\ref{fig-ONC-big}. $\theta^1$ Ori C is indicated with a five point star.  \label{fig-ONC-sec0}}
\end{figure}

The Orion A objects in this paper were selected based on the identification of young stars with disks by IRAC/2MASS color-color diagrams \citep{Megeath_OriAB_survey_2012}. We observed them using \textit{Spitzer}/IRS during campaigns 36, 39, 40, and 44 between 2006 November and 2007 October.  We include 303 Class II objects which we classified based on their Spitzer 4.5 to 22 $\micron$ spectral index.  Of these, we observed 241 with full IRS wavelength coverage of 5-37 $\micron$ (SL and LL modules). In addition, 62 objects located close to Trapezium region were observed with partial wavelength coverage (5-14 $\micron$; SL only) because the detectors for the longer wavelengths would have been saturated by bright background emission. To this group we added 16 additional objects (5 in ONC; 11 in L1641) which were reclassified as Class II from Class 0/I sources observed in the Orion A protostar survey by \citet[][in preparation]{Poteet2016}. 14 of these 16 were observed during campaigns 39 and 40, but 2 sources were observed in campaign 56. This gives us a total of 319 sources with 62 near the Trapezium, 132 in the ONC, and 125 in L1641 in this paper. Table~\ref{table:IRSlog} gives the IRS observation log, including coordinates, observation date, observing mode, AOR ID, and campaign numbers.  The positions of these IRS targets in ONC, L1641, and the Trapezium are indicated in Figure~\ref{fig-ONC-big}, Figure~\ref{fig-L1641-big}, and Figure~\ref{fig-ONC-sec0}, respectively.

We began with version S15.3 of the basic calibrated data (BCD) product from the Spitzer Science Center IRS pipeline for both SL and LL. As described in detail by \citet{khkim2013-OriATD}, we first fix bad, hot and rogue detector-array pixels before extracting objects from the 2D spectral images. Since the spectral extraction of Orion objects needs very careful background subtraction due to the complex and strong sky emission varying spatially and the high stellar density, we used four different source extraction methods to optimize the rejection of emission from background sky and nearby sources. The first method is to extract the spectral source using an automated extractor (``auto") with two versions of sky subtractions: sky subtraction of the two nods (``off-nod") and sky subtraction of the sky spectrum in each grating order obtained while the target was being observed in the other order (``off-order"). If the sky subtracted 2D images and the reduced spectra have artifacts attributable to sky-subtraction and/or contamination from nearby sources, we re-extract the source by removing the sky with 0th or 1st order polynomial fitting to the background emission on either side of the target. We call this method ``man" because we selected the sky manually. In both ``auto" and ``man", we extracted sources from the uniformly-weighted signal along a tapered column, 3-5 pixels wide using the Spectral Modeling, Analysis and Reduction Tool \citep[SMART;][]{higdon04}. 

The other two methods are to reduce spectra from optimal point-source extraction. One is the AdOpt routine in SMART \citep{Lebouteiller_2010_AdOpt}. It uses an empirical point response function (PRF) and can fit multiple objects along the slit. The other script is OPSE, developed separately from SMART \citep[in preparation]{opse}. OPSE uses an analytical PRF and accounts for pointing errors along the slit. The sky is modeled as a linear function of distance in the 5-9 pixel long extraction window.

We calibrated the flux of spectra by multiplication with relative spectral response functions (RSRFs), which were generated by dividing a template of a calibration star's intrinsic spectrum by the calibrator's spectrum obtained from its raw spectral extraction, extracted in the same way as the target spectrum. Our photometric standards were $\alpha$ Lac (A1V; M. Cohen 2003, private communication) for SL, $\xi$ Dra (K2III; M. Cohen 2003, private communication) for the part of LL at  wavelength less than 32 $\micron$, and Mrk 231 (J.Marshall 2007, private communication) for LL beyond 32 $\micron$.

After comparing the spectra reduced from the different source extraction methods for each object, we selected the final spectrum with the fewest artifacts. To obtain such a spectrum, we combined spectra extracted in different ways to create the spectrum adopted here, if necessary. For example, we use the combined spectrum for OriA-19, which is mixed and matched with SL1 and SL2 from AdOpt and LL1 and LL2 from OPSE. We list the reduction choices for the final selection of spectra in Table~\ref{table:IRSlog}.

Not all of our sources have been cited in the literature. We list some selected identifications of the targets in Table~\ref{table:id}. In many cases, objects are primarily identified by their coordinates, and it is convenient to name the object according to its coordinates. However, to discuss the sources more easily, we cite the the ID number given in Table~\ref{table:IRSlog} throughout this paper (e.g., OriA-123 or 123 if the list of objects is long).

\subsection{IRTF/SpeX}

Of our 257 IRS targets observed in both SL and LL modules in Orion A with \textit{Spitzer}/IRS, we observed 120 at near-IR (0.8--2.4 $\micron$) wavelengths with the medium resolution spectrograph SpeX \citep{SpeX_Rayner}, on the NASA Infrared Telescope Facility (IRTF) on Mauna Kea during the 2010A, 2011A , and 2011B semesters.

We observed all of the SpeX targets with the Short-wavelength Cross-Dispersed mode (SXD). We obtained spectra with various slit widths of 0.3", 0.5" and 0.8" for observations depending on the seeing conditions of each night. We used only the 0.8" slit width for the observations of 2011 February because the weather and seeing were generally poor. 

During the SpeX observations, we discovered that some of the IRS targets are multiple systems. Among the SpeX targets with close companions, the companions of four objects, OriA-38, OriA-173, OriA-208, and OriA-290, were observed separately from the primary targets. For targets with very close neighbors, OriA-4, OriA-26, OriA-47, OriA-98, and OriA-280, we oriented the slit so as to observe both simultaneously and extracted the spectra of primary and secondary separately during the data reduction process. We also found that OriA-22, OriA-125, OriA-154, OriA-213, and OriA-233 were potential binary systems because they appeared elongated and peanut-shaped in the K-band guider images. The components of these systems were not clearly resolved, so we present their SpeX spectra as combined spectrum from both primary and secondary of the binary. Among those close and/or suspicious binaries observed with SpeX, OriA-26 is the only one identified as a spectroscopic binary based on radial velocity measurement by \citet{Tobin_2009_ONC_SB, Tobin_ONCSB_2013ApJ...773...81T}. 

We observed an A0V star, HD 37887, every 30--60 minutes for flux calibration and telluric absorption correction \citep{SpeX_tellcor_Vacca2003}. To allow for changing the slit orientation to avoid including nearby sources, we also observed the calibrator source with the same slit position angle as the target's slit position angle. The IRTF/SpeX observation log is given in Table~\ref{table:SpeXlog}. We reduced our spectra with the Spextool package \citep{SpeXtool}, and flux calibration and telluric absorption correction \citep{SpeX_tellcor_Vacca2003} were done with a spectrum of HD 37887, observed near in time and air mass to each object. In a few exceptional cases, we used a telluric calibration spectrum taken later in time and made a light loss correction to correct a target spectrum for slit losses relative to the flux calibrator, HD 37887.

\subsection{Ancillary Photometry}

We compiled broadband photometry to construct SEDs in addition to IRS and SpeX spectra. We gathered optical photometry from the Naval Observatory Merged Astrometric Dataset (NOMAD: \citet{NOMAD_2005}) with B, V, and R bands at 0.44, 0.55, and 0.64 $\micron$, respectively. For ONC sources, we acquired additional data from \citet{Dario2009_ubvri} at U, B, V, TiO, and I bands centered at 0.347, 0.454, 0.538, 0.6217, 0.862 $\micron$, respectively. More optical and near-IR photometry were added from the Sloan Digital Sky Survey (SDSS) photometric catalog (release 8, \citet{SDSS-dr8}) at u, g, r, i, and z bands centered on 0.31, 0.48, 0.62, 0.76, and 0.91 $\micron$, respectively. We also take I, J, and K bands at 0.8, 1.25, and 2.16 $\micron$, respectively, from DENIS database \citep{DENIS_2005}.  We collected 2MASS J (1.25 $\micron$), H (1.65 $\micron$), and K$_s$ (2.17 $\micron$) magnitudes from 2MASS catalog \citep{2MASS_catalog_2006}.  Most of our targets (except OriA-146, OriA-158, OriA-302, and OriA-306) have 2MASS photometry. We adopt J, H, and K magnitudes from \citet{Prisinzano_2008ApJ...677..401P} for OriA-306.
Additionally, photometry at bands Z (0.88 $\micron$), Y (1.03 $\micron$), J (1.25 $\micron$), H (1.63 $\micron$), and K (2.20 $\micron$)  from the UKIRT Infrared Deep Sky Survy (UKIDSS DR8, \citet{UKIDSS-dr8}) were collected. 
IRAC (3.6, 4.5, 5.8, and 8.0 $\micron$) and MIPS (24 $\micron$)  photometry of our sources had been acquired prior to the IRS observations \citep{Megeath_OriAB_survey_2012}, and we included them. 
We also include near- to mid-IR observations at 3.4, 4.6, 12, and 22 $\micron$ from the Wide-field Infrared Survey Explorer (WISE) all-sky data release \citep{WISE2012}. It is recommended to use WISE photometry with caution due to the source extraction methods used in the pipeline \citep[e.g.,][]{Koenig_Leisawitz_WISEcaution_2014}. We include WISE photometry only if a corresponding observation has a ``ex" tag value of zero, i.e., not an extended source, to avoid including neighboring objects because WISE has relatively coarse resolution. We find these criteria eliminate WISE data which show spurious jump compared to IRAC and MIPS photometry in their SEDs.

In case there are multiple (mostly two or three) sets of UKIDSS and SDSS for an object within a distance of 3 arcsec from our target position, we include the combined photometric points of multiple sets for the corresponding SEDs, as we cannot ignore the possibility of multiple system. \cite{Tobin_2009_ONC_SB, Tobin_ONCSB_2013ApJ...773...81T} identified 89 spectroscopic binaries (SBs) in the ONC with high/moderate confidence. Among those 89 SBs, 13 objects match with our objects in ONC. The corresponding objects are OriA-5, 26, 37, 45, 60, 88, 91, 111, 142, 148, 180, 189, and 205. Five of them are double-line binaries (OriA-37, 45, 60, 91, and 111), and the remaining sources are single-lined binaries. We have checked if 13 SBs have multiple UKIDSS and/or SDSS photometry. OriA-26 is the only object with multiple sets both by SDSS and UKIDSS. OriA-37 and OriA-60 have multiple sets of UKIDSS.  
However, we do not discuss multiplicity further here. A comprehensive discussion of the multiplicity of these objects awaits higher spatial and spectral resolution observations.

\section{BASIC CHARACTERISTICS OF THE STARS}

\subsection{Spectral Types and Effective Temperatures}

For the most part, we gathered the spectral type (SpT) of our targets from the literature \citep{Hillenbrand97_ONCpopulation_age, RHS2000,Allen_thesis_L1641,Dario2009_ubvri,Fang2009,Parihar2009_ONC_variables_SpT4V1498Ori, Hsu_L1641SpT_2012ApJ...752...59H}. We also added unpublished results kindly provided by John Tobin and Jes\'{u}s Hern\'{a}ndez. 
We utilized SpeX spectra to determine spectral types for objects without spectral type information from the literature.  As we have described \citep{khkim2013-OriATD},  to determine spectral type from SpeX spectra, we compared absorption features of Na I, Al I, Mg I, Ca I and CO in source spectra to those of template spectra \citep{IRTF_spectral_Library_2009,IRTF_SpeX_MLTdwarf_Cushing_2005}. In addition, we have updated spectral types adopted in \cite{khkim2013-OriATD} with spectral types for L1641 young stars from \citet{Hsu_L1641SpT_2012ApJ...752...59H}. Spectral types of OriA-230, OriA-271, and OriA-294 derived previously from SpeX spectra (M0.0, M2.0, and M2.5 respectively) were replaced with the spectral types of \citet{Hsu_L1641SpT_2012ApJ...752...59H} (M3.5, M0.2, and M2.0 respectively). These modest differences in spectral types between our determination and  \citet{Hsu_L1641SpT_2012ApJ...752...59H}  ($\Delta$(SpT) = 3.5, 1.8, and 0.5, respectively) prove that our spectral type derivation using SpeX spectra is sufficient to narrow the spectral type range down to 2--3 sub-types. Hence, we still adopt spectral types derived from SpeX spectra for 7 objects (OriA-11, 16, 29, 34, 249, and 302) because there is no alternative in the literature for them. The spectral types reported in the literature and the adopted spectral types are listed in Table~\ref{table:SpTypeAv}. Most cases, the reported spectral types for an object agree in few sub-types. Less than 10$\%$ of the objects with known spectral types have broad range of reported spectral types.

\begin{figure}[!htbp]
\epsscale{0.8}
\plotone{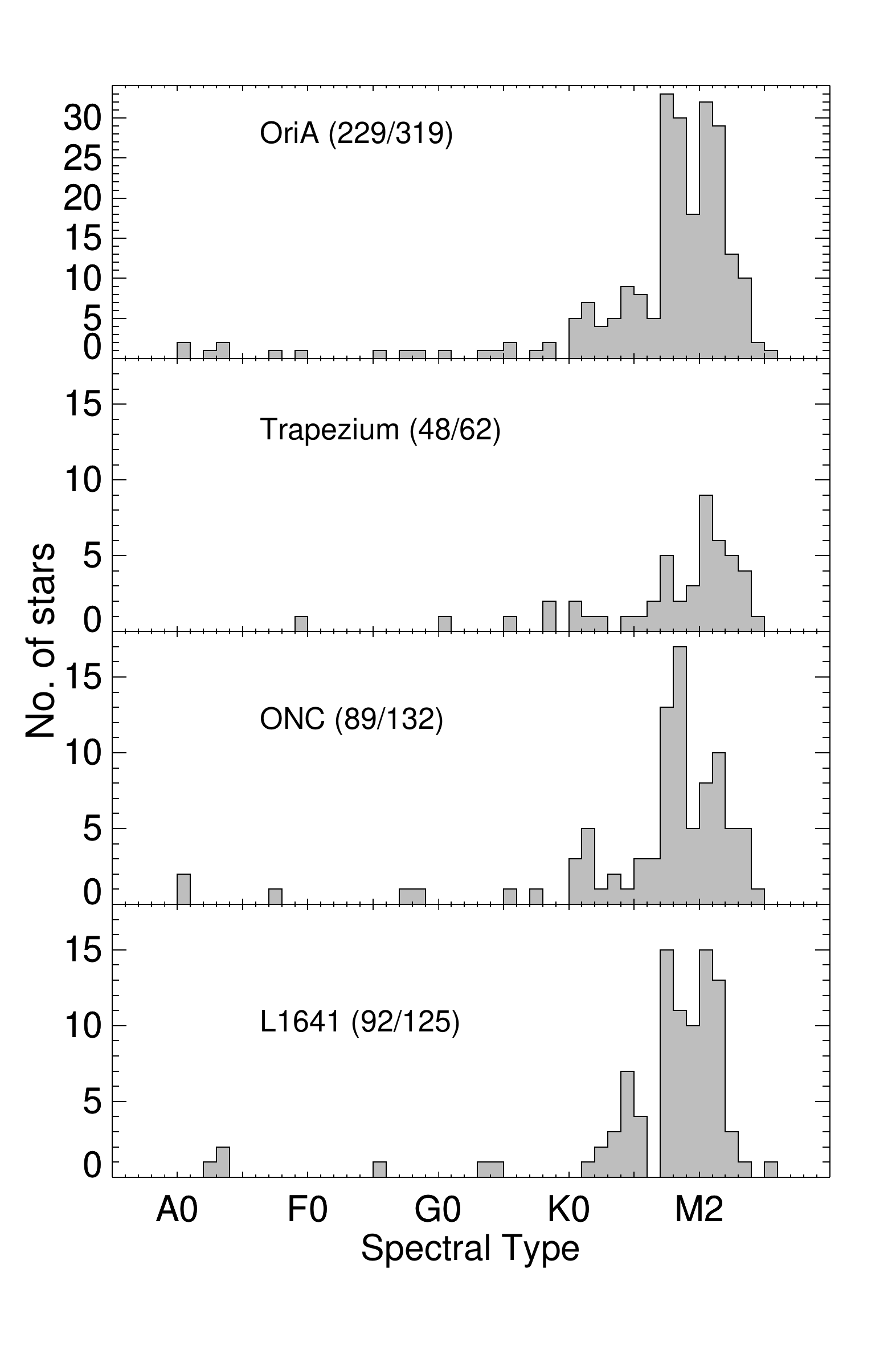}
\caption{Spectral type distribution of Class II objects in Orion A that have available spectral types among those observed with IRS. (note: OriA-227 is classified as K7.9 and it is the only one we adopted as K8. We merged bin of K8 into the bin of M0 in order to plot the histogram in the same binning scheme in literature.)   \label{fig-OriA-knownSpT}}
\end{figure}

The spectral type distribution for objects with available spectral type information is shown in Figure~\ref{fig-OriA-knownSpT}. The distribution shape of the histograms generally agree with the previous studies of objects in ONC~\citep{Hillenbrand_2013_ONC} and in L1641~\citep{Fang_L1641_2013ApJS..207....5F, Hsu_2013_IMFOriA}, even though we do not have complete spectral type information. The total number of objects with known spectral type is 229: 48 objects in Trapezium with partial IRS spectra, and 181 objects for which we have the full range of IRS spectra in ONC (89 objects) and L1641 (92 objects). The fractions of our IRS targets for which we have spectral types are 77$\%$, 67$\%$, and 74$\%$ for the Trapezium, ONC and L1641, respectively. The median spectral type of Class II objects in Orion A is M0: M1 for objects in the Trapezium; M0 for objects in the ONC; and M1 for objects in L1641 (see Figure~\ref{fig-OriA-knownSpT}).

We adopted the effective temperature ($T_{eff}$) and bolometric correction (BC) scale for pre-main sequence (MS) stars derived by \citet{Pecaut_TeffSpT_2013ApJS..208....9P}. They found that pre-MS $T_{eff}$ are $\sim$250 K cooler than MS $T_{eff}$ for G5 through K6 types while other spectral types $T_{eff}$ agree in $\sim$100 K between pre-MS and dwarfs. The accounted systematic plus statistical uncertainty of pre-MS $T_{eff}$ is less than 1$\%$ of $T_{eff}$ \citep{Pecaut_TeffSpT_2013ApJS..208....9P}. 
These new scales for pre-MS stars are fully available only for spectral types F0 to M5 in Table 5 in \citet{Pecaut_TeffSpT_2013ApJS..208....9P}. For spectral types earlier than F0 and later than M5, we consulted with Pecaut $\&$ Mamajek (private communication) to find the best way, and we adopted the following combination. For spectral types earlier than F0, we took the $T_{eff}$ and BC scales for dwarfs in Table 5 in it. For spectral types later than M5, we adopted the $T_{eff}$ of dwarfs for spectral types later than M5, and we used the BC scale of pre-MS. 
For the objects of unknown spectral types, we adopted an effective temperature of 3770 K, which is the temperature of the mean spectral type (M0) of Class II sources with known SpT in Orion A.
The uncertainty of $T_{eff}$ lies within the typical uncertainty of spectral types, which can be translated into about 345 K as a median temperature difference between few ($\sim$3) sub-spectral type differences. We adopt it as a typical uncertainty of $T_{eff}$.

\subsection{Extinction Correction}

\begin{figure}[!htbp]
\epsscale{1}
\plotone{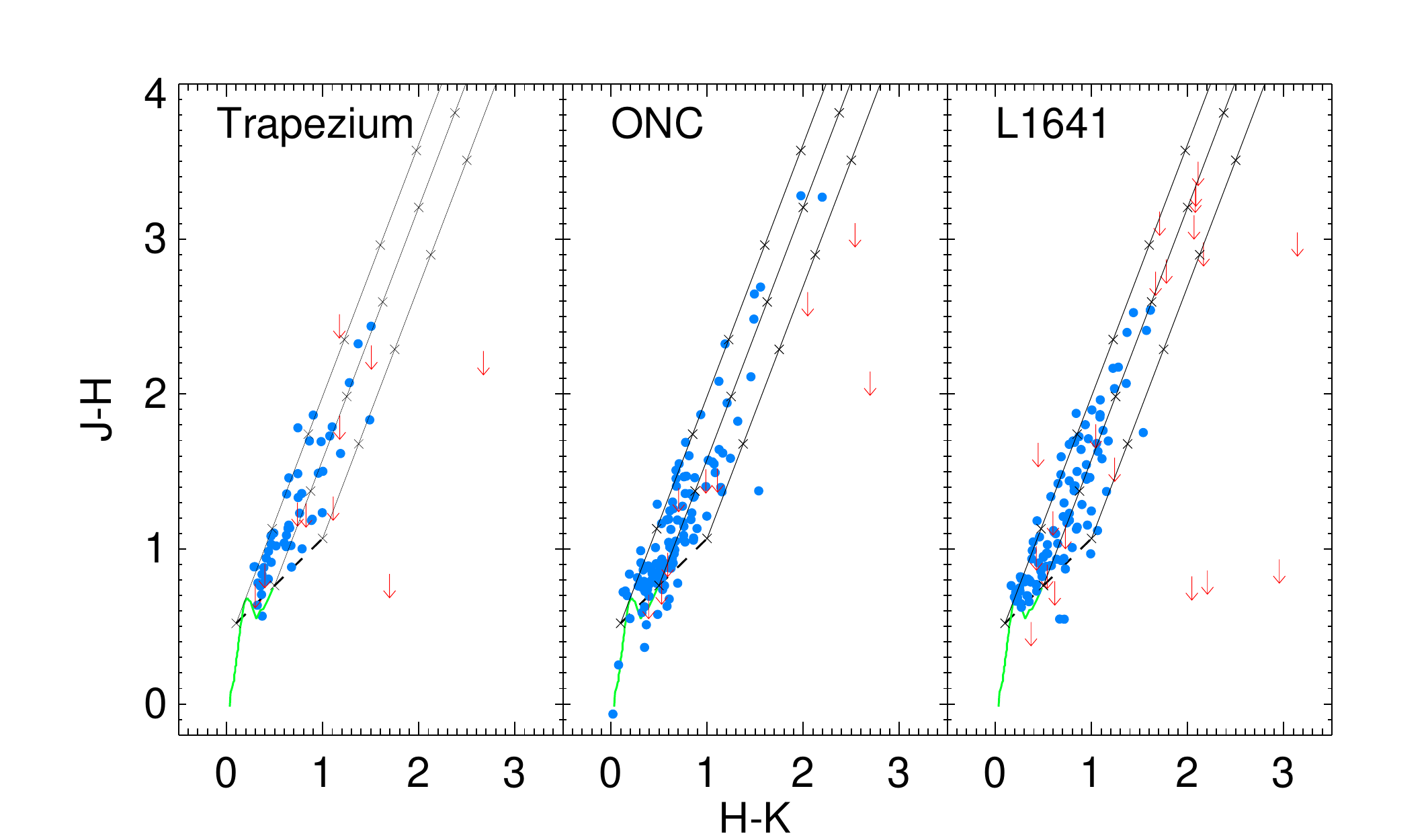}
\caption{$J-H$ versus $H-K$ color-color diagram of Class II objets observed with IRS in the sub-regions of Orion A: Trapezium (the left panel), ONC (the middle panel), and L1641 (the right panel). $J$, $H$, and $K$ photometry is from 2MASS. The CTTS locus is indicated with the dashed line. The colors of main-sequence giants and dwarf (Bessell $\&$ Brett 1998) are also plotted with green solid line. The solid lines started from the CTTS locus indicate the increasing extinction. The increments of $A_{V}$ are denoted by crosses by the interval of $A_{V}$=5. The downward arrows are used for objects which their 2MASS photometry have any bad flags.  \label{fig-JHHK-OriA}}
\end{figure}

We correct for extinction toward our sources to minimize the extinction effects on disk classification and interpretation of our data. The way of estimating visual extinction ($A_{V}$) toward our targets is described in detail in \citet{khkim2013-OriATD}. Basically, we use the relationship 
\begin{equation}\label{eq:Av}
{A_V}={\frac{\frac{A_V}{A_{\lambda_2}}}{\frac{A_{\lambda_1}}{A_{\lambda_2}}-1}}\times{E({\lambda_1}-{\lambda_2})},
\end{equation}

where $E({\lambda_1}-{\lambda_2})$ is the color excess between two wavelengths ($\lambda_1$ and $\lambda_2$) which is equal to ${([{\lambda_1}]-[{\lambda_2}])_{obs}}-{([{\lambda_1}]-[{\lambda_2}])_{int}}$.
To derive the color excess, $E({\lambda_1}-{\lambda_2})$, we take J, H, and K bands from 2MASS and I and J bands from DENIS and adopt  $I-J$, $J-H$, and $H-K$ photospheric colors from \citet{Pecaut_TeffSpT_2013ApJS..208....9P}. For sources without available spectral type information, we adopt  an intrinsic color from $J-H$ and $H-K$ of the Classical T Tauri Star (CTTS) locus of colors from \citet{cttslocus97}. The CTTS locus is indicated over the color-color diagram of our sample using 2MASS photometry in Figure~\ref{fig-JHHK-OriA}.  In Equation~\ref{eq:Av}, $A_{\lambda}$ is the extinction at a wavelength of $\lambda$. We assume the optical total-to-selective extinction ratio, $R_V$, to be 5 because (1) $R_V$ values of dense clouds are usually between 4 and 6 \citep{mathis90} and (2)  measures of  $R_{V}$ for lines of sight in Orion have confirmed that $R_{V}\sim$5 \citep{CardelliClaytonMathis_1989ApJ...345..245C}.
Therefore, we make use of the extinction curve with $R_V$ = 5.0  from \citet{mathis90} to obtain $A_\lambda$ at the wavelengths of the I, J, H, and K band.

We employ extinction laws depending on the $A_V$ values inferred from Equation~\ref{eq:Av}. We take the \citet{mathis90} extinction curve with $R_V$ = 5.0 in the case of $A_V$ $<$ 3. If $A_V$ $>$ 3, we used the extinction curve corrected for standard-star photospheric SiO absorption near 8 $\micron$ (S. Fogerty, in private commnunication) from the empirical extinction curves of \citet{McClure2009} which present two composite extinction curves, one for $3 < A_{V} < 8 $ and one for $A_{V} > 8$.

To select a final $A_V$ for an object, we examine the SEDs which are extinction corrected with the derived $A_V$s from each intrinsic color choice. We make a judgment based on freedom from artifacts of the correction (e.g. artificial CO$_2$ ice features or spurious structure in the silicate features) and good agreement with the photospheric spectrum of the star's type, to find the most reasonable $A_V$. In spite of our best effort of judgment, we have some ($\sim$10$\%$) objects which show modest or strong UV/optical excess in the extinction corrected SEDs. There are several possible explanations for the UV/optical excess, including contributions from background nebulosity,  jets, over-correction of extinction, and/or less accurate photometry, if not all.  The over-correction of reddening seems to be frequent in case the spectral type of an object is not known. We speculate that the following objects are over-corrected: OriA-31, 54, 65, 162, 207, 251, 255, and 297 due to the limit of spectral type information; and OriA-221 and 231 due to some other reason, since none of the various values of $A_V$s are reasonable. We find frequently that photometry data from NOMAD show a sudden increment compared to other photometry, and the objects suffering this problem are OriA-30, 58, 66, 84, 97, 103, 127, 128, 183, 190, 192, 194, 206, and 222. They are usually located in nebulous regions. Another group for which we consider nebulosity and jets as possible contributors are OriA-3, 30, 37, 78, 81, 86, 90, 91, 122, 125, 217, 242, and 259. Among them, OriA-37, OriA-90, and OriA-91 are related with HH 505, HH 885, and HH 888, respectively. Even though we do not know the spectral type of OriA-207, it could be an edge-on disk because it shows up as an hourglass shape in the SDSS9 scattered light image. As can be seen from the preceding discussion, there is difficulty in measuring accurate extinction. We present final $A_V$s and undertainties along with the $A_V$ method selected for each object: $I-J$, $J-H$, and $H-K$ in Table~\ref{table:SpTypeAv}. 

The estimation of extinction bears several sources of uncertainty: the uncertainties of photometry, intrinsic color, spectral type, and extinction curve. We do not account for all of those uncertainties because those uncertainties are not larger than the range of $A_V$ values estimated through different choice of color excess described above. We indicate the range of $A_V$s for each object with the finally selected $A_V$ in Table~\ref{table:SpTypeAv}. The range of $A_V$s for most objects lies in few magnitudes: the average offset of the smallest from the adopted $A_V$ is 1.65 for objects which have the lower offset is less than 3 magnitudes (77$\%$); the average difference between the highest and the adopted $A_V$ is 1.67 for objects which have the offset is less than 3 magnitudes (43$\%$). Only 2$\%$ of our sample have the lower or upper offset greater than 10 magnitude from the adopted $A_V$.

\begin{figure}[!htbp]
\epsscale{0.65}
\plotone{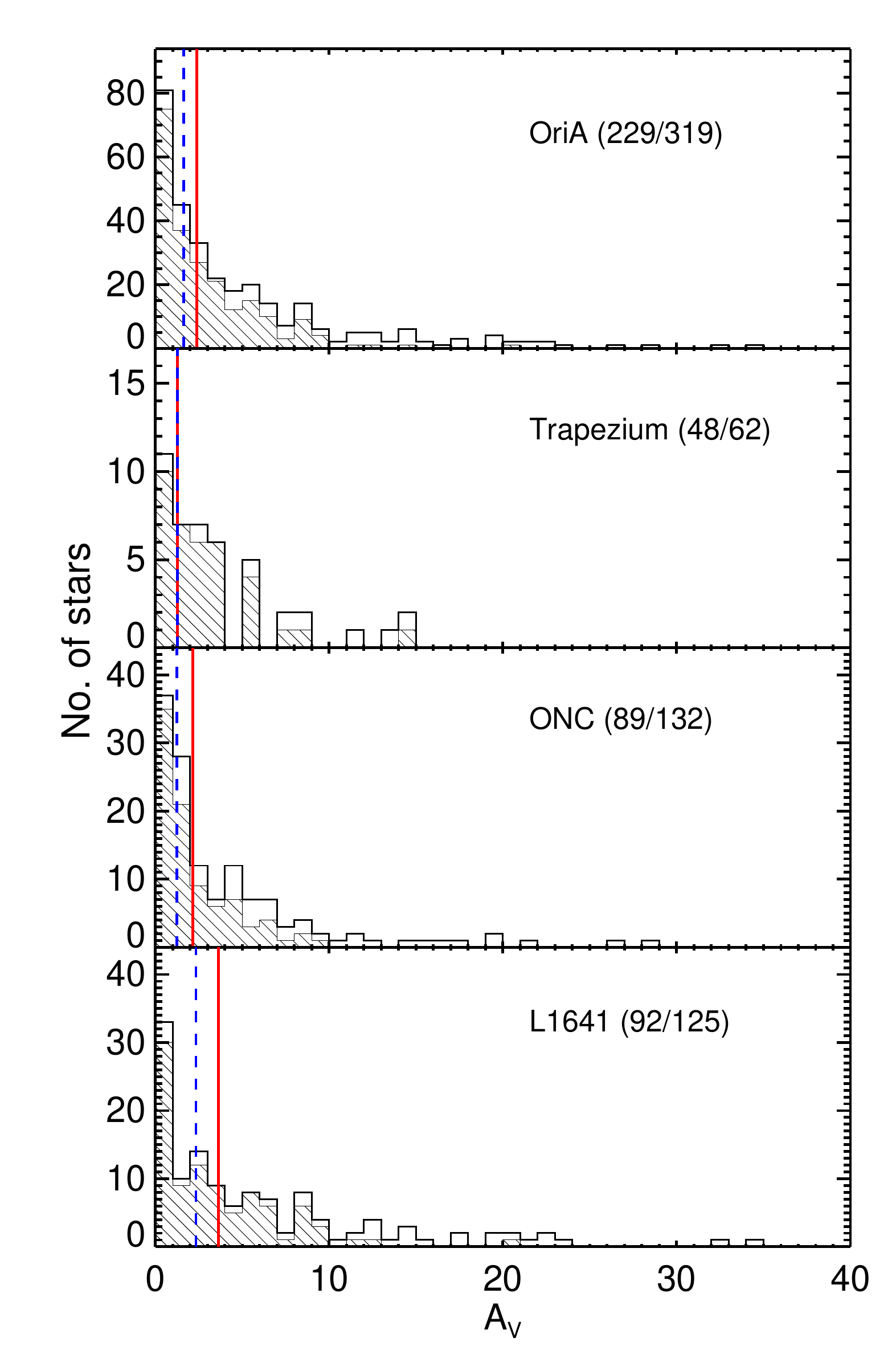}
\caption{Distribution of visual extinction, $A_V$, for 319 Class II disks in Orion A observed with IRS. Hashed lines indicate the $A_V$ distribution of the objects with known spectral type information. The solid vertical lines indicate the median $A_V$ of all objects. The dashed lines are for the median $A_V$ of the objects with available spectra types. The numbers in each pannels are for the number of objects with available spectral types and the number of objects without regard to the availability of spectral type.  \label{fig-OriA-Av}}
\end{figure}

The distributions of the visual extinctions according to the sub-regions of Orion A are shown in Figure~\ref{fig-OriA-Av}. When taking only objects with known spectral types, the median $A_{V}$ of L1641 ($A_V$=2.34) is higher than that of the Trapezium ($A_V$=1.26) or ONC ($A_V$=1.25). We performed Kolmogorov-Smirnov (K-S) tests on the distributions of $A_{V}$ of pairs of sub-regions and list the results in Table~\ref{table:KStest-Av}. From the K-S test, where $D$ is the maximum difference between the empirical distribution function of two groups and $p$ indicates the probability that there is no significant difference between the distributions, we confirm that the difference of $A_{V}$ distributions by sub-regions is real and significant. The differences are caused by the following mechanisms: (1) less extinction for the objects located in a cavity in the Trapezium and the ONC due to the strong UV field from OB stars in the Trapezium, which has blown out small dust grains~\citep{Hillenbrand97_ONCpopulation_age}; (2) large extinction for the objects located inside of several local dark clouds in L1641. We verified that the objects with the highest visual extinction values reside in denser clouds within this region. The trend of the $A_V$ medians in the Trapezium, ONC, and L1641 imply that most of the extinction is local, within the star-forming region and not along the largest part of the line of sight between Orion and the Earth (observers).

\subsection{Spectral Energy Distributions}

We constructed SEDs of our sample with the broad band photometry that we assembled in Section 2.3 and SpeX and IRS spectra.
We plot SEDs of 257 Class II objects observed in full IRS wavelength coverage from 5.2--37 $\micron$ in Figure~\ref{fig-SLLL-SED}.  We only present the SEDs of 44 of 62 objects  observed with partial IRS wavelength coverage of 5.2--14 $\micron$  in Figure~\ref{fig-SLonly-SED}, because there are 18 objects in the Trapezium for which IRS spectral data was saturated at some wavelengths and hence unusable. 

There are ten spectra among 257 observed in full IRS wavelength coverage that we do not include our general analysis. We summarize these ten objects below. 
	\begin{itemize}
		\item OriA-7, 88, 136, 148: They have incomplete spectra due to saturation.
		\item OriA-247, 261, 296: They vary substantially over time.
		\item OriA-216:  We identify OriA-216 as a galaxy because the PAH features are located at redshifted wavelengths with z=0.35 (Figure~\ref{fig-216galaxy}). It is also classified as a galaxy by SDSS classification ~\citep{SDSS-dr8}.
		\item OriA-208:  The SED of it includes interesting information: the system is revealed as a binary from SpeX observation, and SDSS and UKIDSS photometry agree with our SpeX spectra of the fainter object. It was not possible to distinguish two sources from IRS because the separation is about 1$\sim$2 arcsec.
		\item OriA-95: it is not a galaxy, but it has very uncertain extinction correction.
	\end{itemize}
 We do not include these 10 objects for the analysis in the present work, except OriA-88 in the analysis related to TDs.

There are many interesting individual objects in our survey. For example, we speculate possible variable sources based on the flux density disagreement of IRS spectrum with the IRAC and WISE photometry. The possible variability candidates are OriA-6, 9, 60, 61, 117, 166, 189, 190, 197, 202, 210, 212, 231, 244, 247, 256, 258, 261, 276, 277, 289, 301, 304, 316, and 318. We also find several objects with possible outer disk truncation based on the steeply decreasing emission beyond 20 $\micron$: OriA-15, 45, 98, 173, 180, and 225. 

A detailed description and discussion of individual objects is deferred to later papers.

\subsection{Stellar Luminosity and Stellar Mass }
The photospheric emission indicated with long dash-dot line in each SED in Figure~\ref{fig-SLLL-SED} and Figure~\ref{fig-SLonly-SED} was derived from the intrinsic colors of \citet{Pecaut_TeffSpT_2013ApJS..208....9P} at temperature $T_{eff}$, scaled to match the de-reddened 2MASS J band photometry (see SEDs in Figures~\ref{fig-SLLL-SED} and \ref{fig-SLonly-SED}). From the scaling factor, which is a solid angle applied to the photosphere model to get the observed flux density at J band and the assumed distance to Orion A ($d$=414 pc), we estimate the stellar radius ($R_\star$).  We derived the stellar luminosity ($L_\star$) of each object from $T_{eff}$ and the stellar radius ($R_{\star}$). The adopted $T_{eff}$ and estimated $L_\star$ are listed in Table~\ref{table:basicprop}. We display the distribution of $L_\star$ of 291 objects from the samples in the ONC and L1641 in Figure~\ref{fig-Lstar-OriA}. Performing a K-S test on the luminosity distributions of the ONC and L1641 regions, we conclude there is no significant difference in the distribution of the luminosity of both regions.

 To infer the stellar mass ($M_{\star}$), we plot the derived $L_{\star}$ and $T_{eff}$ of our targets on evolutionary tracks. In this paper, we use the Siess PMS evolutionary tracks \citep{siess2000}. The adopted $M_\star$ in Table~\ref{table:basicprop} are the average $M_{\star}$ read from the tracks of various metallicities. We take the standard deviation of $M_\star$s measured in various metalicity conditions as the uncertainty of $M_\star$. Figure~\ref{fig-HRD-OriA} shows the objects with known spectral types in an H-R diagram along with the solar-metallicity (Z=0.02) evolutionary tracks as an example. We compared $M_\star$ distribution for the Class II disks in ONC and L1641 in Figure~\ref{fig-Mstar-OriA}. There is no significant difference in the mass distribution of the two regions.
 
Since $L_\star$ and $M_\star$ distributions between ONC and L1641 are not statistically very different, we will assume that any statistical differences between the two populations seen in further analysis with disk properties or star/disk accretion properties originate in disk properties, not in their stellar properties.

\section{DISK PROPERTIES}

\subsection{Spectral Index}
To deduce characteristics of disks, we measure the spectral indices defined as
\begin{equation}
n_{\lambda_{1} - \lambda_{2}} = \frac{\log(\lambda_{2} F_{\lambda_{2}})-\log(\lambda_{1} F_{\lambda_{1}})}{\log(\lambda_{2})-\log(\lambda_{1})}~~,
\end{equation}
 where $\lambda_{1}$ and $\lambda_{2}$ are the anchor wavelengths.
We use spectral indices to infer the radial and vertical distribution of dust. Spectral slopes between near-IR and mid-IR such as $n_{K-25}$, $n_{5-12}$, and $n_{12-20}$ are the spectral indices commonly used for disk classification.  We will discuss our usage of $n_{K-25}$, $n_{5-12}$, and $n_{12-20}$ for disk classification in Section 4.3.

The indices most commonly used to infer the degree of dust settling (vertical structure) and inner/outer disk truncation (radial structure) are $n_{K-6}$, $n_{6-13}$, and $n_{13-31}$~\citep[e.g.][]{McClure2010, manoj2011, Furlan2011Taurus, Arnold2012}.  Using self-consistent disk models of T Tauri stars, \citet{dalessio06} discussed the relation between the depletion ($\epsilon$) of the small grains relative to the standard dust-to-gas mass ratio  and the spatial distribution of the emergent flux, as well as the dependence of $\epsilon$ on other properties such as the radial structure and grains size distributions.  The emitting regions, corresponding to fluxes emitted in certain wavelengths, change depending on dust grain sizes and degree of settling. For example, in a case of a model with ISM dust in the least settled disks with the largest depletion parameter ($\epsilon$ = 1, i.e. no depletion of small grains in the upper disk layers), most ($>$ 90 $\%$) of the emergent flux at 6, 13, and 31 $\mu$m come from less than 0.3, 10, and 75 AU from the host star, respectively, while the most emitted fluxes at 6, 13, and 31 $\mu$m are from less than 0.3, 5, and 10 AU, in the case of the most settled disk ($\epsilon$=0.001, i.e. a factor of 1000 depletion of small grains in the upper disk layers) \citep[][Fig. 9]{dalessio06}.  The stellar temperature also affects the distance of the emitting regions from a host star. Stars with higher effective temperature will have a 6 $\mu$m flux disk-emitting region beyond 0.3 AU, while this region will be closer to the host star if the star is colder than the effective temperature (4000K) of a typical T Tauri star. Regardless of the host-star's temperature or the degree of disk settling, the wavelengths of 6, 13, and 31 $\mu$m are between the silicate emission features around 10 and 20 $\mu$m. Therefore, the spectral indices between two adjacent wavelength regions of them ($n_{6-13}$ and $n_{13-31}$) are considered continuum spectral indices that can probe different regions of disks. In addition, we make use of $n_{K-6}$ in order to measure characteristics of the innermost parts of disks.

The anchor positions for measuring spectral indices have been selected to be useful in comparison of radial properties of disks from various previous studies \citep{furlan06, McClure2010, manoj2011, Arnold2012}. At each anchor wavelength, we averaged flux values from a small wavelength region to derive a representative flux value; so, for the 5 $\micron$ flux (``5") we used the wavelength region 5.2-5.54 $\micron$; for ``6" we used 5.4--6.0 $\micron$; for ``12" we used 12.7--13.1 $\micron$; for ``13" we used 12.8--14.0 $\micron$; for ``20" we used 19.7--19.95 $\micron$; for ``25" we used 24.5--25.5 $\micron$; for ``31" we used 30.3--31.0 $\micron$. For ``K" we adopted the 2MASS K$_s$ (K$_s$ band: 2.00--2.31 $\micron$.)

Spectral indices measured with any combinations of two wavelengths, $n_{K-6}$, $n_{6-13}$, $n_{13-31}$, $n_{5-12}$, $n_{12-20}$, and $n_{K-25}$  are listed in Table~\ref{table:sptidx}. We report $n_{20-31}$, which is used as an index of outer disk truncation/evaporation in Section 6, in Table~\ref{table:table-n2031-Distance}. The uncertainties reported with these spectral indices in Table~\ref{table:sptidx} and~\ref{table:table-n2031-Distance} are the propagated uncertainties from their spectra without including other uncertainties of determination of effective temperatures and extinction.

\subsection{Parameters of the Degree of Dust Processing}
Along with the continuum spectral indices, which are sensitive to the dust distribution within the disk, we can utilize some parameters that can be extracted from silicate features at 10, 20, and/or 33 $\micron$\footnote{We note that the crystalline olivine features at 33 $\micron$ is not reliable for most objects because there were many noisy pixels over this wavelength range. Therefore, we do not use EW33 for analysis in this work.} to acquire clues of how much small dust grains are processed.

A parameter commonly used for understanding the strength of absorption or emission features is the equivalent width. The equivalent width for the silicate emission features is a measure of optically thin emission per unit area of underlying optically thick continuum:
\begin{equation} \label{eq:EW10}
EW(\lambda)=\ \int_{\lambda_{1}}^{\lambda_{2}} {\frac{F_{\lambda}-F_{\lambda,cont}}{F_{\lambda,cont}}} \,d\lambda\,~~,
\end{equation}
where $F_{\lambda,cont}$ is the continuum emission determined from a polynomial fit to certain wavelength regions where the silicate emission features do not exist.

In the limit of small silicate feature optical depth, the equivalent width is proportional to the optical depth itself, or similarly the column density of dust in the inner disk \citep{dmw_2009}.
We compute EW10 and EW20 for amorphous silicate emission features around 10 $\micron$ and 20 $\micron$. The wavelength ranges for integrating are 8--13 $\micron$ and 16--28 $\micron$ for 10 and 20 $\micron$ features, respectively.  We list EW10 and EW20 in Table~\ref{table:ew1020}.

Another useful parameter is the integrated flux of the feature, $F({\lambda})$.
\begin{equation} \label{eq:F10}
F({\lambda})=\ \int_{\lambda_{1}}^{\lambda_{2}} ({F_{\lambda}-F_{\lambda,cont}}) \,d\lambda\,~~,
\end{equation}
$F({\lambda})$, the integrated flux after continuum subtraction, is a probe of the mass of dust lying in the uppermost surface of the disk, while EW($\lambda$) indicates the relative strength of the optically thin emission feature to the continuum emission from optically thick dust in the disk. We calculate $F{10}$ and $F{20}$ in the same wavelength limits used for the measurement of EW10 and EW20.

A parameter generally adopted as an index of the degree of dust processing is the ratio of  the continuum-subtracted and normalized flux at 11.3 $\micron$ and 9.8 $\micron$. This index, $F_{11.3}/F_{9.8}$, has been used to probe the degree of grain growth and crystallinity in the inner (1--2 AU) parts of disks \citep{Przygodda+2003,vanBoekel+2005,Kessler-Silacci+2006ApJ,Honda+2006ApJ,Bouwman+2008ApJ,Olofsson+2009A&A}. When grains are still unprocessed and in their pristine state, amorphous silicate grains show a broad emission peak at 9.8 $\micron$. As small grains become crystallized, a strong feature at 11.2 $\micron$, due to forsterite, appears. Also, as grains grow, the broad amorphous silicate feature becomes flat-topped. Therefore, $F_{11.3}/F_{9.8}$ is also known as a shape index because the shape of the silicate feature at 10 $\micron$ gets broader and flatter as the degree of crystallinity and grain growth gets higher with the larger values of $F_{11.3}/F_{9.8}$.
Interstellar grains show narrow silicate profiles indicative of submicron, amorphous grains; protoplanetary-disk grains in addition have profiles with narrow substructure owing to crystallization and widening long-wavelengths due to grain growth. Both forms of processing are captured in the $F_{11.3}/F_{9.8}$ (``shape") index. Previous studies of dust features in Taurus disks ~\citep{dmw_2009, Sargent+2009-GrainGrowth} found that the change in shape is mostly due to crystallization.

The major contributor to the uncertainties of extracting information from the silicate emission features at 10 and 20 $\micron$ is how to determine the underlying continuum. Based on our previous experiences~\citep[e.g.][]{manoj2011, Arnold2012}, the determination of continuum of 10 $\micron$ silicate feature is better than that of 20 $\micron$ feature. We estimate the uncertainties of properties measured with 10 $\micron$ silicate features by assuming the uncertainty of 10$\%$ in the underlying continuum. We assume an uncertainty of 20 $\%$ of the continuum for the measurement of EW20 and other properties measured from 20$\micron$ silicate feature.  We report the parameters related to dust processing and their uncertainties in Table~\ref{table:ew1020}.

\subsection{Disk Classification}
\subsubsection{Classification of Disks Using Continuum Spectral Indices}
One assessment of the evolutionary state of a young stellar object (YSO) comes from the slope of its SED~\citep{Lada1987-YSO-classes}. Two wavelengths at near- and at mid-IR, typically K ($\sim$2 $\micron$) and 25 $\micron$, have been used for YSO classification. $n_{K-25}$ is used to define the SED classes: Class I if $n_{K-25}$ $>$ 0.3; Flat Spectrum if -0.3 $< n_{K-25} <$ 0.3; Class II if -1.6 $< n_{K-25} <$ -0.3; Class III for $n_{K-25} <$ -1.6. However, as is well known, SED classification can also give a wrong impression of evolutionary state for objects viewed close to face-on or edge-on~\citep{Robitaille+2006ApJS, Carpsi+2008A&A, McClure2010}.

To minimize the extinction effect on evolutionary classification based on spectral indices, \citet{McClure2010} made good use of the ``extinction-free" indices, $n_{5-12}$ and $n_{12-20}$, which are independent of extinction because $A_{\lambda}$ at the anchors are same in the~\citet{McClure2009} extinction curves, to classify the evolutionary stages of the YSOs in the Ophiuchus star-forming region (Oph). They calibrated their classification scheme with Class I, II, and III samples in Taurus~\citep{furlan06, furlan2008ClassI} and applied to Oph objects. It divides objects into photospheres ($n_{5-12} <$ -2.25), disks (-2.25 $< n_{5-12} <$ -0.2), and envelope ($n_{5-12} >$ -0.2). The second extinction-free index, $n_{12-20}$, includes not only continuum but also the 20 $\micron$ silicate feature. This index is used to determine roughly how much the disk is cleared, i.e., to suggest candidate debris or transitional disks. These extinction-free indices have been used in the literature~\citep[e.g.][]{Furlan2011Taurus, manoj2011, Arnold2012} as an initial filter to classify evolutionary stage of objects in the format of $n_{K-25}$ versus $n_{5-12}$ and $n_{12-20}$ versus $n_{5-12}$. 

In Figure~\ref{fig-n512-nk25} and \ref{fig-n512-n1220} we apply $n_{K-25}$, $n_{5-12}$, and $n_{12-20}$ spectral indices of the objects in our Orion A sample to classify their evolutionary stages. We see that our sample is mostly placed into Class II (from $n_{K-25}$) and disks (from $n_{5-12}$ and $n_{12-20}$) area in  Figure~\ref{fig-n512-nk25} and Figure~\ref{fig-n512-n1220}. Some objects, however, are located in the area for ``transitional disks", ``FS", or ``envelopes".  We confirm that all Orion A objects that fall into ``transitional disks" region in Figure~\ref{fig-n512-n1220} are the transitional disks previously identified by~\citet{khkim2013-OriATD}.  Almost half of the objects in the ``envelopes" area are also previously identified transitional disks, as indicated with open circles in Figure~\ref{fig-n512-n1220}. 

We notice that TDs that lie in the ``envelopes" region due to their $n_{5-12}$ values have $n_{12-20} \gtrsim$ 0 for both ONC and L1641 (see Figure~\ref{fig-n512-n1220}). The list of objects which are not identified as TDs but located in the ``envelopes" area is as follows, from high to low $n_{12-20}$: OriA-123, 159, 190, 21, 86, and 305 in ONC; OriA-311, 289, 191, 266, 209, 231, 258, 272, and 312 in L1641. We examine their SEDs and do not find strong evidence for envelopes, such as silicate absorption at 10 $\micron$, CO$_2$ ice absorption at 15 $\micron$ or steeply increasing flux after 20 $\micron$. Instead, the common characteristics of their SEDs are that they show strong 10 $\micron$ and 20 $\micron$ silicate emission features with rather flatter and redder spectral index between 5 $\micron$ and 12 $\micron$ while having low/less excess over photosphere at near-IR wavelengths ($<$ 5 $\micron$). We investigate the possibility of introducing a strong 10 $\micron$ silicate feature by over-correcting of extinction. We find that most of them are not highly extincted except two objects, OriA-209 and OriA-231, which have $A_{V}$ greater than 10. Therefore, the over-correction of extinction of deeply embedded protostars cannot be the major reason for the objects with $n_{5-12} \gtrsim$ -0.2 in  Orion sample. Even though these objects lie in the ``envelopes"  area in Figure~\ref{fig-n512-n1220}, we do not classify them  as ``envelopes-dominant" objects.

In order to figure out the differences of classification and sample distribution among star-forming regions, we also plot Taurus and NGC 1333 objects studied in \citet{Furlan2011Taurus} and \citet{Arnold2012} in Figure~\ref{fig-n512-n1220}. We chose Taurus and NGC 1333 because Taurus is a fiducial region in many studies and NGC 1333 is one of the youngest star-forming region. In the upper panel of Figure~\ref{fig-n512-n1220} we find that two TDs in Taurus are located in the ``envelopes" area and that most of Taurus objects having large $n_{5-12}$ ($>$-0.2) have $n_{12-20} > 0$. We also notice that there are many objects distributed in $-2 < n_{12-20} < 2$ in NGC 1333 in the ``envelops" area. To better demonstrate how the Orion sample and Taurus/NGC 1333 samples in the ``envelopes" region in Figure~\ref{fig-n512-n1220} are different, we plot $n_{12-20}$ versus $n_{K-25}$ for them in Figure~\ref{fig-n1220-nK25}. The objects located in the ``envelopes" area in Figure~\ref{fig-n512-n1220} are indicated with colored symbols: magenta for ONC and L1641 and blue for Taurus and NGC 1333. We find that YSOs in our sample of Orion A falling in the ``envelope" region as marked in magenta lie in ``Class II" by $n_{K-25}$ and ``disks" by $n_{12-20}$, while most Taurus and NGC 1333 objects marked by blue squares are located in ``Class I" or ``FS" area in Figure~\ref{fig-n1220-nK25}.  

The classification scheme by $n_{5-12}$ versus $n_{12-20}$ has worked well to reclassify as disks some disk-dominant objects that fall into Class I region according to $n_{K-25}$.  Conversely, we would like to understand why so many YSOs in Orion A that lie in the ``envelopes" are of the $n_{5-12}$ index are classified as Class II based on $n_{K-25}$. Considering that the Orion sample was selected as exclusively disk candidates from IRAC/2MASS color-color diagrams from \cite{Megeath_OriAB_survey_2012}, the possible dominant reason of this different classification is because we have several TDs with a steep rise between 5 and 12 $\micron$ but more moderate changes between 12 and 20 $\micron$ than envelope sources. If an object is in an initial stage of inner disk dissipation, it would have little excess around 5 $\micron$, but stlll a strong excess at longer wavelengths, so its 5--12 $\micron$ spectral index would be large and thus in the ``envelopes" area, while its K--25 $\micron$ index would be more typical of Class II objects.

From the comparison of sample distributions and source classifications done in Figure~\ref{fig-n512-nk25}, \ref{fig-n512-n1220} and \ref{fig-n1220-nK25}, we reaffirm that the disk classification schemes based on spectral indices ($n_{K-25}$, $n_{5-12}$ and $n_{12-20}$) work well as a first classification filter and they are complementary to each other, but one needs to check the SEDs carefully  to avoid ambiguous classifications.

\subsubsection{Transitional Disks}
Transitional disks (TDs) have radial gap(s) or central holes, so their SEDs are different from the SEDs of radially continuous full disks (FDs). Thus TDs are distributed in a separate region from the FDs on the spectral indices diagrams. We have described how spectral indices and EW10 can be used to find TD candidates in \citet{khkim2013-OriATD}.  Here, we revisit the selection criteria for transitional disks that we described previously \citep{khkim2013-OriATD}: $n_{K-6}$ $\leq$ -2.1; $n_{13-31}$ $\geq$ 0.5; EW10 $\geq$ 4.3. 
After updating spectral type information and adding 16 objects in Orion A Class II disk sources, we recalculate the selection criteria to check if any significant changes are needed for the thresholds of $n_{K-6}$, $n_{13-31}$, and EW10. We find that the lowest 12.5$\%$ (octile) of $n_{K-6}$ and the highest 12.5$\%$ of EW10 for Class II disks in Ori A, Tau, Cha I, Ophiuchus, and NGC 1333 are very similar to the previous values. The highest 12.5$\%$ of $n_{13-31}$ has somewhat noticeably changed from 0.5 to 0.57, but this change is still uncertain and does not affect much on the TDs already identified. Therefore, we keep the selection criteria used by \citet{khkim2013-OriATD}. We indicate how TDs are separated from the distribution of FDs in ONC and L1641 in Figure~\ref{fig-td-nk6-ew10-n1331}. We identify three more TDs from the 16 Class II disks added in our sample.

There are three subtypes of TDs depending on the disk's radial structure inferred by the morphology of their SED and self-consistent disk modeling: CTD, PTD, and WTD. CTD is an acronym for ``classical transitional disk'' -- a TD with a central hole. The characteristics of an SED of a CTD is no/negligible disk excess over 2--6 $\micron$ and a steep flux increment after 13 $\micron$. A ``pre-transitional disk'' (PTD) shows a strong disk excess similar to optically thick disk emission in the near-IR, a dip, and redder emission after 13 $\micron$, which is explained with a radial gap between the optically thick inner and outer disk \citep{espaillat07b}. In case there is weak excess somewhat between that of CTD and PTD, the excess may be due to optically thin inner disk emission. We call these TDs as ``weak-excess transitional disk" (WTD). \citet{khkim2013-OriATD} defined Inner Disk Excess Fraction (IDEF) to classify three subtypes of TDs from their disk excess in 2--6 $\micron$.  
 
 Among three newly identified TDs, OriA-306 and OriA-307 belong to the ONC, and they are subclassified as a PTD and a CTD, respectively. OriA-314 belongs to L1641 and is classified as a WTD. Thus, we have a total of 65 transitional disks: 34 in ONC and 31 in L1641.

\subsection{Mass Accretion Rates}
We observed 120 Class II disks with IRTF/SpeX in SXD mode from 0.8-2.4 $\micron$. We utilize Pa$\gamma$ (1.094 $\micron$), Pa$\beta$ (1.282 $\micron$), and Br$\gamma$ (2.166 $\micron$) which are in the wavelength coverage of SpeX SXD spectra in order to measure mass accretion rates from the hydrogen recombination lines. We start with the de-reddened SpeX spectra with the $A_{V}$ determined as described in Section 3.2. We obtain mass accretion rates of 113 objects among the 120 observed. The objects excluded from the determination of mass accretion rates are OriA-21, 45, 52, 88, 135, 170, and 266, which are mostly earlier spectral type objects with strong hydrogen absorption lines. We also analyze the reduced SpeX spectra of companions of nine objects, OriA-4, 26, 38, 47, 98, 173, 208, 280, and 290. Among them, six objects show emission of the hydrogen recombination lines, and three companions show absorption lines. We do not include the mass accretion rates of companions in the following analysis due to the lack of information about the companions' properties. In addition, we do not include the mass accretion rate of OriA-208 and OriA-247 in our analysis because they belong to the ten objects that have incomplete IRS spectra, are variable, and/or are a galaxy as described in Section 3.3. Therefore, we use mass accretion rates of 111 objects in ONC and L1641 for any analysis related to disk-stellar mass accretion properties in this work.

To get the mass accretion rates from three hydrogen recombination lines, we first fit each hydrogen recombination line with a gaussian function plus a local continuum to get the line luminosity: $L_{Pa \gamma}$, $L_{Pa \beta}$, and $L_{Br \gamma}$. Then we calculate the accretion luminosity, $L_{acc}$, from the line luminosity of each hydrogen recombination line. To do so, we adopt the empirical correlations between $L_{acc}$ and $L_{line}$ derived by \citet{muzerolle98Brgamma} for Pa$\beta$ and Br$\gamma$ and \citet{gatti2008pagamma} for Pa$\gamma$ to convert to the line luminosity to the accretion luminosity as follows: 
\begin{equation}
log(L_{acc}/L_{\odot}) = 1.36 \times log(L_{Pa \gamma}/L_{\odot}) + 4.1
\end{equation}
\begin{equation}
log(L_{acc}/L_{\odot}) = 1.14 \times log(L_{Pa \beta}/L_{\odot}) + 3.15
\end{equation}
\begin{equation}
log(L_{acc}/L_{\odot}) = 1.26 \times log(L_{Br \gamma}/L_{\odot}) + 4.43
\end{equation}

Finally, we obtain the disk-star mass accretion rate, $\dot{M}$, from the relation:
\begin{equation}
{\dot{M}}={\frac{L_{acc}R_{\star}}{GM_{\star}}},
\label{Eq:Mdot}
\end{equation} 	
where $R_{\star}$ and $M_{\star}$ are stellar radius and stellar mass which we calculated as described in Section 3.4, and $L_{acc}$ is the accretion luminosity from Equation (5), (6), and (7).

The accretion rate estimates from three recombination lines in our spectra are generally within a factor of 2--3 of each other. We adopt the average $\dot{M}$ and report it in Table~\ref{table:basicprop}.  We regard the resulting average $\dot{M}$ as an upper limit when fewer than three lines were observed with poor signal-to-noise ratio or when the fitting of three lines are not reliable due to low signal-to-noise. We adopt the average $\dot{M}$ measured from two lines as a detection for some cases if two lines are prominent. 

We find that the distributions of mass accretion rates of Class II objects in the ONC and L1641 are not very different when we include all possible $\dot{M}$ in the distributions, either by visual inspection or by the K-S test (Figure~\ref{fig-Mdot-OriA}). We compare $\dot{M}$ distributions of two sub-groups, FDs and TDs, in Figure~\ref{fig-Mdot-OriA-TDvsFD}, and check that $\dot{M}$ of TDs are decreased significantly compared to that of FDs. The median values of $log\dot{M}$ of FDs are -7.81 for ONC and -7.99 for L1641. For TDs, they are -8.71 and -8.79 for ONC and L1641, respectively. We confirm that the difference of the median $\dot{M}$ between FD and TDs in ONC and L1641 is almost a factor of 10, as seen in the Taurus-Aurigae association by \citet{najita07} and their extended study including Ophiuchus \citep{Najita_2015MNRAS.450.3559N}. 

We compare the median values of $\dot{M}$ between FDs and TDs in the narrower spectral type ranges, K--M, K type, M type, and M3 or later, to investigate any impact by the $\dot{M}$ dependence on spectral types to the different mass accretion distribution between FDs and TDs because TDs are more weighted to the later types (K--M). The number of objects in the K--M group are 56 for FDs and 50 for TDs, which are comparable size of sample between two groups. The median $\dot{M}$ of FDs and TDs for the K--M group are $1.03 \times 10^{-8}$ $M_{\odot}/yr$ and $1.76 \times 10^{-9}$ $M_{\odot}/yr$, respectively. When we take only K type objects, the median $\dot{M}$ are $4.54 \times 10^{-8}$ $M_{\odot}/yr$ for FDs with 23 objects and $4.77 \times 10^{-9}$ $M_{\odot}/yr$ for TDs with 13 objects. We have 33 FDs and 37 TDs in the M type group, and the median $\dot{M}$ is $3.51 \times 10^{-9}$ $M_{\odot}/yr$ and $1.5 \times 10^{-9}$ $M_{\odot}/yr$, respectively. For the M3 or later types with 11 objects of FDs and 18 objects of TDs, the median values are $1.93 \times 10^{-9}$ $M_{\odot}/yr$  and $1.34 \times 10^{-9}$ $M_{\odot}/yr$, respectively. We note that the median values of $\dot{M}$ decreases from K through M/M3 or later types both in FDs and TDs.  We find that the median $\dot{M}$ of the K type objects for TDs are almost one magnitude lower than that for FDs. Even though the differences of the median $\dot{M}$ between FDs and TDs for the M type and M3 or later groups are not as large as seen for the K type, the median values of TDs are lower than the median values of FDs in the corresponding sub-groups. Since the number of objects in the each sub-group divided by the spectral type ranges between TDs and FDs are not very different, we confirm that the displacement of the mass accretion rate distribution of TDs from that of FDs seen in Figure~\ref{fig-Mdot-OriA-TDvsFD} are not the result of the intrinsic dependency on the spectral type of the mass accretion rates. The median $\dot{M}$ of the sub-groups are summarized in Table~\ref{table:prop-median-Mdot-TDFD}.

\section{COMPARISION TO DISKS IN TAURUS}
In this section, we compare the distribution of spectral indices of the Orion A Class II samples to those of the Class II sample in Taurus. We first compare, in Section 5.1,  the whole sample without classifying sources, whether their disk is radially continuous (FDs) or not (TDs). More detailed comparisons of samples of Orion A and Taurus broken down by radial structures (FDs/TDs) follow in Section 5.2. 

Our goal is to provide the measured properties of disks in Orion A as well as to discern any differences with disks in Taurus. Therefore, results from an exhaustive study of the characteristics of Orion A Class II disks and correlations between properties will be discussed in the next paper (Kim et al. 2016b, in preparation). Here, to measure the quantitative difference between the distributions of a given parameter among our sub-region (OriA, Trapezium, ONC, and L1641) and Taurus, we perform a K-S test and measure the median of the property for each group.

We consider a K-S difference $D$ between two groups to be significant if $p$, the probability that $D$ could result from two random selections from the same distribution, is less than 0.05. If $p$ is less than 0.01, we take the deviation to be a highly significant. With small $p$, the value of $D$ indicates a significant maximum difference between the cumulative distributions. In our data $D$ varies all the way from nearly unity to about 0.1. The largest values indicate completely distinct, non-overlapping distributions; the smallest values indicate largely overlapping distributions with maximum differences consistent with Poisson statistics and the total numbers in our samples. We rank the significant differences as large ($D$ $>$ 0.5), medium (0.25 $\lesssim D \lesssim$ 0.5), and small ($D$ $<$ 0.25). $D$ and $p$ for all pairs appear along with the histograms in Figure~\ref{fig-sl-histogram-comp}--\ref{fig-tdfd-n1331-OirATau}.

\subsection{Index from 5-14 $\micron$ spectrum: Trapezium, ONC, L1641 and Taurus}
As we described in Section 2, Class II objects in the Trapezium could only be observed with the SL module (5--14 $\micron$). We show the distributions of $n_{K-6}$, $n_{6-13}$, EW10, and $F_{11.3}/F_{9.8}$, which are the properties taken from IRS SL spectra of objects in three sub-regions of Orion A and Taurus, in Figure~\ref{fig-sl-histogram-comp}, without separating objects by their radial structures. 

The distributions of $n_{K-6}$ from the disks in the three sub-regions of Orion A in the upper left panel of Figure~\ref{fig-sl-histogram-comp} show that $n_{K-6}$ of ONC and L1641 is biased toward higher values than Taurus, even though this difference is not statistically significant.

The upper right panel of Figure~\ref{fig-sl-histogram-comp} shows that the distribution of $n_{6-13}$ in ONC is shifted to higher values of $n_{6-13}$ compared to Taurus. A K-S test result for ONC versus Taurus shows that this displacement is statistically highly significant. The $n_{6-13}$ distribution of L1641 also tends to higher values than that of Taurus, but the displacement is not noticeably large. 

The distributions of EW10 of the three sub-regions of Orion A in the lower left panel of Figure~\ref{fig-sl-histogram-comp} are all statistically significantly different from that of Taurus: all three are skewed toward larger EW10.
The $F_{11.3}/F_{9.8}$ distributions of Orion A disks in the lower right panel of Figure~\ref{fig-sl-histogram-comp} show different distributions by subregion of Orion A. Comparing $F_{11.3}/F_{9.8}$ values of the Trapezium, the ONC, and L1641 to those of Taurus, the D values from K-S tests decrease and the p values increase from Trapezium through ONC to L1641 in the lower right panel. The $F_{11.3}/F_{9.8}$ distribution of ONC is concentrated around smaller values than that of Taurus. Considering the median age differences of ONC, L1641, and Taurus, we may infer that the $F_{11.3}/F_{9.8}$ distribution difference between ONC and Taurus, and the smaller --- probably insignificant --- difference between L1641 and Taurus, indicate increased processing of dust as time goes on.

On the other hand, despite lots of arguments in ages, the young stars in the Trapezium (i.e., the center of ONC) are probably somewhat younger~\citep{Getman_2014ApJ, Megeath_2016AJ....151....5M} than the rest of the ONC, L1641, and Taurus, but its $F_{11.3}/F_{9.8}$ distribution is broad and skewed toward higher values than that of Taurus. This large shape difference between Trapezium and Taurus, which goes in the direction of larger degrees of dust processing, is possibly a disk-evolutionary difference rather than a dust-evolutionary difference: less-processed material at somewhat larger radii, which is still warm enough to contribute significantly to the 10 $\micron$ silicate feature, may have been selectively removed from the systems, by radiation from the Trapezium O/B type stars. We discuss the outer disk evolution in Trapezium in detail in Section 6. We also compare subdivisions of $F_{11.3}/F_{9.8}$ distribution separated by the disk radial structures in the following subsection.

\subsection{Disk and Dust Processing Indicators from Full IRS Spectra: ONC, L1641 and Taurus}
Now, we consider the properties of Class II samples in ONC and L1641 observed with full IRS spectrum, 5--37 $\micron$. Here we look into how the distributions of disk and dust properties are different not only by the star-forming region among ONC, L1641, and Taurus but also by the radial structure of disks between TDs and FDs. We compare the distributions of $n_{K-6}$, $n_{6-13}$, $n_{13-31}$, EW10, and $F_{11.3}/F_{9.8}$. A caveat concerning the comparison of the grouped sub-samples in Orion A and Taurus is the small sample size of TDs in Taurus. The K-S test is powerful two-sample non-parametric test that is reliable even for small number of sample ($<$10; \citet{Wall_Jenkins_StatisticsBook_2003}). Therefore, we perform the K-S test for the sub-samples and discuss their similarities or differences based on the performance. The output from the statistical tests, median and $D$ and $p$ from K-S test, are listed in Table~\ref{table:prop-median-KS-TDFD-OriATau}. 

We find that the distributions of TDs in Orion A or the subsets by the regions, ONC and L1641, are not much different from the distributions of TDs in Taurus for the properties, $n_{K-6}$, $n_{6-13}$, EW10, $F_{11.3}/F_{9.8}$, and $n_{13-31}$,  by checking through histograms in the lower panels of Figure~\ref{fig-tdfd-nk6-OirATau} through Figure~\ref{fig-tdfd-n1331-OirATau}. 

In the case of FDs, we notice that the properties of FDs in Orion A tend to have higher values than that in Taurus, in general, except $n_{13-31}$ and $F_{11.3}/F_{9.8}$. Even though $D$ is not large in both cases of OriA-Tau and L1641-Tau, $D$ values are larger in the case of the ONC-Tau comparison of FDs than in the case of the L1641-Tau comparison, and $p$ values indicate that the differences between ONC and Taurus are significant for $n_{K-6}$, $n_{6-13}$, and EW10.

The spectral index, $n_{K-6}$ in Figure~\ref{fig-tdfd-nk6-OirATau} and $n_{6-13}$ in Figure~\ref{fig-tdfd-n613-OirATau}, are measures of the optically thick disk continuum structure in the inner radius of a disk. EW10 in Figure~\ref{fig-tdfd-ew10-OirATau} is related to the optically thin small dust grains. Therefore, we can infer that disks in the ONC are less processed and still have more flared disks and more small grains in vertically optically thin regions than disks in Taurus. The inner disks in L1641 seem to be more processed than disks in the ONC and less processed than those in Taurus because their $n_{K-6}$, $n_{6-13}$, and EW10 are distributed somewhat in the middle of ONC and Taurus. 

Among the properties compared here, EW10 of FDs between ONC and Taurus show the most significant and largest difference. EW10 measures amounts of small dust relative to the underlying dust continuum. From the higher EW10, in spite of higher $n_{K-6}$ and $n_{6-13}$ indicating the less continuum depletion in the ONC, we infer large amounts of small dust in optically thin regions in disks of ONC.   

We look into the distribution of $F_{11.3}/F_{9.8}$ to learn more about dust properties in Figure~\ref{fig-tdfd-f113f98-OirATau}. The results from the K-S test for $F_{11.3}/F_{9.8}$ do not support any significantly different distribution between disks in Orion A and disks in Taurus, at a first glance. However, we note the peak and median shift of $F_{11.3}/F_{9.8}$ of FDs: ONC has the smallest median; L1641 has a median larger than that of ONC, but it is smaller than Taurus. These median shifts in $F_{11.3}/F_{9.8}$ along the median ages of star-forming region may be a clue regarding dust evolution (e.g. growth/crystallization). When we include our analysis to include the star-forming regions, NGC 1333 and Chamaeleon I, we observe an interesting evolution of $F_{11.3}/F_{9.8}$ along the median age of star-forming regions: 0.5 (NGC 1333) 0.52 (ONC) 0.53 (L1641) 0.57 (Taurus) 0.59 (Chamaeleon I). 
However, the dust evolution is very complicated, and a detailed analysis to understand it is beyond the scope of this work. 

In contrast to the comparison of the inner disk and grain processing indicators, the comparisons of the distribution of $n_{13-31}$ of ONC, L1641, and Taurus do not show significant differences (Figure~\ref{fig-tdfd-n1331-OirATau}). This index is sensitive to the degree of sedimentation in the outer disk \citep[e.g.][]{furlan06}; thus we find no difference in the settling of dust to the disk mid-plane among these three regions.

Combining our findings from the distributions of $n_{K-6}$, $n_{6-13}$, $n_{13-31}$, EW10, and $F_{11.3}/F_{9.8}$, we suggest that the inner disk evolves faster than the outer disk and dust grain processing (growth and/or crystallization) occurs faster with inner disk evolution while the outer disk is less processed and sedimented at 1-3 Myr old.

\subsection{Median Spectra}
Analysis with a median SED gives a general insight on how the protoplanetary disks in a star-forming region evolve. \citet{dalessio_1999ApJ527} applied a median T Tauri star SED to compare their disk models and observational data of T Tauri stars in Taurus. Their choice of a median SED was with K5--M2 stars in~\citet{kh95} because the majority of spectral types of the disk sample in Taurus lies in spectral types of K5--M2 and the selection of narrow spectral type ranges can reduce the large variation in fluxes by restricting the range of stellar effective temperatures~\citep{dalessio_1999ApJ527} . In this vein, \cite{furlan06} generated a K5--M2 median spectrum in 5--36 $\micron$ range with the available IRS spectra of Class II disks in Taurus. After assembling a more complete sample of disks in Taurus, \citet{Furlan2011Taurus} updated the K5--M2 median spectrum of Taurus disks. They were also able to generate median spectra of M3--M5 and M6--M9 with large number of samples. 
The analysis with the median IRS spectra of Class II disks observed in nearby star-forming regions is a broadly adopted method to evaluate generally the evolutionary state of disks by comparing median SEDs from region to region~\citep{furlan06, Furlan2009TauChaOph, Furlan2011Taurus, McClure2010, manoj2011, Arnold2012}. Especially the Taurus K5--M2 median has been widely used as a fiducial reference to examine the status of disk evolution in many other works~\cite[e.g.][]{Fang_L1641_2013ApJS..207....5F}.

Therefore, in order to figure out the general state of evolution of Class II disks in ONC and L1641, we generate the median spectra of disks in ONC and L1641, and compare them to the median spectrum of Class II disks in Taurus taken from \citet{Furlan2011Taurus}.
To generate median spectrum, we first select the spectra of disks that do not show evidence of time variability or evidence for a radial gap or central hole. We exclude the spectra without host-star spectral type information in the selection for a median. Then, we group the spectra in three spectral type ranges: A0--K4, K5--M2, and M3--M5(M7). We made the group of K5--M2 because it is a prevailing selection as explained in the previous paragraph. To compare with the median SED of M3--M5 of Taurus, we calculated a median SED of ONC and L1641 with objects having spectral type of M3 or later than M3. The spectral types of all objects used for the ONC median lie between M3 and M5. We include one M7.5 object to generate the L1641 median because the number of L1641 objects in the spectral type ranges (M3 or later) is small. The A0--K4 group has not been used previously with Taurus sample due to the broad range of spectral types and low fraction of objects belong to the group. Even though we hold the similar limitation to calculate A0--K4 median SEDs with Orion A sample, we created the group with A0--K4 objects to check the evolution situation of objects with spectral types earlier than K5 because we have frequently noticed diminishing fluxes in longer wavelength from objects with earlier spectral types in Orion A sample. 
 
With three separated groups, we first normalized the $H$-band flux of each spectrum in each category to the median $H$-band flux of the corresponding group and then we calculate the median flux at individual wavelength, as described in \citet{dalessio_1999ApJ527}, to minimize the effect of different stellar luminosity in a given group. The displayed median spectra in Figure~\ref{fig-median-metric-2regions} were normalized to the corresponding median $H$-band flux of Taurus disks in each panel for the comparison to the median spectrum of Taurus. In Figure~\ref{fig-median-metric-2regions}, the Taurus median spectra of K5--M2 and M3--M5 are adopted from \citet{Furlan2011Taurus}. The Taurus median of A0--K4 is what we calculated in the same manner applied for other median spectra in order to compare with the median spectrum of ONC and L1641.

In the top panels of Figure~\ref{fig-median-metric-2regions}, the median of the A0--K4 group for the ONC and L1641  differ noticeably from the median spectrum of Taurus. The median spectrum of the A0--K4 group in ONC has slightly higher flux levels of disk emission than the A0--M4 median spectrum of Taurus over the wavelength range of SL coverage, but the ONC median spectrum shows steeply decreasing fluxes beyond 13 $\micron$ with prominent crystalline-silicate features. The median spectrum of the A0--K4 group in L1641 has high fluxes compared to the A0--K4 median of Taurus over the wavelengths of IRS SL spectral coverage, similarly to the case of ONC median with stronger excess at 5-13 $\micron$, and the L1641 median spectrum also shows lower flux in LL coverage compared to that of Taurus. We also note that the flux level of the L1641 A0--K4 median is higher than the ONC A0--K4 median. There are several possible causes contributing to the steep SEDs beyond 13 $\micron$: disk settling as grains grow and settle toward midplane; outer disk truncation by the gravitational effect of companions; ablation of disk atmosphere by the strong external radiation from nearby OB stars.
 The objects contributing to the A0--K4 ONC median are OriA-3, 21, 35, 36, 40, 45, 112, 117, 125, 131, 135, 141, 170, 173, and 205. For the A0--K4 L1641 median, the contributing objects are OriA-191, 196, 199, 201, 224, 225, 226, 235, 236, 262, 266, 269, 280, 284, and 295. Among them, OriA-45 and OriA-205 are reported as a double lined and a single lined spectroscopic binary, respectively. We also discussed that OriA-125 is a potential binary system from our SpeX observation as we discussed in Section 2. OriA-280 has a very close neighbor based on the SpeX observation, too. Even though there is still a lack of complete mutiplicity information for objects in Orion, we can speculate that it is not rare that the outer disk truncation happens around A0--K4 host stars due to the gravitational effect by neighbor sources in ONC and L1641. 
Again, a caveat of the A0--K4  median spectra comparison is that the spectral type range for A0--K4 median is very broad compared to the K5-M2 and M3-M5(M7) median spectra.  However, the fractions of non-K type objects used for A0--K4 median for ONC and L1641 are 40$\%$ which is comparable to 37.5$\%$ of that for Taurus. Therefore, we consider that the samples for A0--K4 median spectra of ONC, L1641, and Taurus are still comparable, even though the median spectra do not represent the disk properties in a narrow spectral type range.

We compare the median spectra of the  K5--M2 groups of ONC, L1641, and Taurus in the middle panels of Figure~\ref{fig-median-metric-2regions}. In the case of ONC versus Taurus, the median spectrum from ONC disks has more excess over 5--35 $\micron$, which indicates less evolution of inner disk and less depletion of small (micron-sized) grains than that of Taurus. We notice that fluxes of 50$\%$ of objects used for the K5--M2 ONC median are higher than the Taurus K5--M2 median by examining the quartiles indicated with the dotted lines in the panel. On the other hand, the K5--M2 median spectra of L1641 and Taurus are generally similar. Even though some details like grain properties may show somewhat different characteristics, the general degree of dust settling between Class II disks with K5--M2 spectral types in L1641 and Taurus appear to be similar. The quartiles indicated in the plot with the L1641 K5-M2 median support it.

The bottom row in Figure~\ref{fig-median-metric-2regions} contains the comparison of the M3--M5(M7) median spectra. Due to the small numbers in the group, we included the spectra of M3 and later than M3 to compare to the M3--M5 median of Taurus. In both L1641 and ONC, the M3--M5(M7) median spectra appear to have higher excess emission than Taurus. A possible explanation of this higher emission of disks around host-stars of M3 or later spectral type in Orion A is the younger ages of the systems with M3 or later spectral types in Orion A resulting in less time for disk evolution than the systems in Taurus. Alternatively, the M3--M5(M7) systems in Orion could have had larger initial disk masses than systems with later spectral type in Taurus. We will discuss the M3--M5 median in a future paper, until such time as a complete survey of Orion A including faint objects is carried out.

In Figure~\ref{fig-normflux-pristine} we compare 10 $\micron$ silicate features of the A0--K4, K5--M2, and M3--M5(M7) medians of our targets and Taurus to the pristine, ISM-like, silicate feature which is generated from LkCa 15 and GM Aur. LkCa 15 and GM Aur are protoplanetary disks, and they are transtional disks \citep{calvet05} in Taurus and have a silicate features that are most similar to those of ISM dust grains~\citep{dmw_2009, Sargent+2009-GrainGrowth}. In the figure we plot the continuum-subtracted and -normalized flux. From that, we can check how the silicate feature strength varies from  median to median and how the feature shape differs from median to median. The profile of the 10 $\micron$ silicate feature of the K5--M2 median of the ONC is very close to the profile of the pristine silicate feature. However, the profile of L1641 for the K5--M2 median is broader than the width of the profile of the pristine silicate feature, and the height is also decreased compared to that of K5--M2 median of ONC. This indicates that grains grow, and there are smaller amounts of small dust grains compared to the ONC. The K5--M2 median profile shape of L1641 is similar to that of Taurus, even though detailed features and dust compositions might be different. For the M3 and later spectral types, we also see a somewhat similar pattern of profile changes in the case of K5--M2: the median profiles become broader from ONC to Taurus. However, we suspect a greater degree of dust processing in the M3--M5 median for ONC, because the M3--M5 profile of the ONC is neither as smooth nor as narrow as the pristine profile or the K5--M2 profile of the ONC. We notice also that the flatness of the median profile of M3 and later types is increased from ONC toward Taurus, which supports our finding of the evolution trend of $F_{11.3}/F_{9.8}$ as we discussed before. On the other hand, for the K4 and earlier spectral types, the continuum-subtracted and -normalized 10 micron silicate feature looks more evolved for ONC and L1641 than Taurus. As we have seen and discussed in the median comparison in Figure~\ref{fig-median-metric-2regions}, the faster outer disk evolution by dynamical effects and photoevaporative effects in Orion environments, if they are dominant and effective, seem to affect the dust processing in the inner disk and disk surface area that gives rise to the 10 $\micron$ silicate emission feature and the 10 $\micron$ features to show evidence of more processed dust. 

\citet{Shuping_2006ApJ644L} discusses the significant degree of grain processing by the UV radiation field of $\theta^1$ Ori C in 10 $\micron$ silicate feature of eight proplyds in the vicinity ($<$ 30 arcsec) of $\theta^1$ Ori C. We looked up the spectral types of the proplyds in \citep{Shuping_2006ApJ644L} and found that only four objects have known spectral type information. Three objects have spectral types of early K and one with M type. It looks like our experiments with K5--M2 and M3 later spectral types show the opposite results if we consider only the distance from $\theta^1$ Ori C. However, there is an obstacle in comparing our results directly with their result. The objects in \citet{Shuping_2006ApJ644L}  are located in much harsher environments ($<$0.1 pc from $\theta^1$ Ori C) while our objects are located further than 0.7 pc from $\theta^1$ Ori C. Even though it is difficult to make a clear comparison between our findings regarding silicate dust processing in the A0--K4 median spectra in ONC/L1641 and results from \citet{Shuping_2006ApJ644L} due to the limited information, we speculate there are similar causes of dust processing in the inner disk regions of the objects which shows the evidences of outer disk dissipation, i.e., the flux reduction at longer wavelengths suffered by proplyds due to outer disk erosion by the strong UV radiation. 
However, we are cautious with any further interpretation because the spectral type range in A0--K4 median is broad and the median cannot represent general characteristics of objects in a  narrow range of spectral types.

\section{ENVIRONMENTAL EFFECTS AND EVOLUTIONARY FEATURES OF THE OUTER DISKS}

The Orion A star-forming region is a much richer environment than Taurus and Cha I, in which the stellar density is small throughout, and the UV radiation field is not much larger than the interstellar field. However, most stars form in denser clusters containing high mass stars \citep{Lada_1991ApJ_L1630, Carpenter_2000AJ_2MASS, LadaLada2003, Evans_2009_c2dLegacy, Koenig_Leisawitz_WISEcaution_2014, Megeath_2016AJ....151....5M}, like Orion. Not only the high stellar density in Orion but also the strong ionizing radiation from OB associations affect the evolution of protoplanetary disks. The evidence of action of photoevaporative winds from the Trapezium cluster over the outer disks of protoplanetary disks within or near the Orion Nebula ---``proplyds"--- were observed by the \textit{Hubble Space Telescope} (HST) \citep{Odell-Wan_1994ApJ...436..194O_Proplyds}: comet-shaped ionized gas clouds surrounding each disk, and smaller-than-usual disk sizes measured directly from the disk silhouette. The outward truncation of proplyds in ONC has been detected at a submillimeter wavelength with Atacama Large Millimeter/submillimeter Array (ALMA)~\citep{Mann_2014ApJ_proplydsALMA}.

To test better this hypothesis of outer disk ablation by UV radiation from the Trapezium with IRS spectra, we construct a spectral index sensitive to the optically thick continuum of the outer disk (r $>$ 5 AU from 5--37 $\micron$ coverage of IRS) and the optically thin small dust grains in the surface of the outer disk, $n_{20-31}$. The anchor points for 20 $\micron$ and 31 $\micron$ are same as indicated in the Section 4.2, 19.7--19.95 $\micron$ and 30.3--31 $\micron$, respectively. The measured $n_{20-31}$ and the projected distances ($d$) of objects from $\theta^1$ Ori C are listed in Table~\ref{table:table-n2031-Distance}.

We present the variation of $n_{20-31}$ over finer scale of distance by plotting $n_{20-31}$ and the projected distance of each target from $\theta^1$ Ori C in Figure~\ref{fig-enveff-n2031-dis}. In the figure, we use three different symbols to break the objects into three sub-groups of spectral type ranges as defined in Section 5.3 for the median spectra discussion. This is to examine patterns that are dependent on properties of host stars, if any exist. At first glance, we notice that the objects of A0--K4 group have generally low $n_{20-31}$ while the objects of M3--M6 group are located in rather higher side of $n_{20-31}$.
We measure the K5--M2 median and the median of all objects of $n_{20-31}$ of each sub-region to see the general pattern better with the scattered data points in Figure~\ref{fig-enveff-n2031-dis}.  From the medians and the distribution of scattered data points, we see that $n_{20-31}$ of FDs in the region B increases as the projected distance, $d$, increases. At greater distance (d$>$1.5 pc), we do not see the steep variation pattern of $n_{20-31}$ as seen in the region B, through region C to L1641.

The distribution of $n_{13-31}$ with the projected distance is shown in Figure~\ref{fig-enveff-n1331-dis}. Even though the declining pattern of $n_{13-31}$ closer to $\theta^1$ Ori C is not as remarkable as shown for $n_{20-31}$, there is also a weak declining tendency for $n_{13-31}$ in B and C. We also notice that some objects show significant drop of $n_{13-31}$ in B region.

It seems that the effect of UV illumination by the Trapezium stars, if that is the dominant one, is largest within 1.5 pc ($\sim$ 730 arcsec) from $\theta^1$ Ori C, based on the observed variation patterns of $n_{20-31}$ and $n_{13-31}$. Combining the clues acquired from the observed variation pattern in the distribution of $n_{20-31}$ and $n_{13-31}$ along the projected distance, we suggest a possible solution to explain our observations. Considering that $n_{20-31}$ is the index combining optically thin small grains and the optically thick continuum while $n_{13-31}$ is more directly represents continuum variation, we infer that evaporation of small particles and gases on the surface of the outer disk may be dominant and faster than the outward truncation of the optically-thick disk itself. This may be consistent with outward truncation of the disks by the the photoevaporative disk erosion and small dust removal as seen in stars closer ($<$ 0.5 pc) to O type stars in NGC 2244 \citep{Balog_2007_ApJ...660.1532B, Balog_2008_ApJ...688..408B}. The ablation of the outer disks are also seen in the disks closer to the central star (O9) in  $\sigma$ Ori cluster from the smaller infrared excess at 70 and 160 $\mu$m from PACS observation than disk bearing objects located far from the massive star (J. Hern\'{a}ndez et al. 2014, personal communication).

We examine the parameters useful to infer the evolutionary status of inner disks (r $<$ 1 AU), $n_{K-6}$, EW10, $F_{10}$, and $F_{11.3}/F_{9.8}$ via distance from $\theta^1$ Ori C. We do not see any prominent dependence of the parameters to a target's separation from $\theta^1$ Ori C. An interpretation is that UV radiation can affect the outer disks where the thermal velocities of the ionized heated gas exceeds the escape velocity from the centra stars, but does not affect the inner disk much. On the other hand, as we have discussed in Section 5.2, if an object is located in much closer to the UV source, the inner disk can be influenced by the UV radiation. However, we do not make any detailed inference about the environmental effect on the inner disk evolution based on our observations of $n_{K-6}$, EW10, $F{10}$, and $F_{11.3}/F_{9.8}$.

\section{SUMMARY AND CONCLUSION}

We present IRS/Spitzer observations of 319 Class II disks in Orion A. We also present SpeX spectra of 120 objects and report mass accretion rates measured from hydrogen recombination lines in SpeX spectra. We analyzed the distributions of stellar, disk, and dust properties of our objects. We compared their distributions separated by sub-regions in Orion A, and we also compared the properties of Orion A disks to those of disks in the Taurus star-forming region. We draw the following conclusions from the analysis done in this work:

1. The median visual extinction $A_{V}$ is larger in L1641 than ONC and the Trapezium. It confirms the effect of strong UV radiation from OB stars in the Trapezium to blow out small dust grains near them.

2. We confirm that the $\dot{M}$ distribution of TDs in ONC and L1641 is about a factor of 10 lower than that of FDs. The median $\dot{M}$ of FDs in ONC and L1641 are $1.55 \times 10^{-8}$ $M_{\odot}/yr$ and $1.01 \times 10^{-8}$ $M_{\odot}/yr$, respectively. The median $\dot{M}$ of TDs in ONC and L1641 are $1.95 \times 10^{-9}$ $M_{\odot}/yr$ and $1.62 \times 10^{-9}$ $M_{\odot}/yr$, respectively.  When we compared $\dot{M}$ for disks in ONC and that of L1641 separately, the distribution of $\dot{M}$ in ONC is slightly skewed toward a higher value, but there is no statistically significant difference in $\dot{M}$ between ONC and L1641 when the same disk groups (FD vs. FD; TD vs. TD) in two sub-regions are compared.

3. We have compared disk properties between FDs in Orion A sub-regions and FDs in Taurus. We found that $n_{K-6}$ and $n_{6-13}$, which probe the inner regions of disks, tend to be statistically-significantly higher in the ONC compared with Tau.  We also notice that the distribution of EW10 in ONC is statistically-significantly shifted from that of Taurus toward higher values and that in L1641 is distributed in between those of the ONC and Taurus. We also detected possible evolution of the degrees of grain growth and crystallization from $F_{11.3}/F_{9.8}$ in different star-forming regions following their median ages (NGC 1333, ONC, L1641, Tau,  Cha I). By comparing the 10 $\micron$ silicate features of median spectra after subtracting continuum and normalization to the continuum, we confirmed that the ONC profile of 10 $\micron$ features are much smoother and narrower, similar to the profile of pristine silicate features, while the profile of L1641 indicates grain growth and crystallization, similar to Taurus. However, unlike the indicators of different properties of small dust grains, EW10 and $F_{11.3}/F_{9.8}$, we found no significant differences in the ``sedimentation index" $n_{13-31}$ among ONC, L1641, and Taurus. All of these findings support that the disk evolution occurs faster in the inner regions and the dust processing time scale is faster than the disk sedimentation time scale. Considering the median age of the ONC ($<$1 Myr), L1641 ($\sim$ 1 Myr),  and Taurus (2 Myr), grain growth and crystallization are increased in the 1-2 Myr age range.

4. We have searched for trends between disk parameters and the strength of the UV illumination upon the disks. From the examination of median spectra in subsections of Orion A, we found the median spectra of the subsections of ONC are blue at long wavelengths beyond 20 $\micron$, which is consistent with outward truncation of the disks due to UV radiation field from the Trapezium, as in the case for the proplyds. We compared the distributions of spectral indices, equivalent widths, and integrated fluxes to the variation of the distance of objects from $\theta^1$ Ori C.
We observed a remarkable decline of $n_{20-31}$ toward to the center of Trapezium. The distribution of $n_{13-31}$ shows a similar declining pattern. These decreasing trends of $n_{20-31}$ and $n_{13-31}$ are dominantly observed within 1.5 pc from $\theta^1$ Ori C. Considering the definitions and the implications of the parameters studied in this work, we suggest a depletion of optically thin surface material of outer disk (r $\gtrsim$ 1 AU--10 AU).

The IRS survey of protoplanetary disks in Orion A presented in this work opens a new catalog of protoplanetary disks by offering disk properties measured from IRS spectra. We will discuss how the properties are interactively related each other in the next paper. Future observations of the objects in this work with instruments capable of greater angular resolution and sensitivity in multi-wavelength ranges from IR to submm/mm will enhance our understanding of the process of disk evolution and planet formation.

\acknowledgments
This work is based on observations made with the Spitzer Space Telescope, which is operated by the Jet Propulsion Laboratory, California Institute of Technology under NASA contract 1407. Support for this work was provided by NASA through Contract Number 1257184 issued by JPL/Caltech, and Cornell subcontracts 31419-5714 to the University of Rochester. 
This publication makes use of data products from the Two Micron All Sky Survey, which is a joint project of the University of Massachusetts and the Infrared Processing and Analysis Center/California Institute of Technology, funded by the National Aeronautics and Space Administration and the National Science Foundation. 
This research has made use of the SIMBAD database, operated at CDS, Strasbourg, France. This research has made use of the VizieR catalogue access tool, CDS, Strasbourg, France. The original description of the VizieR service was published in A$\&$AS 143, 23.  
This publication makes use of data products from the Wide-field Infrared Survey Explorer, which is a joint project of the University of California, Los Angeles, and the Jet Propulsion Laboratory/California Institute of Technology, funded by the National Aeronautics and Space Administration. 
Funding for SDSS-III has been provided by the Alfred P. Sloan Foundation, the Participating Institutions, the National Science Foundation, and the U.S. Department of Energy Office of Science. SDSS-III is managed by the Astrophysical Research Consortium for the Participating Institutions of the SDSS-III Collaboration including the University of Arizona, the Brazilian Participation Group, Brookhaven National Laboratory, University of Cambridge, Carnegie Mellon University, University of Florida, the French Participation Group, the German Participation Group, Harvard University, the Instituto de Astrofisica de Canarias, the Michigan State/Notre Dame/JINA Participation Group, Johns Hopkins University, Lawrence Berkeley National Laboratory, Max Planck Institute for Astrophysics, Max Planck Institute for Extraterrestrial Physics, New Mexico State University, New York University, Ohio State University, Pennsylvania State University, University of Portsmouth, Princeton University, the Spanish Participation Group, University of Tokyo, University of Utah, Vanderbilt University, University of Virginia, University of Washington, and Yale University.  
The UKIDSS project is defined in Lawrence et al (2007). UKIDSS uses the UKIRT Wide Field Camera (WFCAM; Casali et al, 2007).  The photometric system is described in Hewett et al (2006), and the calibration is described in Hodgkin et al. (2009). The pipeline processing and science archive are described in Irwin et al (2009, in prep) and Hambly et al (2008).
K.H.Kim was a visiting Astronomer at the Infrared Telescope Facility, which is operated by the University of Hawaii under contract NNH14CK55B with the National Aeronautics and Space Administration.

{\it Facilities:} \facility{Spitzer (IRS)}, \facility{IRTF (SpeX)}.


\bibliography{references-1}

\begin{thebibliography}{114}
\expandafter\ifx\csname natexlab\endcsname\relax\def\natexlab#1{#1}\fi

\bibitem[{{Adams} {et~al.}(1988){Adams}, {Lada}, \&
  {Shu}}]{AdamsLadaShu_1988ApJ...326..865A_FS}
{Adams}, F.~C., {Lada}, C.~J., \& {Shu}, F.~H. 1988, \apj, 326, 865

\bibitem[{{Adelman-McCarthy} \& {et al.}(2011)}]{SDSS-dr8}
{Adelman-McCarthy}, J.~K., \& {et al.} 2011, VizieR Online Data Catalog, 2306,
  0

\bibitem[{{Allen} \& {Mosby}(2008)}]{Allen_SpT_pc}
{Allen}, L., \& {Mosby}, G. 2008, personal communication

\bibitem[{{Allen}(1995)}]{Allen_thesis_L1641}
{Allen}, L.~E. 1995, PhD thesis, School of Physics, University of New South
  Wales, Sydney, NSW 2052, Australia <EMAIL>lea@newt.phys.unsw.edu.au</EMAIL>

\bibitem[{{Andre} {et~al.}(1993){Andre}, {Ward-Thompson}, \&
  {Barsony}}]{Andre_1993_class0}
{Andre}, P., {Ward-Thompson}, D., \& {Barsony}, M. 1993, \apj, 406, 122

\bibitem[{{Andrews} {et~al.}(2011){Andrews}, {Wilner}, {Espaillat}, {Hughes},
  {Dullemond}, {McClure}, {Qi}, \& {Brown}}]{andrews2011_sma_CTDs}
{Andrews}, S.~M., {Wilner}, D.~J., {Espaillat}, C., {Hughes}, A.~M.,
  {Dullemond}, C.~P., {McClure}, M.~K., {Qi}, C., \& {Brown}, J.~M. 2011, \apj,
  732, 42

\bibitem[{{Arnold} {et~al.}(2012){Arnold}, {Watson}, {Kim}, {Manoj}, {Remming},
  {Sheehan}, {Adame}, {Forrest}, {Furlan}, {Mamajek}, {McClure}, {Espaillat},
  {Ausfeld}, \& {Rapson}}]{Arnold2012}
{Arnold}, L.~A., {Watson}, D.~M., {Kim}, K.~H., {Manoj}, P., {Remming}, I.,
  {Sheehan}, P., {Adame}, L., {Forrest}, W.~J., {Furlan}, E., {Mamajek}, E.,
  {McClure}, M., {Espaillat}, C., {Ausfeld}, K., \& {Rapson}, V.~A. 2012,
  \apjs, 201, 12

\bibitem[{{Balog} {et~al.}(2007){Balog}, {Muzerolle}, {Rieke}, {Su}, {Young},
  \& {Megeath}}]{Balog_2007_ApJ...660.1532B}
{Balog}, Z., {Muzerolle}, J., {Rieke}, G.~H., {Su}, K.~Y.~L., {Young}, E.~T.,
  \& {Megeath}, S.~T. 2007, \apj, 660, 1532

\bibitem[{{Balog} {et~al.}(2008){Balog}, {Rieke}, {Muzerolle}, {Bally}, {Su},
  {Misselt}, \& {G{\'a}sp{\'a}r}}]{Balog_2008_ApJ...688..408B}
{Balog}, Z., {Rieke}, G.~H., {Muzerolle}, J., {Bally}, J., {Su}, K.~Y.~L.,
  {Misselt}, K., \& {G{\'a}sp{\'a}r}, A. 2008, \apj, 688, 408

\bibitem[{{Bertout} {et~al.}(2007){Bertout}, {Siess}, \&
  {Cabrit}}]{bertout07disktime}
{Bertout}, C., {Siess}, L., \& {Cabrit}, S. 2007, \aap, 473, L21

\bibitem[{{Bouwman} {et~al.}(2008){Bouwman}, {Henning}, {Hillenbrand}, {Meyer},
  {Pascucci}, {Carpenter}, {Hines}, {Kim}, {Silverstone}, {Hollenbach}, \&
  {Wolf}}]{Bouwman+2008ApJ}
{Bouwman}, J., {Henning}, T., {Hillenbrand}, L.~A., {Meyer}, M.~R., {Pascucci},
  I., {Carpenter}, J., {Hines}, D., {Kim}, J.~S., {Silverstone}, M.~D.,
  {Hollenbach}, D., \& {Wolf}, S. 2008, \apj, 683, 479

\bibitem[{{Calvet} {et~al.}(2005){Calvet}, {D'Alessio}, {Watson},
  {Franco-Hern{\'a}ndez}, {Furlan}, {Green}, {Sutter}, {Forrest}, {Hartmann},
  {Uchida}, {Keller}, {Sargent}, {Najita}, {Herter}, {Barry}, \&
  {Hall}}]{calvet05}
{Calvet}, N., {D'Alessio}, P., {Watson}, D.~M., {Franco-Hern{\'a}ndez}, R.,
  {Furlan}, E., {Green}, J., {Sutter}, P.~M., {Forrest}, W.~J., {Hartmann}, L.,
  {Uchida}, K.~I., {Keller}, L.~D., {Sargent}, B., {Najita}, J., {Herter},
  T.~L., {Barry}, D.~J., \& {Hall}, P. 2005, \apjl, 630, L185

\bibitem[{{Caratti O Garatti} {et~al.}(2012){Caratti O Garatti}, {Garcia
  Lopez}, {Antoniucci}, {Nisini}, {Giannini}, {Eisl{\"o}ffel}, {Ray},
  {Lorenzetti}, \& {Cabrit}}]{CarattiOGaratti2012}
{Caratti O Garatti}, A., {Garcia Lopez}, R., {Antoniucci}, S., {Nisini}, B.,
  {Giannini}, T., {Eisl{\"o}ffel}, J., {Ray}, T.~P., {Lorenzetti}, D., \&
  {Cabrit}, S. 2012, \aap, 538, A64

\bibitem[{{Cardelli} {et~al.}(1989){Cardelli}, {Clayton}, \&
  {Mathis}}]{CardelliClaytonMathis_1989ApJ...345..245C}
{Cardelli}, J.~A., {Clayton}, G.~C., \& {Mathis}, J.~S. 1989, \apj, 345, 245

\bibitem[{{Carpenter}(2000)}]{Carpenter_2000AJ_2MASS}
{Carpenter}, J.~M. 2000, \aj, 120, 3139

\bibitem[{{Casassus} {et~al.}(2013){Casassus}, {van der Plas}, {M}, {Dent},
  {Fomalont}, {Hagelberg}, {Hales}, {Jord{\'a}n}, {Mawet}, {M{\'e}nard},
  {Wootten}, {Wilner}, {Hughes}, {Schreiber}, {Girard}, {Ercolano}, {Canovas},
  {Rom{\'a}n}, \& {Salinas}}]{Casassus_2013Natur.493..191C}
{Casassus}, S., {van der Plas}, G., {M}, S.~P., {Dent}, W.~R.~F., {Fomalont},
  E., {Hagelberg}, J., {Hales}, A., {Jord{\'a}n}, A., {Mawet}, D.,
  {M{\'e}nard}, F., {Wootten}, A., {Wilner}, D., {Hughes}, A.~M., {Schreiber},
  M.~R., {Girard}, J.~H., {Ercolano}, B., {Canovas}, H., {Rom{\'a}n}, P.~E., \&
  {Salinas}, V. 2013, \nat, 493, 191

\bibitem[{{Crapsi} {et~al.}(2008){Crapsi}, {van Dishoeck}, {Hogerheijde},
  {Pontoppidan}, \& {Dullemond}}]{Carpsi+2008A&A}
{Crapsi}, A., {van Dishoeck}, E.~F., {Hogerheijde}, M.~R., {Pontoppidan},
  K.~M., \& {Dullemond}, C.~P. 2008, \aap, 486, 245

\bibitem[{{Currie} {et~al.}(2009){Currie}, {Lada}, {Plavchan}, {Robitaille},
  {Irwin}, \& {Kenyon}}]{Currie_2009_homodep}
{Currie}, T., {Lada}, C.~J., {Plavchan}, P., {Robitaille}, T.~P., {Irwin}, J.,
  \& {Kenyon}, S.~J. 2009, \apj, 698, 1

\bibitem[{{Cushing} {et~al.}(2005){Cushing}, {Rayner}, \&
  {Vacca}}]{IRTF_SpeX_MLTdwarf_Cushing_2005}
{Cushing}, M.~C., {Rayner}, J.~T., \& {Vacca}, W.~D. 2005, \apj, 623, 1115

\bibitem[{{Cushing} {et~al.}(2004){Cushing}, {Vacca}, \& {Rayner}}]{SpeXtool}
{Cushing}, M.~C., {Vacca}, W.~D., \& {Rayner}, J.~T. 2004, \pasp, 116, 362

\bibitem[{{Cutri} \& {et al.}(2012)}]{WISE2012}
{Cutri}, R.~M., \& {et al.} 2012, VizieR Online Data Catalog, 2311, 0

\bibitem[{{da Rio} {et~al.}(2010){da Rio}, {Robberto}, {Soderblom}, {Panagia},
  {Hillenbrand}, {Palla}, \& {Stassun}}]{Dario2009_ubvri}
{da Rio}, N., {Robberto}, M., {Soderblom}, D.~R., {Panagia}, N., {Hillenbrand},
  L.~A., {Palla}, F., \& {Stassun}, K. 2010, VizieR Online Data Catalog, 218,
  30261

\bibitem[{{Daemgen} {et~al.}(2012){Daemgen}, {Correia}, \&
  {Petr-Gotzens}}]{Daemgen+2012}
{Daemgen}, S., {Correia}, S., \& {Petr-Gotzens}, M.~G. 2012, \aap, 540, A46

\bibitem[{{D'Alessio} {et~al.}(2006){D'Alessio}, {Calvet}, {Hartmann},
  {Franco-Hern{\'a}ndez}, \& {Serv{\'{\i}}n}}]{dalessio06}
{D'Alessio}, P., {Calvet}, N., {Hartmann}, L., {Franco-Hern{\'a}ndez}, R., \&
  {Serv{\'{\i}}n}, H. 2006, \apj, 638, 314

\bibitem[{{D'Alessio} {et~al.}(1999){D'Alessio}, {Calvet}, {Hartmann},
  {Lizano}, \& {Cant{\'o}}}]{dalessio_1999ApJ527}
{D'Alessio}, P., {Calvet}, N., {Hartmann}, L., {Lizano}, S., \& {Cant{\'o}}, J.
  1999, \apj, 527, 893

\bibitem[{{DENIS Consortium}(2005)}]{DENIS_2005}
{DENIS Consortium}. 2005, VizieR Online Data Catalog, 2263

\bibitem[{{Espaillat} {et~al.}(2007){Espaillat}, {Calvet}, {D'Alessio},
  {Hern{\'a}ndez}, {Qi}, {Hartmann}, {Furlan}, \& {Watson}}]{espaillat07b}
{Espaillat}, C., {Calvet}, N., {D'Alessio}, P., {Hern{\'a}ndez}, J., {Qi}, C.,
  {Hartmann}, L., {Furlan}, E., \& {Watson}, D.~M. 2007, \apjl, 670, L135

\bibitem[{{Espaillat} {et~al.}(2012){Espaillat}, {Ingleby}, {Hern{\'a}ndez},
  {Furlan}, {D'Alessio}, {Calvet}, {Andrews}, {Muzerolle}, {Qi}, \&
  {Wilner}}]{Espaillat_TDclass_2012}
{Espaillat}, C., {Ingleby}, L., {Hern{\'a}ndez}, J., {Furlan}, E., {D'Alessio},
  P., {Calvet}, N., {Andrews}, S., {Muzerolle}, J., {Qi}, C., \& {Wilner}, D.
  2012, \apj, 747, 103

\bibitem[{{Espaillat} {et~al.}(2014){Espaillat}, {Muzerolle}, {Najita},
  {Andrews}, {Zhu}, {Calvet}, {Kraus}, {Hashimoto}, {Kraus}, \&
  {D'Alessio}}]{Espaillat_obsTDs_2014arXiv1402.7103E}
{Espaillat}, C., {Muzerolle}, J., {Najita}, J., {Andrews}, S., {Zhu}, Z.,
  {Calvet}, N., {Kraus}, S., {Hashimoto}, J., {Kraus}, A., \& {D'Alessio}, P.
  2014, ArXiv e-prints

\bibitem[{{Evans} {et~al.}(2009){Evans}, {Dunham}, {J{\o}rgensen}, {Enoch},
  {Mer{\'{\i}}n}, {van Dishoeck}, {Alcal{\'a}}, {Myers}, {Stapelfeldt},
  {Huard}, {Allen}, {Harvey}, {van Kempen}, {Blake}, {Koerner}, {Mundy},
  {Padgett}, \& {Sargent}}]{Evans_2009_c2dLegacy}
{Evans}, II, N.~J., {Dunham}, M.~M., {J{\o}rgensen}, J.~K., {Enoch}, M.~L.,
  {Mer{\'{\i}}n}, B., {van Dishoeck}, E.~F., {Alcal{\'a}}, J.~M., {Myers},
  P.~C., {Stapelfeldt}, K.~R., {Huard}, T.~L., {Allen}, L.~E., {Harvey}, P.~M.,
  {van Kempen}, T., {Blake}, G.~A., {Koerner}, D.~W., {Mundy}, L.~G.,
  {Padgett}, D.~L., \& {Sargent}, A.~I. 2009, \apjs, 181, 321

\bibitem[{{Fang} {et~al.}(2013){Fang}, {Kim}, {van Boekel}, {Sicilia-Aguilar},
  {Henning}, \& {Flaherty}}]{Fang_L1641_2013ApJS..207....5F}
{Fang}, M., {Kim}, J.~S., {van Boekel}, R., {Sicilia-Aguilar}, A., {Henning},
  T., \& {Flaherty}, K. 2013, \apjs, 207, 5

\bibitem[{{Fang} {et~al.}(2009){Fang}, {van Boekel}, {Wang}, {Carmona},
  {Sicilia-Aguilar}, \& {Henning}}]{Fang2009}
{Fang}, M., {van Boekel}, R., {Wang}, W., {Carmona}, A., {Sicilia-Aguilar}, A.,
  \& {Henning}, T. 2009, \aap, 504, 461

\bibitem[{{F{\H u}r{\'e}sz} {et~al.}(2008){F{\H u}r{\'e}sz}, {Hartmann},
  {Megeath}, {Szentgyorgyi}, \& {Hamden}}]{FHM2008}
{F{\H u}r{\'e}sz}, G., {Hartmann}, L.~W., {Megeath}, S.~T., {Szentgyorgyi},
  A.~H., \& {Hamden}, E.~T. 2008, \apj, 676, 1109

\bibitem[{{Furlan} {et~al.}(2006){Furlan}, {Hartmann}, {Calvet}, {D'Alessio},
  {Franco-Hern{\'a}ndez}, {Forrest}, {Watson}, {Uchida}, {Sargent}, {Green},
  {Keller}, \& {Herter}}]{furlan06}
{Furlan}, E., {Hartmann}, L., {Calvet}, N., {D'Alessio}, P.,
  {Franco-Hern{\'a}ndez}, R., {Forrest}, W.~J., {Watson}, D.~M., {Uchida},
  K.~I., {Sargent}, B., {Green}, J.~D., {Keller}, L.~D., \& {Herter}, T.~L.
  2006, \apjs, 165, 568

\bibitem[{{Furlan} {et~al.}(2011){Furlan}, {Luhman}, {Espaillat}, {D'Alessio},
  {Adame}, {Manoj}, {Kim}, {Watson}, {Forrest}, {McClure}, {Calvet}, {Sargent},
  {Green}, \& {Fischer}}]{Furlan2011Taurus}
{Furlan}, E., {Luhman}, K.~L., {Espaillat}, C., {D'Alessio}, P., {Adame}, L.,
  {Manoj}, P., {Kim}, K.~H., {Watson}, D.~M., {Forrest}, W.~J., {McClure},
  M.~K., {Calvet}, N., {Sargent}, B.~A., {Green}, J.~D., \& {Fischer}, W.~J.
  2011, \apjs, 195, 3

\bibitem[{{Furlan} {et~al.}(2008){Furlan}, {McClure}, {Calvet}, {Hartmann},
  {D'Alessio}, {Forrest}, {Watson}, {Uchida}, {Sargent}, {Green}, \&
  {Herter}}]{furlan2008ClassI}
{Furlan}, E., {McClure}, M., {Calvet}, N., {Hartmann}, L., {D'Alessio}, P.,
  {Forrest}, W.~J., {Watson}, D.~M., {Uchida}, K.~I., {Sargent}, B., {Green},
  J.~D., \& {Herter}, T.~L. 2008, \apjs, 176, 184

\bibitem[{{Furlan} {et~al.}(2009){Furlan}, {Watson}, {McClure}, {Manoj},
  {Espaillat}, {D'Alessio}, {Calvet}, {Kim}, {Sargent}, {Forrest}, \&
  {Hartmann}}]{Furlan2009TauChaOph}
{Furlan}, E., {Watson}, D.~M., {McClure}, M.~K., {Manoj}, P., {Espaillat}, C.,
  {D'Alessio}, P., {Calvet}, N., {Kim}, K.~H., {Sargent}, B.~A., {Forrest},
  W.~J., \& {Hartmann}, L. 2009, \apj, 703, 1964

\bibitem[{{G{\^a}lfalk} \& {Olofsson}(2008)}]{galfalk_olofsson_L1641N}
{G{\^a}lfalk}, M., \& {Olofsson}, G. 2008, \aap, 489, 1409

\bibitem[{{Gatti} {et~al.}(2008){Gatti}, {Natta}, {Randich}, {Testi}, \&
  {Sacco}}]{gatti2008pagamma}
{Gatti}, T., {Natta}, A., {Randich}, S., {Testi}, L., \& {Sacco}, G. 2008,
  \aap, 481, 423

\bibitem[{{Getman} {et~al.}(2014){Getman}, {Feigelson}, \&
  {Kuhn}}]{Getman_2014ApJ}
{Getman}, K.~V., {Feigelson}, E.~D., \& {Kuhn}, M.~A. 2014, \apj, 787, 109

\bibitem[{{Glebocki} \& {Gnacinski}(2005)}]{Gleboki+2005}
{Glebocki}, R., \& {Gnacinski}, P. 2005, VizieR Online Data Catalog, 3244, 0

\bibitem[{{Grankin} {et~al.}(2007){Grankin}, {Melnikov}, {Bouvier}, {Herbst},
  \& {Shevchenko}}]{Grankin+2007}
{Grankin}, K.~N., {Melnikov}, S.~Y., {Bouvier}, J., {Herbst}, W., \&
  {Shevchenko}, V.~S. 2007, \aap, 461, 183

\bibitem[{{Greene} {et~al.}(1994){Greene}, {Wilking}, {Andre}, {Young}, \&
  {Lada}}]{Greene1994_flat}
{Greene}, T.~P., {Wilking}, B.~A., {Andre}, P., {Young}, E.~T., \& {Lada},
  C.~J. 1994, \apj, 434, 614

\bibitem[{{Herbig} \& {Bell}(1988)}]{HBC1988Orion}
{Herbig}, G.~H., \& {Bell}, K.~R. 1988, {Third Catalog of Emission-Line Stars
  of the Orion Population : 3 : 1988}

\bibitem[{{Hernandez}(2008)}]{Hernandez_SpT_MDM_pc}
{Hernandez}, J. 2008, personal communication

\bibitem[{{Hern{\'a}ndez} {et~al.}(2007){Hern{\'a}ndez}, {Calvet},
  {Brice{\~n}o}, {Hartmann}, {Vivas}, {Muzerolle}, {Downes}, {Allen}, \&
  {Gutermuth}}]{Hernandez_2007ApJ...671.1784H_OriOB1}
{Hern{\'a}ndez}, J., {Calvet}, N., {Brice{\~n}o}, C., {Hartmann}, L., {Vivas},
  A.~K., {Muzerolle}, J., {Downes}, J., {Allen}, L., \& {Gutermuth}, R. 2007,
  \apj, 671, 1784

\bibitem[{{Hernandez} \& {Tobin}(2009)}]{Hernandez_SpT_hecfast_pc}
{Hernandez}, J., \& {Tobin}, J. 2009, personal communication

\bibitem[{{Higdon} {et~al.}(2004){Higdon}, {Devost}, {Higdon}, {Brandl},
  {Houck}, {Hall}, {Barry}, {Charmandaris}, {Smith}, {Sloan}, \&
  {Green}}]{higdon04}
{Higdon}, S.~J.~U., {Devost}, D., {Higdon}, J.~L., {Brandl}, B.~R., {Houck},
  J.~R., {Hall}, P., {Barry}, D., {Charmandaris}, V., {Smith}, J.~D.~T.,
  {Sloan}, G.~C., \& {Green}, J. 2004, \pasp, 116, 975

\bibitem[{{Hillenbrand}(1997)}]{Hillenbrand97_ONCpopulation_age}
{Hillenbrand}, L.~A. 1997, \aj, 113, 1733

\bibitem[{{Hillenbrand} \& {Carpenter}(2000)}]{HC2000}
{Hillenbrand}, L.~A., \& {Carpenter}, J.~M. 2000, \apj, 540, 236

\bibitem[{{Hillenbrand} {et~al.}(2013){Hillenbrand}, {Hoffer}, \&
  {Herczeg}}]{Hillenbrand_2013_ONC}
{Hillenbrand}, L.~A., {Hoffer}, A.~S., \& {Herczeg}, G.~J. 2013, \aj, 146, 85

\bibitem[{{Honda} {et~al.}(2006){Honda}, {Kataza}, {Okamoto}, {Yamashita},
  {Min}, {Miyata}, {Sako}, {Fujiyoshi}, {Sakon}, \& {Onaka}}]{Honda+2006ApJ}
{Honda}, M., {Kataza}, H., {Okamoto}, Y.~K., {Yamashita}, T., {Min}, M.,
  {Miyata}, T., {Sako}, S., {Fujiyoshi}, T., {Sakon}, I., \& {Onaka}, T. 2006,
  \apj, 646, 1024

\bibitem[{{Houk} \& {Swift}(2000)}]{Houk+1999}
{Houk}, N., \& {Swift}, C. 2000, VizieR Online Data Catalog, 3214, 0

\bibitem[{{Hsu} {et~al.}(2012){Hsu}, {Hartmann}, {Allen}, {Hern{\'a}ndez},
  {Megeath}, {Mosby}, {Tobin}, \&
  {Espaillat}}]{Hsu_L1641SpT_2012ApJ...752...59H}
{Hsu}, W.-H., {Hartmann}, L., {Allen}, L., {Hern{\'a}ndez}, J., {Megeath},
  S.~T., {Mosby}, G., {Tobin}, J.~J., \& {Espaillat}, C. 2012, \apj, 752, 59

\bibitem[{{Hsu} {et~al.}(2013){Hsu}, {Hartmann}, {Allen}, {Hern{\'a}ndez},
  {Megeath}, {Tobin}, \& {Ingleby}}]{Hsu_2013_IMFOriA}
{Hsu}, W.-H., {Hartmann}, L., {Allen}, L., {Hern{\'a}ndez}, J., {Megeath},
  S.~T., {Tobin}, J.~J., \& {Ingleby}, L. 2013, \apj, 764, 114

\bibitem[{{Kenyon} \& {Hartmann}(1995)}]{kh95}
{Kenyon}, S.~J., \& {Hartmann}, L. 1995, \apjs, 101, 117

\bibitem[{{Kessler-Silacci} {et~al.}(2006){Kessler-Silacci}, {Augereau},
  {Dullemond}, {Geers}, {Lahuis}, {Evans}, {van Dishoeck}, {Blake}, {Boogert},
  {Brown}, {J{\o}rgensen}, {Knez}, \& {Pontoppidan}}]{Kessler-Silacci+2006ApJ}
{Kessler-Silacci}, J., {Augereau}, J.-C., {Dullemond}, C.~P., {Geers}, V.,
  {Lahuis}, F., {Evans}, II, N.~J., {van Dishoeck}, E.~F., {Blake}, G.~A.,
  {Boogert}, A.~C.~A., {Brown}, J., {J{\o}rgensen}, J.~K., {Knez}, C., \&
  {Pontoppidan}, K.~M. 2006, \apj, 639, 275

\bibitem[{{Kim} {et~al.}(2016){Kim}, {Watson}, \& {Forrest}}]{kim2015c}
{Kim}, K.~H., {Watson}, D.~M., \& {Forrest}, W.~J. 2016, in preparation

\bibitem[{{Kim} {et~al.}(2013){Kim}, {Watson}, {Manoj}, {Forrest}, {Najita},
  {Furlan}, {Sargent}, {Espaillat}, {Muzerolle}, {Megeath}, {Calvet}, {Green},
  \& {Arnold}}]{khkim2013-OriATD}
{Kim}, K.~H., {Watson}, D.~M., {Manoj}, P., {Forrest}, W.~J., {Najita}, J.,
  {Furlan}, E., {Sargent}, B., {Espaillat}, C., {Muzerolle}, J., {Megeath},
  S.~T., {Calvet}, N., {Green}, J.~D., \& {Arnold}, L. 2013, \apj, 769, 149

\bibitem[{{Koenig} \& {Leisawitz}(2014)}]{Koenig_Leisawitz_WISEcaution_2014}
{Koenig}, X.~P., \& {Leisawitz}, D.~T. 2014, \apj, 791, 131

\bibitem[{{Kukarkin} {et~al.}(1971){Kukarkin}, {Kholopov}, {Pskovsky},
  {Efremov}, {Kukarkina}, {Kurochkin}, \& {Medvedeva}}]{GCVS1971}
{Kukarkin}, B.~V., {Kholopov}, P.~N., {Pskovsky}, Y.~P., {Efremov}, Y.~N.,
  {Kukarkina}, N.~P., {Kurochkin}, N.~E., \& {Medvedeva}, G.~I. 1971, in
  General Catalogue of Variable Stars, 3rd ed. (1971), 0--+

\bibitem[{{Lada}(1987)}]{Lada1987-YSO-classes}
{Lada}, C.~J. 1987, in IAU Symposium, Vol. 115, Star Forming Regions, ed.
  M.~{Peimbert} \& J.~{Jugaku}, 1--17

\bibitem[{{Lada} \& {Lada}(2003)}]{LadaLada2003}
{Lada}, C.~J., \& {Lada}, E.~A. 2003, \araa, 41, 57

\bibitem[{{Lada} {et~al.}(1991){Lada}, {Depoy}, {Evans}, \&
  {Gatley}}]{Lada_1991ApJ_L1630}
{Lada}, E.~A., {Depoy}, D.~L., {Evans}, II, N.~J., \& {Gatley}, I. 1991, \apj,
  371, 171

\bibitem[{{Lebouteiller} {et~al.}(2010){Lebouteiller}, {Bernard-Salas},
  {Sloan}, \& {Barry}}]{Lebouteiller_2010_AdOpt}
{Lebouteiller}, V., {Bernard-Salas}, J., {Sloan}, G.~C., \& {Barry}, D.~J.
  2010, \pasp, 122, 231

\bibitem[{{Lubow} \& {D'Angelo}(2006)}]{Lubow_Dangelo_2006}
{Lubow}, S.~H., \& {D'Angelo}, G. 2006, \apj, 641, 526

\bibitem[{{Mann} {et~al.}(2014){Mann}, {Di Francesco}, {Johnstone}, {Andrews},
  {Williams}, {Bally}, {Ricci}, {Hughes}, \&
  {Matthews}}]{Mann_2014ApJ_proplydsALMA}
{Mann}, R.~K., {Di Francesco}, J., {Johnstone}, D., {Andrews}, S.~M.,
  {Williams}, J.~P., {Bally}, J., {Ricci}, L., {Hughes}, A.~M., \& {Matthews},
  B.~C. 2014, \apj, 784, 82

\bibitem[{{Manoj} {et~al.}(2006){Manoj}, {Bhatt}, {Maheswar}, \&
  {Muneer}}]{Manoj2006ApJ}
{Manoj}, P., {Bhatt}, H.~C., {Maheswar}, G., \& {Muneer}, S. 2006, \apj, 653,
  657

\bibitem[{{Manoj} {et~al.}(2011){Manoj}, {Kim}, {Furlan}, {McClure}, {Luhman},
  {Watson}, {Espaillat}, {Calvet}, {Najita}, {D'Alessio}, {Adame}, {Sargent},
  {Forrest}, {Bohac}, {Green}, \& {Arnold}}]{manoj2011}
{Manoj}, P., {Kim}, K.~H., {Furlan}, E., {McClure}, M.~K., {Luhman}, K.~L.,
  {Watson}, D.~M., {Espaillat}, C., {Calvet}, N., {Najita}, J.~R., {D'Alessio},
  P., {Adame}, L., {Sargent}, B.~A., {Forrest}, W.~J., {Bohac}, C., {Green},
  J.~D., \& {Arnold}, L.~A. 2011, \apjs, 193, 11

\bibitem[{{Mathis}(1990)}]{mathis90}
{Mathis}, J.~S. 1990, \araa, 28, 37

\bibitem[{{McClure}(2009)}]{McClure2009}
{McClure}, M. 2009, \apjl, 693, L81

\bibitem[{{McClure} {et~al.}(2010){McClure}, {Furlan}, {Manoj}, {Luhman},
  {Watson}, {Forrest}, {Espaillat}, {Calvet}, {D'Alessio}, {Sargent}, {Tobin},
  \& {Chiang}}]{McClure2010}
{McClure}, M.~K., {Furlan}, E., {Manoj}, P., {Luhman}, K.~L., {Watson}, D.~M.,
  {Forrest}, W.~J., {Espaillat}, C., {Calvet}, N., {D'Alessio}, P., {Sargent},
  B., {Tobin}, J.~J., \& {Chiang}, H.-F. 2010, \apjs, 188, 75

\bibitem[{{Megeath} {et~al.}(2010){Megeath}, {Fischer}, {Ali}, {Allen},
  {Poteet}, {Watson}, {Wilson}, \& {HOPS Team}}]{Megeath_HOPS_2010AAS}
{Megeath}, S.~T., {Fischer}, W., {Ali}, B., {Allen}, L., {Poteet}, C.,
  {Watson}, D., {Wilson}, T., \& {HOPS Team}. 2010, in Bulletin of the American
  Astronomical Society, Vol.~42, American Astronomical Society Meeting
  Abstracts \#215, \#440.13

\bibitem[{{Megeath} {et~al.}(2012){Megeath}, {Gutermuth}, {Muzerolle},
  {Kryukova}, {Flaherty}, {Hora}, {Allen}, {Hartmann}, {Myers}, {Pipher},
  {Stauffer}, {Young}, \& {Fazio}}]{Megeath_OriAB_survey_2012}
{Megeath}, S.~T., {Gutermuth}, R., {Muzerolle}, J., {Kryukova}, E., {Flaherty},
  K., {Hora}, J.~L., {Allen}, L.~E., {Hartmann}, L., {Myers}, P.~C., {Pipher},
  J.~L., {Stauffer}, J., {Young}, E.~T., \& {Fazio}, G.~G. 2012, \aj, 144, 192

\bibitem[{{Megeath} {et~al.}(2016){Megeath}, {Gutermuth}, {Muzerolle},
  {Kryukova}, {Hora}, {Allen}, {Flaherty}, {Hartmann}, {Myers}, {Pipher},
  {Stauffer}, {Young}, \& {Fazio}}]{Megeath_2016AJ....151....5M}
{Megeath}, S.~T., {Gutermuth}, R., {Muzerolle}, J., {Kryukova}, E., {Hora},
  J.~L., {Allen}, L.~E., {Flaherty}, K., {Hartmann}, L., {Myers}, P.~C.,
  {Pipher}, J.~L., {Stauffer}, J., {Young}, E.~T., \& {Fazio}, G.~G. 2016, \aj,
  151, 5

\bibitem[{{Menten} {et~al.}(2007){Menten}, {Reid}, {Forbrich}, \&
  {Brunthaler}}]{Menten_2007_Oriondistance}
{Menten}, K.~M., {Reid}, M.~J., {Forbrich}, J., \& {Brunthaler}, A. 2007, \aap,
  474, 515

\bibitem[{{Meyer} {et~al.}(1997){Meyer}, {Calvet}, \&
  {Hillenbrand}}]{cttslocus97}
{Meyer}, M.~R., {Calvet}, N., \& {Hillenbrand}, L.~A. 1997, \aj, 114, 288

\bibitem[{{Muzerolle} {et~al.}(1998){Muzerolle}, {Hartmann}, \&
  {Calvet}}]{muzerolle98Brgamma}
{Muzerolle}, J., {Hartmann}, L., \& {Calvet}, N. 1998, \aj, 116, 2965

\bibitem[{{Najita} {et~al.}(2015){Najita}, {Andrews}, \&
  {Muzerolle}}]{Najita_2015MNRAS.450.3559N}
{Najita}, J.~R., {Andrews}, S.~M., \& {Muzerolle}, J. 2015, \mnras, 450, 3559

\bibitem[{{Najita} {et~al.}(2007){Najita}, {Strom}, \& {Muzerolle}}]{najita07}
{Najita}, J.~R., {Strom}, S.~E., \& {Muzerolle}, J. 2007, \mnras, 378, 369

\bibitem[{{Ochsenbein}(1980)}]{SAOcatalogue}
{Ochsenbein}, F. 1980, Bulletin d'Information du Centre de Donnees Stellaires,
  19, 74

\bibitem[{{O'Dell} \& {Wen}(1994)}]{Odell-Wan_1994ApJ...436..194O_Proplyds}
{O'Dell}, C.~R., \& {Wen}, Z. 1994, \apj, 436, 194

\bibitem[{{Olofsson} {et~al.}(2009){Olofsson}, {Augereau}, {van Dishoeck},
  {Mer{\'{\i}}n}, {Lahuis}, {Kessler-Silacci}, {Dullemond}, {Oliveira},
  {Blake}, {Boogert}, {Brown}, {Evans}, {Geers}, {Knez}, {Monin}, \&
  {Pontoppidan}}]{Olofsson+2009A&A}
{Olofsson}, J., {Augereau}, J.-C., {van Dishoeck}, E.~F., {Mer{\'{\i}}n}, B.,
  {Lahuis}, F., {Kessler-Silacci}, J., {Dullemond}, C.~P., {Oliveira}, I.,
  {Blake}, G.~A., {Boogert}, A.~C.~A., {Brown}, J.~M., {Evans}, II, N.~J.,
  {Geers}, V., {Knez}, C., {Monin}, J.-L., \& {Pontoppidan}, K. 2009, \aap,
  507, 327

\bibitem[{{Olofsson} {et~al.}(2011){Olofsson}, {Benisty}, {Augereau}, {Pinte},
  {M{\'e}nard}, {Tatulli}, {Berger}, {Malbet}, {Mer{\'{\i}}n}, {van Dishoeck},
  {Lacour}, {Pontoppidan}, {Monin}, {Brown}, \&
  {Blake}}]{Olofsson+2011_TCha_VLTI}
{Olofsson}, J., {Benisty}, M., {Augereau}, J.-C., {Pinte}, C., {M{\'e}nard},
  F., {Tatulli}, E., {Berger}, J.-P., {Malbet}, F., {Mer{\'{\i}}n}, B., {van
  Dishoeck}, E.~F., {Lacour}, S., {Pontoppidan}, K.~M., {Monin}, J.-L.,
  {Brown}, J.~M., \& {Blake}, G.~A. 2011, \aap, 528, L6

\bibitem[{{Parenago}(1954)}]{Parenago}
{Parenago}, P.~P. 1954, Trudy Gosudarstvennogo Astronomicheskogo Instituta, 25,
  1

\bibitem[{{Parihar} {et~al.}(2009{\natexlab{a}}){Parihar}, {Messina},
  {Distefano}, {Shantikumar}, \&
  {Medhi}}]{Parihar2009_ONC_variables_SpT4V1498Ori}
{Parihar}, P., {Messina}, S., {Distefano}, E., {Shantikumar}, N.~S., \&
  {Medhi}, B.~J. 2009{\natexlab{a}}, \mnras, 400, 603

\bibitem[{{Parihar} {et~al.}(2009{\natexlab{b}}){Parihar}, {Messina},
  {Distefano}, {Shantikumar}, \& {Medhi}}]{Parihar+2009MNRAS}
---. 2009{\natexlab{b}}, \mnras, 400, 603

\bibitem[{{Pecaut} \& {Mamajek}(2013)}]{Pecaut_TeffSpT_2013ApJS..208....9P}
{Pecaut}, M.~J., \& {Mamajek}, E.~E. 2013, \apjs, 208, 9

\bibitem[{{Poteet} {et~al.}(2016){Poteet}, {Watson}, {Megeath}, \&
  {Muzerolle}}]{Poteet2016}
{Poteet}, C., {Watson}, D.~M., {Megeath}, S.~T., \& {Muzerolle}, J. 2016, in
  preparation

\bibitem[{{Prisinzano} {et~al.}(2008){Prisinzano}, {Micela}, {Flaccomio},
  {Stauffer}, {Megeath}, {Rebull}, {Robberto}, {Smith}, {Feigelson}, {Grosso},
  \& {Wolk}}]{Prisinzano_2008ApJ...677..401P}
{Prisinzano}, L., {Micela}, G., {Flaccomio}, E., {Stauffer}, J.~R., {Megeath},
  T., {Rebull}, L., {Robberto}, M., {Smith}, K., {Feigelson}, E.~D., {Grosso},
  N., \& {Wolk}, S. 2008, \apj, 677, 401

\bibitem[{{Przygodda} {et~al.}(2003){Przygodda}, {van Boekel},
  {{\`A}brah{\`a}m}, {Melnikov}, {Waters}, \& {Leinert}}]{Przygodda+2003}
{Przygodda}, F., {van Boekel}, R., {{\`A}brah{\`a}m}, P., {Melnikov}, S.~Y.,
  {Waters}, L.~B.~F.~M., \& {Leinert}, C. 2003, \aap, 412, L43

\bibitem[{{Rayner} {et~al.}(2009){Rayner}, {Cushing}, \&
  {Vacca}}]{IRTF_spectral_Library_2009}
{Rayner}, J.~T., {Cushing}, M.~C., \& {Vacca}, W.~D. 2009, \apjs, 185, 289

\bibitem[{{Rayner} {et~al.}(2003){Rayner}, {Toomey}, {Onaka}, {Denault},
  {Stahlberger}, {Vacca}, {Cushing}, \& {Wang}}]{SpeX_Rayner}
{Rayner}, J.~T., {Toomey}, D.~W., {Onaka}, P.~M., {Denault}, A.~J.,
  {Stahlberger}, W.~E., {Vacca}, W.~D., {Cushing}, M.~C., \& {Wang}, S. 2003,
  \pasp, 115, 362

\bibitem[{{Rebull}(2001)}]{Rebull2001AJ}
{Rebull}, L.~M. 2001, \aj, 121, 1676

\bibitem[{{Rebull} {et~al.}(2000){Rebull}, {Hillenbrand}, {Strom}, {Duncan},
  {Patten}, {Pavlovsky}, {Makidon}, \& {Adams}}]{RHS2000}
{Rebull}, L.~M., {Hillenbrand}, L.~A., {Strom}, S.~E., {Duncan}, D.~K.,
  {Patten}, B.~M., {Pavlovsky}, C.~M., {Makidon}, R., \& {Adams}, M.~T. 2000,
  \aj, 119, 3026

\bibitem[{{Riaz} {et~al.}(2006){Riaz}, {Gizis}, \& {Harvin}}]{Riaz+2006}
{Riaz}, B., {Gizis}, J.~E., \& {Harvin}, J. 2006, \aj, 132, 866

\bibitem[{{Robberto} {et~al.}(2010){Robberto}, {Soderblom}, {Scandariato},
  {Smith}, {Da Rio}, {Pagano}, \& {Spezzi}}]{Robberto_ONC_JHK_2010}
{Robberto}, M., {Soderblom}, D.~R., {Scandariato}, G., {Smith}, K., {Da Rio},
  N., {Pagano}, I., \& {Spezzi}, L. 2010, \aj, 139, 950

\bibitem[{{Robitaille} {et~al.}(2006){Robitaille}, {Whitney}, {Indebetouw},
  {Wood}, \& {Denzmore}}]{Robitaille+2006ApJS}
{Robitaille}, T.~P., {Whitney}, B.~A., {Indebetouw}, R., {Wood}, K., \&
  {Denzmore}, P. 2006, \apjs, 167, 256

\bibitem[{{Sargent} {et~al.}(2009){Sargent}, {Forrest}, {Tayrien}, {McClure},
  {Watson}, {Sloan}, {Li}, {Manoj}, {Bohac}, {Furlan}, {Kim}, \&
  {Green}}]{Sargent+2009-GrainGrowth}
{Sargent}, B.~A., {Forrest}, W.~J., {Tayrien}, C., {McClure}, M.~K., {Watson},
  D.~M., {Sloan}, G.~C., {Li}, A., {Manoj}, P., {Bohac}, C.~J., {Furlan}, E.,
  {Kim}, K.~H., \& {Green}, J.~D. 2009, \apjs, 182, 477

\bibitem[{{Shuping} {et~al.}(2006){Shuping}, {Kassis}, {Morris}, {Smith}, \&
  {Bally}}]{Shuping_2006ApJ644L}
{Shuping}, R.~Y., {Kassis}, M., {Morris}, M., {Smith}, N., \& {Bally}, J. 2006,
  \apjl, 644, L71

\bibitem[{{Siess} {et~al.}(2000){Siess}, {Dufour}, \& {Forestini}}]{siess2000}
{Siess}, L., {Dufour}, E., \& {Forestini}, M. 2000, \aap, 358, 593

\bibitem[{{Skrutskie} {et~al.}(2006){Skrutskie}, {Cutri}, {Stiening},
  {Weinberg}, {Schneider}, {Carpenter}, {Beichman}, {Capps}, {Chester},
  {Elias}, {Huchra}, {Liebert}, {Lonsdale}, {Monet}, {Price}, {Seitzer},
  {Jarrett}, {Kirkpatrick}, {Gizis}, {Howard}, {Evans}, {Fowler}, {Fullmer},
  {Hurt}, {Light}, {Kopan}, {Marsh}, {McCallon}, {Tam}, {Van Dyk}, \&
  {Wheelock}}]{2MASS_catalog_2006}
{Skrutskie}, M.~F., {Cutri}, R.~M., {Stiening}, R., {Weinberg}, M.~D.,
  {Schneider}, S., {Carpenter}, J.~M., {Beichman}, C., {Capps}, R., {Chester},
  T., {Elias}, J., {Huchra}, J., {Liebert}, J., {Lonsdale}, C., {Monet}, D.~G.,
  {Price}, S., {Seitzer}, P., {Jarrett}, T., {Kirkpatrick}, J.~D., {Gizis},
  J.~E., {Howard}, E., {Evans}, T., {Fowler}, J., {Fullmer}, L., {Hurt}, R.,
  {Light}, R., {Kopan}, E.~L., {Marsh}, K.~A., {McCallon}, H.~L., {Tam}, R.,
  {Van Dyk}, S., \& {Wheelock}, S. 2006, \aj, 131, 1163

\bibitem[{{Tayrien} \& {Forrest}(2016)}]{opse}
{Tayrien}, C., \& {Forrest}, W.~J. 2016, in preparation

\bibitem[{{Tobin} {et~al.}(2009){Tobin}, {Hartmann}, {Furesz}, {Mateo}, \&
  {Megeath}}]{Tobin_2009_ONC_SB}
{Tobin}, J.~J., {Hartmann}, L., {Furesz}, G., {Mateo}, M., \& {Megeath}, S.~T.
  2009, \apj, 697, 1103

\bibitem[{{Tobin} {et~al.}(2013){Tobin}, {Hartmann}, {Furesz}, {Mateo}, \&
  {Megeath}}]{Tobin_ONCSB_2013ApJ...773...81T}
---. 2013, \apj, 773, 81

\bibitem[{{UKIDSS Consortium}(2012)}]{UKIDSS-dr8}
{UKIDSS Consortium}. 2012, VizieR Online Data Catalog, 2314, 0

\bibitem[{{Vacca} {et~al.}(2003){Vacca}, {Cushing}, \&
  {Rayner}}]{SpeX_tellcor_Vacca2003}
{Vacca}, W.~D., {Cushing}, M.~C., \& {Rayner}, J.~T. 2003, \pasp, 115, 389

\bibitem[{{van Boekel} {et~al.}(2005){van Boekel}, {Min}, {Waters}, {de Koter},
  {Dominik}, {van den Ancker}, \& {Bouwman}}]{vanBoekel+2005}
{van Boekel}, R., {Min}, M., {Waters}, L.~B.~F.~M., {de Koter}, A., {Dominik},
  C., {van den Ancker}, M.~E., \& {Bouwman}, J. 2005, \aap, 437, 189

\bibitem[{{Wall} \& {Jenkins}(2003)}]{Wall_Jenkins_StatisticsBook_2003}
{Wall}, J.~V., \& {Jenkins}, C.~R. 2003, {Practical Statistics for
  Astronomers}, ed. R.~{Ellis}, J.~{Huchra}, S.~{Kahn}, G.~{Rieke}, \& P.~B.
  {Stetson}

\bibitem[{{Watson} {et~al.}(2004){Watson}, {Kemper}, {Calvet}, {Keller},
  {Furlan}, {Hartmann}, {Forrest}, {Chen}, {Uchida}, {Green}, {Sargent},
  {Sloan}, {Herter}, {Brandl}, {Houck}, {Najita}, {D'Alessio}, {Myers},
  {Barry}, {Hall}, \& {Morris}}]{Watson_2004ApJS..154..391W_ClassI-IRS-DGTauB}
{Watson}, D.~M., {Kemper}, F., {Calvet}, N., {Keller}, L.~D., {Furlan}, E.,
  {Hartmann}, L., {Forrest}, W.~J., {Chen}, C.~H., {Uchida}, K.~I., {Green},
  J.~D., {Sargent}, B., {Sloan}, G.~C., {Herter}, T.~L., {Brandl}, B.~R.,
  {Houck}, J.~R., {Najita}, J., {D'Alessio}, P., {Myers}, P.~C., {Barry},
  D.~J., {Hall}, P., \& {Morris}, P.~W. 2004, \apjs, 154, 391

\bibitem[{{Watson} {et~al.}(2009){Watson}, {Leisenring}, {Furlan}, {Bohac},
  {Sargent}, {Forrest}, {Calvet}, {Hartmann}, {Nordhaus}, {Green}, {Kim},
  {Sloan}, {Chen}, {Keller}, {d'Alessio}, {Najita}, {Uchida}, \&
  {Houck}}]{dmw_2009}
{Watson}, D.~M., {Leisenring}, J.~M., {Furlan}, E., {Bohac}, C.~J., {Sargent},
  B., {Forrest}, W.~J., {Calvet}, N., {Hartmann}, L., {Nordhaus}, J.~T.,
  {Green}, J.~D., {Kim}, K.~H., {Sloan}, G.~C., {Chen}, C.~H., {Keller}, L.~D.,
  {d'Alessio}, P., {Najita}, J., {Uchida}, K.~I., \& {Houck}, J.~R. 2009,
  \apjs, 180, 84

\bibitem[{{Wolff} {et~al.}(2004){Wolff}, {Strom}, \&
  {Hillenbrand}}]{Wolff+2004}
{Wolff}, S.~C., {Strom}, S.~E., \& {Hillenbrand}, L.~A. 2004, \apj, 601, 979

\bibitem[{{Wouterloot} \& {Brand}(1992)}]{Wouterlook+1992}
{Wouterloot}, J.~G.~A., \& {Brand}, J. 1992, \aap, 265, 144

\bibitem[{{Zacharias} {et~al.}(2005){Zacharias}, {Monet}, {Levine}, {Urban},
  {Gaume}, \& {Wycoff}}]{NOMAD_2005}
{Zacharias}, N., {Monet}, D.~G., {Levine}, S.~E., {Urban}, S.~E., {Gaume}, R.,
  \& {Wycoff}, G.~L. 2005, VizieR Online Data Catalog, 1297, 0

\end{thebibliography}


\clearpage
\clearpage

\clearpage

\clearpage

\clearpage

\clearpage

\clearpage
\begin{figure}[!htbp]
\centering
\epsscale{0.8}
\plotone{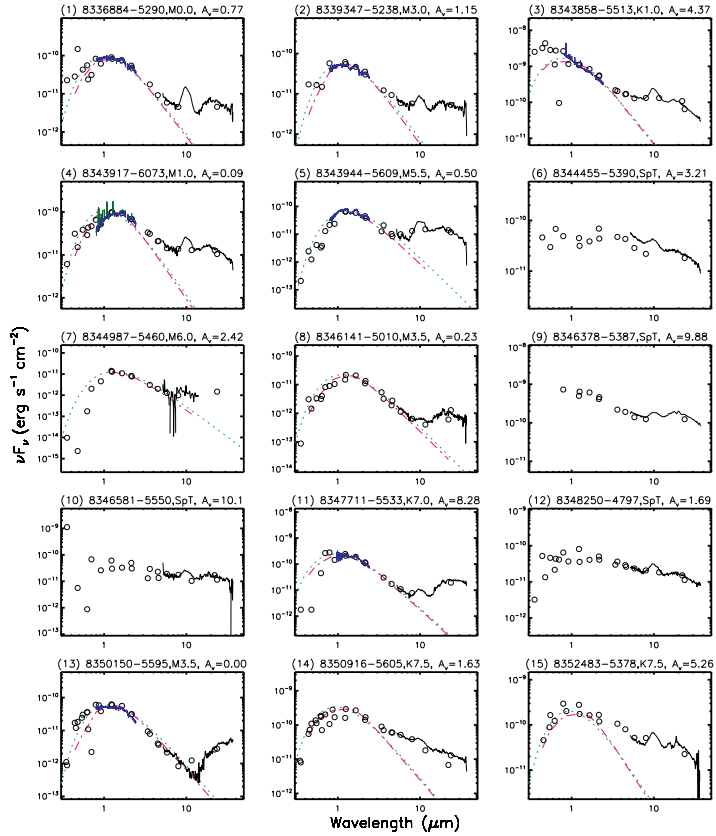}
\caption{De-reddened SEDs of our samples in ONC and L1641, which were observed with IRS SL and LL modules. The SEDs are composed of the following components: IRS (solid line in the wavelength range of 5.2--35 (or 14) \micron); SpeX (solid line in the wavelength range of 0.8--2.4 \micron); photosperic models: a blackbody radiation of the host-star's effective temperature (short dashed line) and a photosphere derived from the intrinsic colors from \citet{Pecaut_TeffSpT_2013ApJS..208....9P} (long dash-dot line); photometric data (open circles): UBVRI from \cite{Dario2009_ubvri}, SDSS, UKIDSS, BVR from NOMAD, DENIS IJH, 2MASS JHK, IRAC, MIPS (24 $\mu$m), and WISE.   \label{fig-SLLL-SED}}
\end{figure}

\clearpage
\begin{figure}[!htbp]
\ContinuedFloat
\centering
\epsscale{0.8}
\plotone{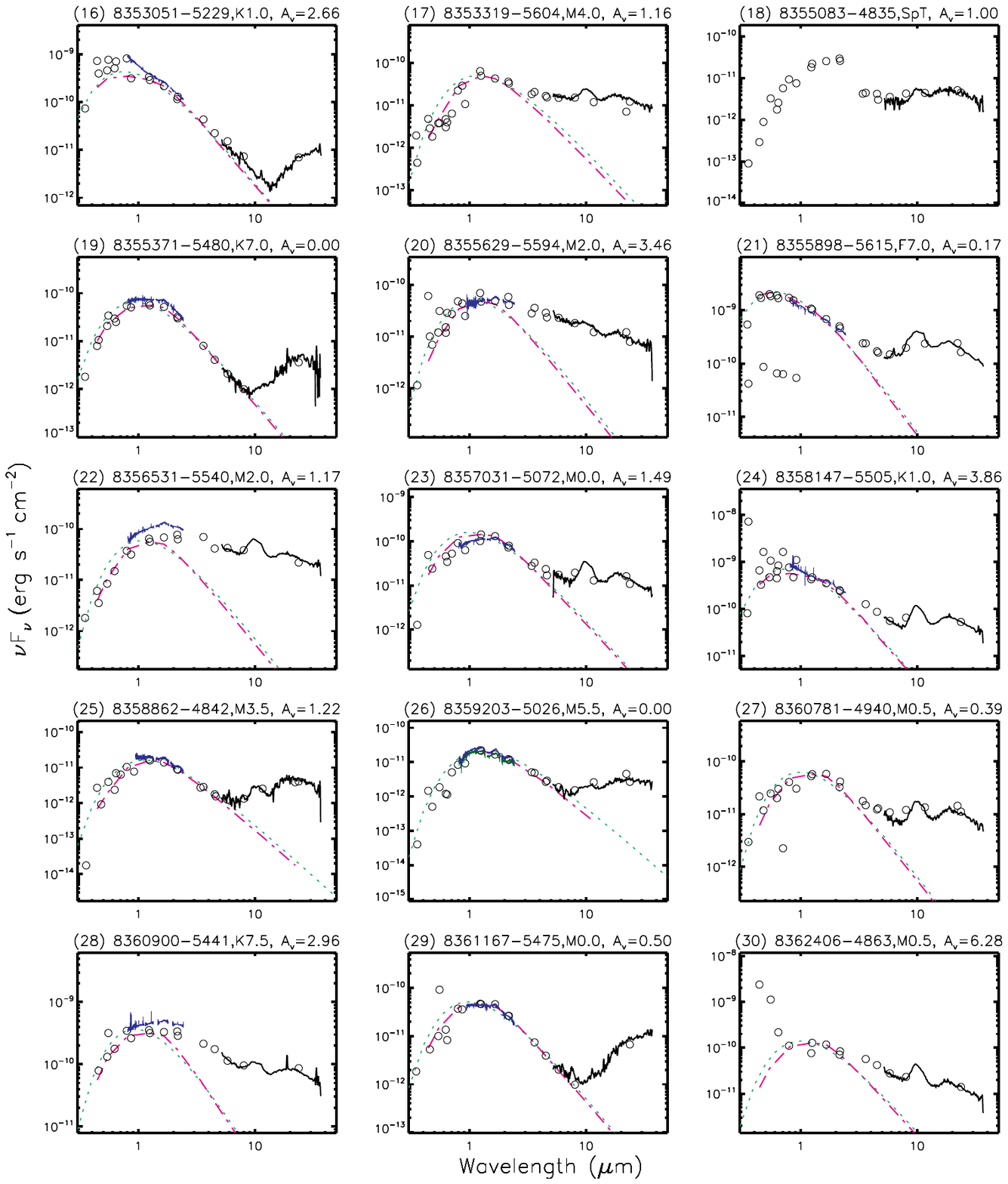}
\caption{continued. }
\end{figure}

\clearpage
\begin{figure}[!htbp]
\ContinuedFloat
\centering
\epsscale{0.9}
\plotone{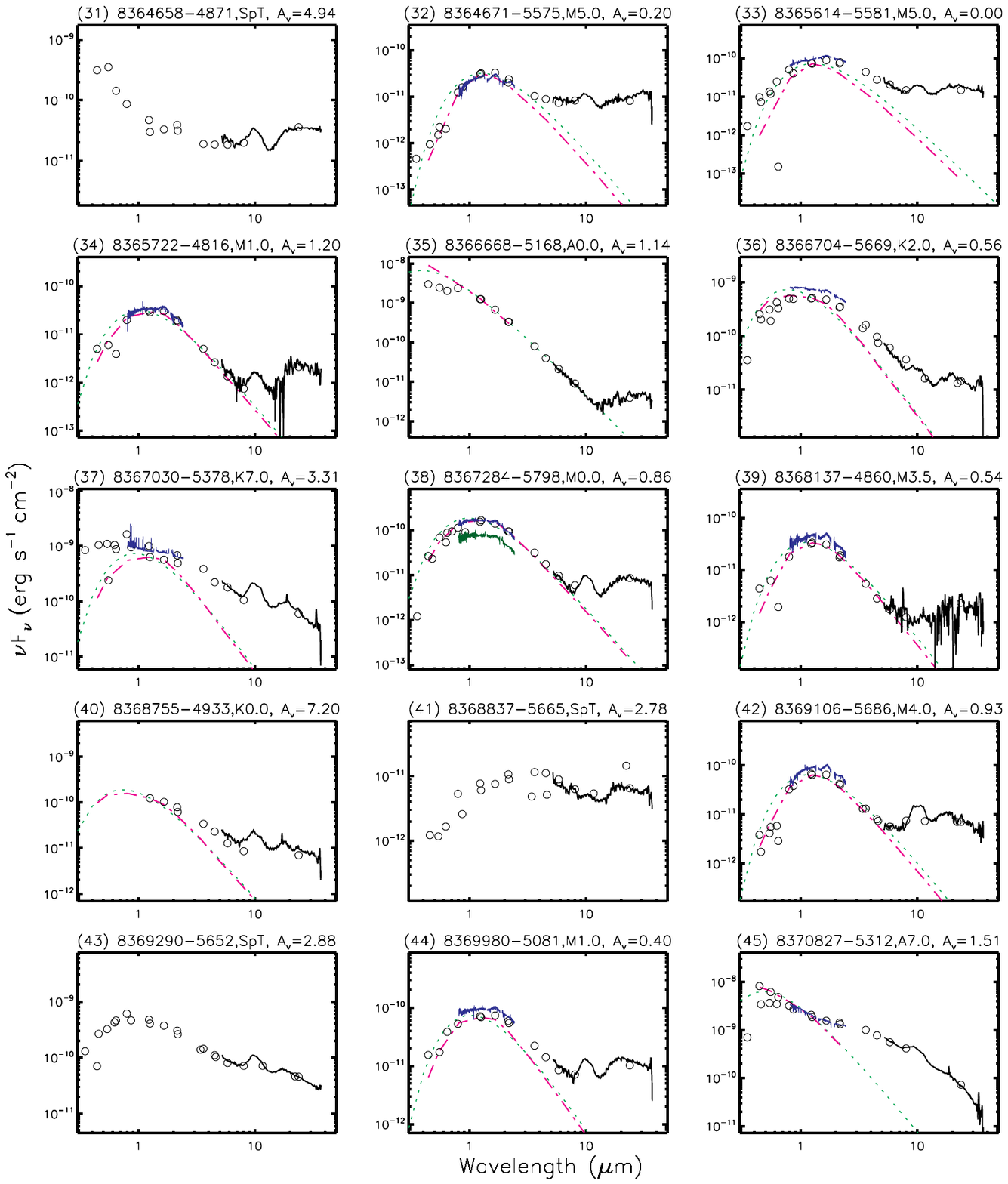}
\caption{continued. }
\end{figure}

\clearpage
\begin{figure}[!htbp]
\ContinuedFloat
\centering
\epsscale{0.9}
\plotone{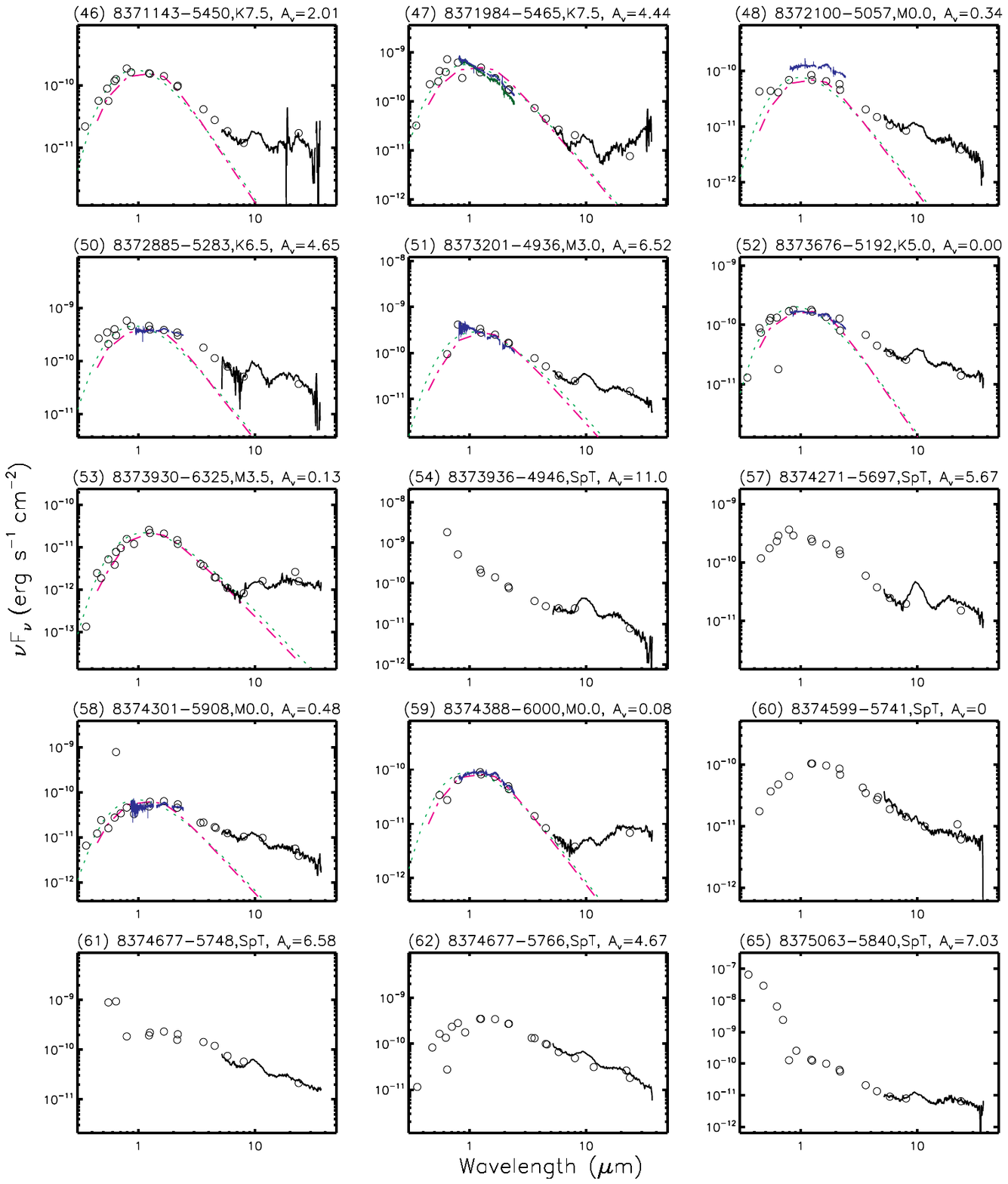}
\caption{continued. }
\end{figure}

\clearpage
\begin{figure}[!htbp]
\ContinuedFloat
\centering
\epsscale{0.9}
\plotone{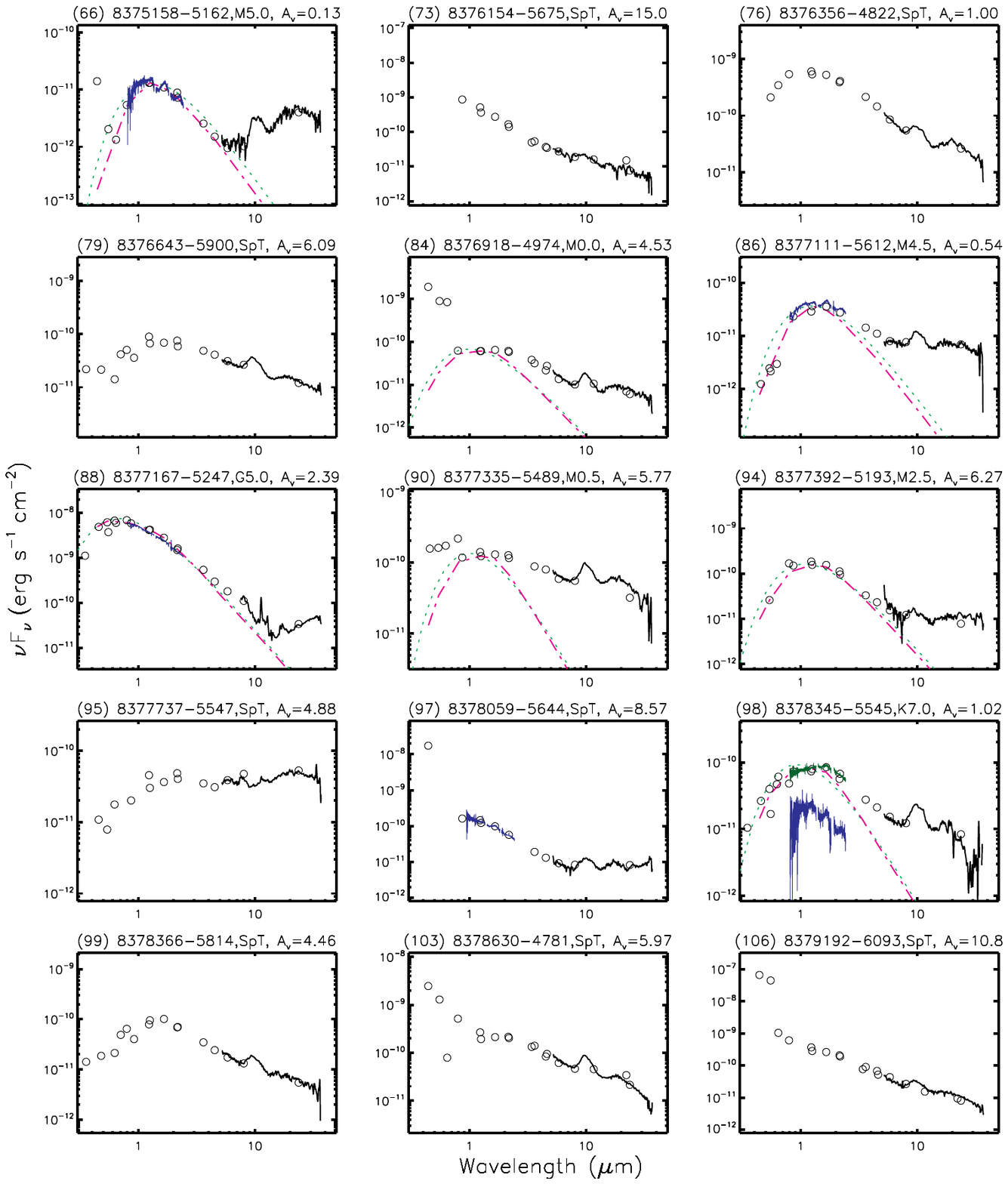}
\caption{continued. }
\end{figure}

\clearpage
\begin{figure}[!htbp]
\ContinuedFloat
\centering
\epsscale{0.9}
\plotone{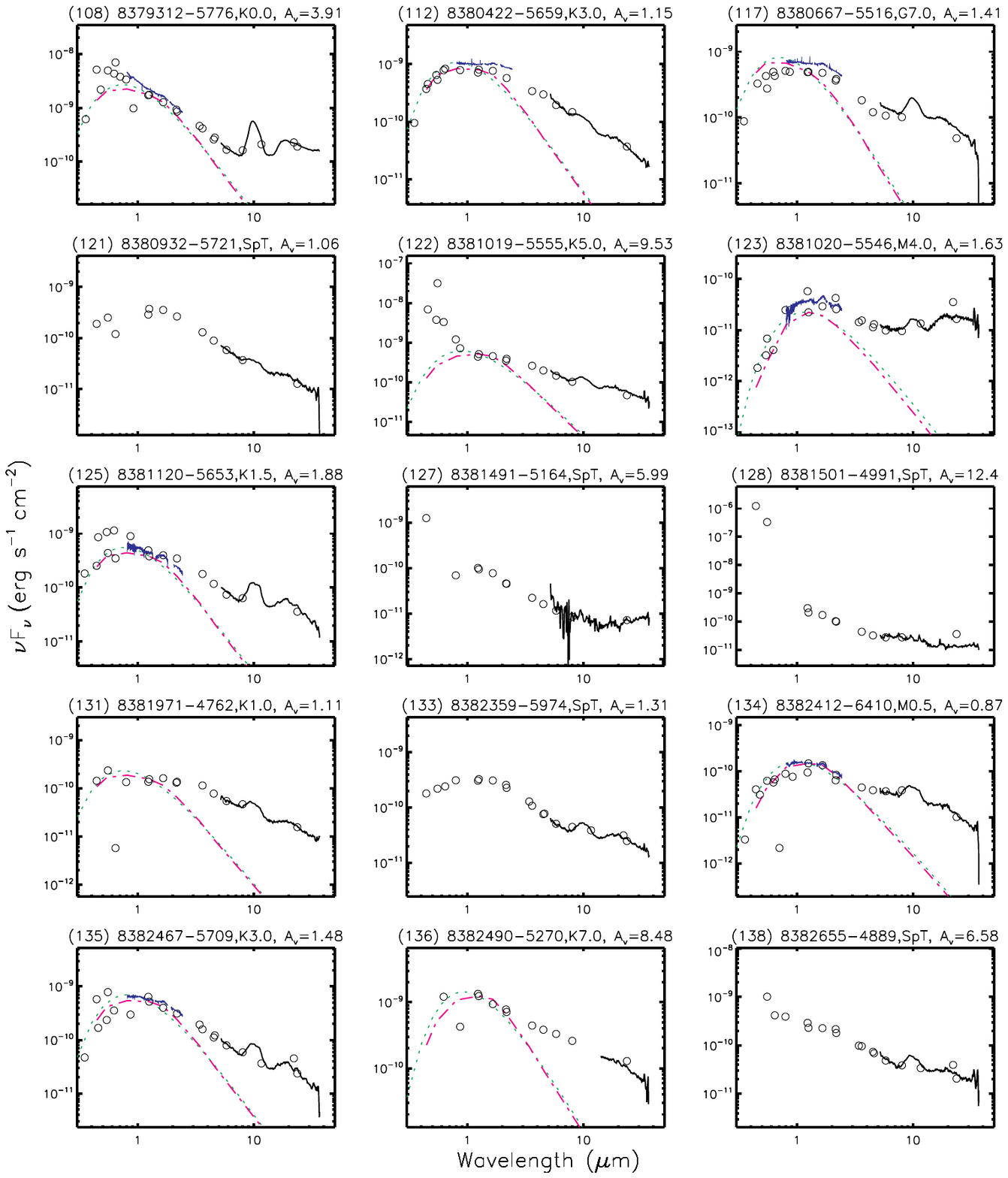}
\caption{continued. }
\end{figure}

\clearpage
\begin{figure}[!htbp]
\ContinuedFloat
\centering
\epsscale{0.9}
\plotone{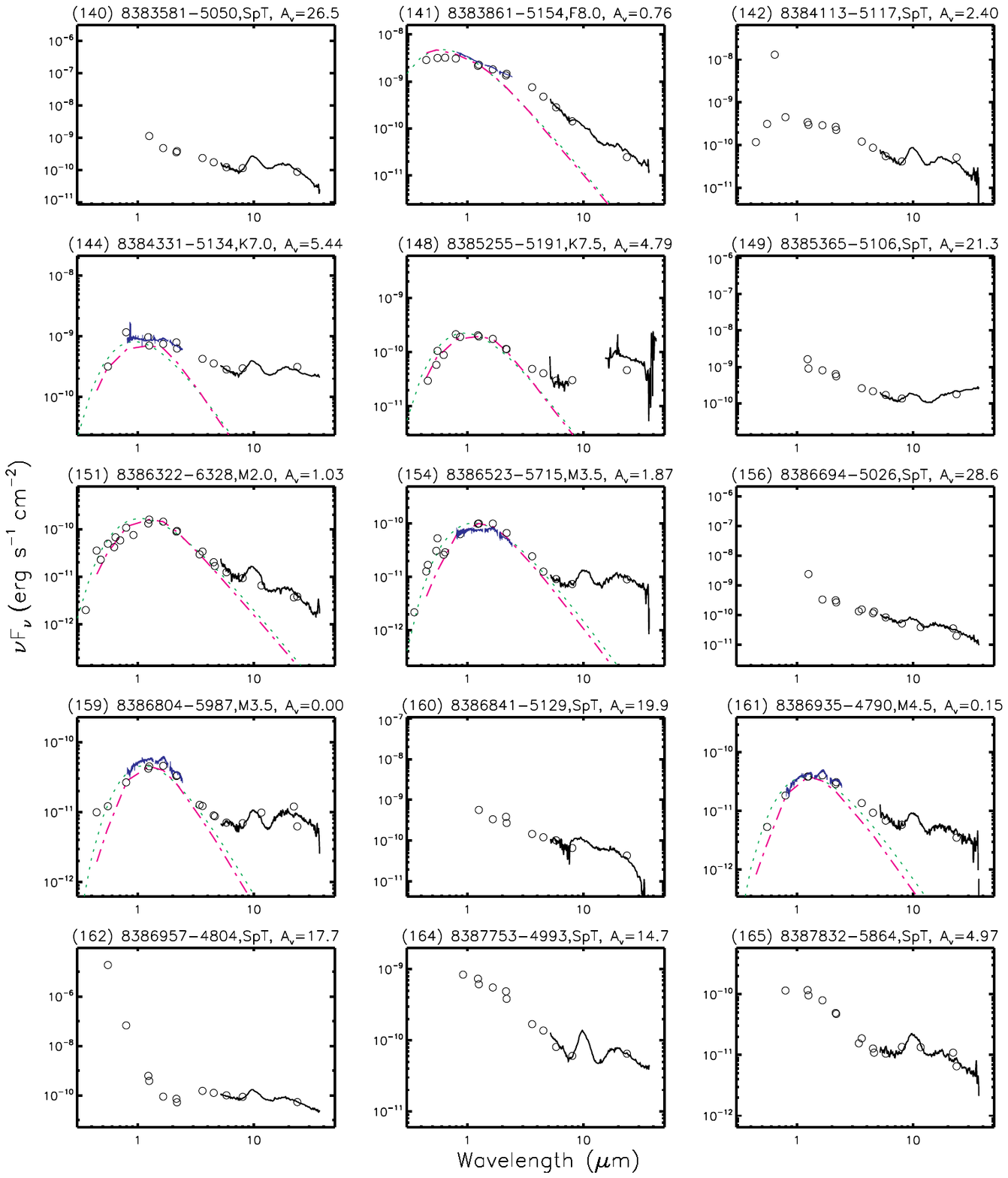}
\caption{continued. }
\end{figure}

\clearpage
\begin{figure}[!htbp]
\ContinuedFloat
\centering
\epsscale{0.9}
\plotone{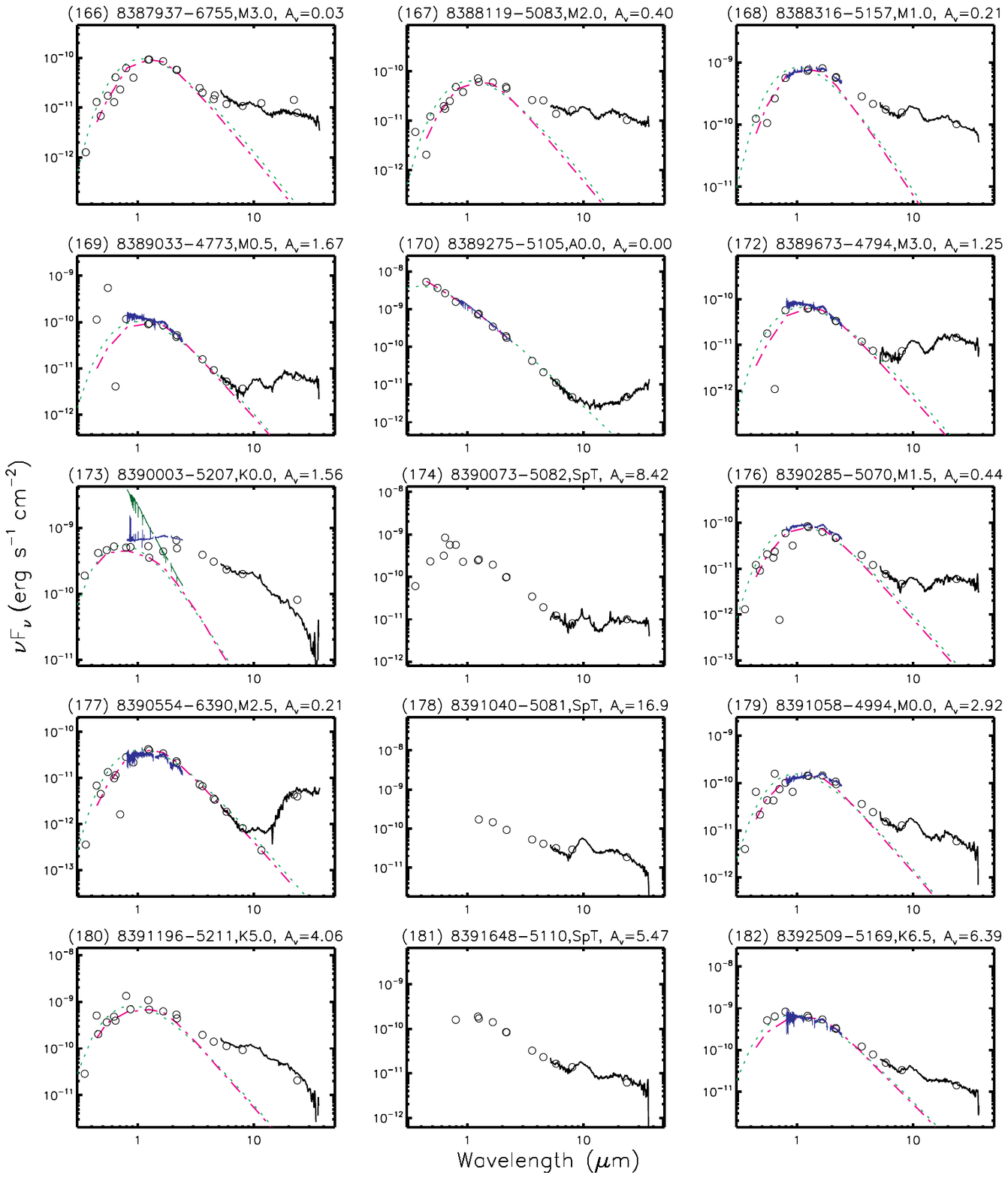}
\caption{continued. }
\end{figure}

\clearpage
\begin{figure}[!htbp]
\ContinuedFloat
\centering
\epsscale{0.9}
\plotone{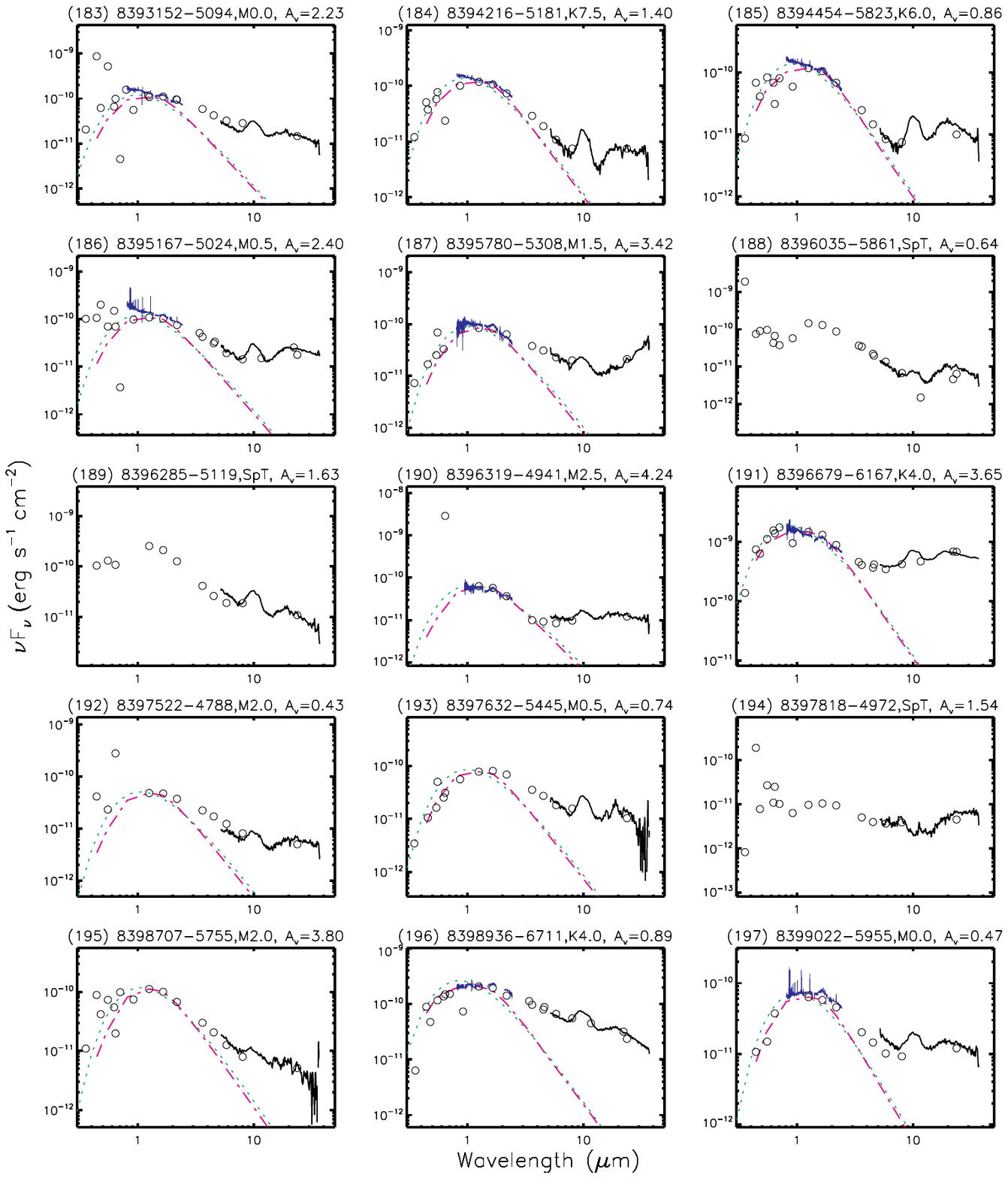}
\caption{continued. }
\end{figure}

\clearpage
\begin{figure}[!htbp]
\ContinuedFloat
\centering
\epsscale{0.9}
\plotone{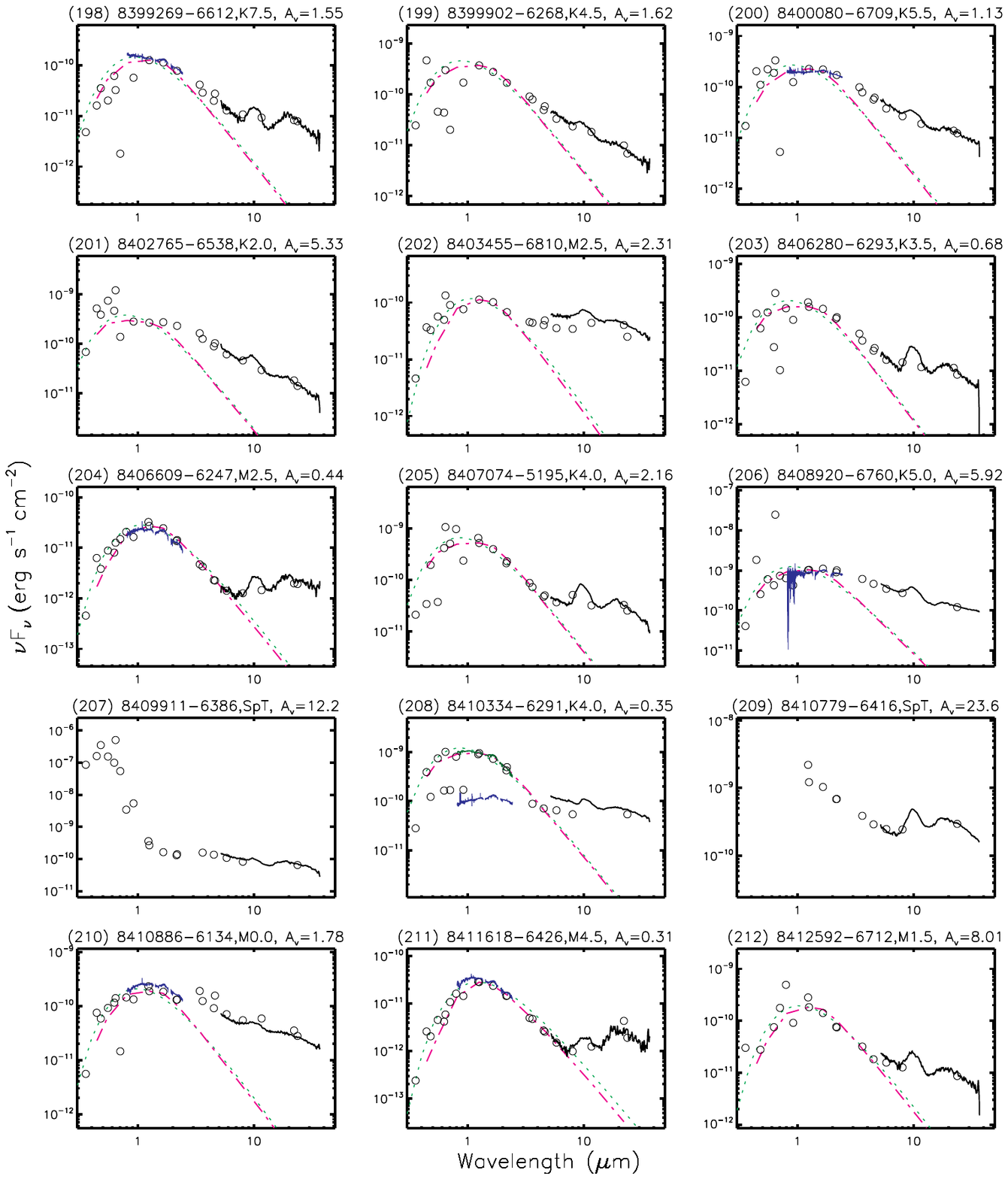}
\caption{continued. }
\end{figure}

\clearpage
\begin{figure}[!htbp]
\ContinuedFloat
\centering
\epsscale{0.9}
\plotone{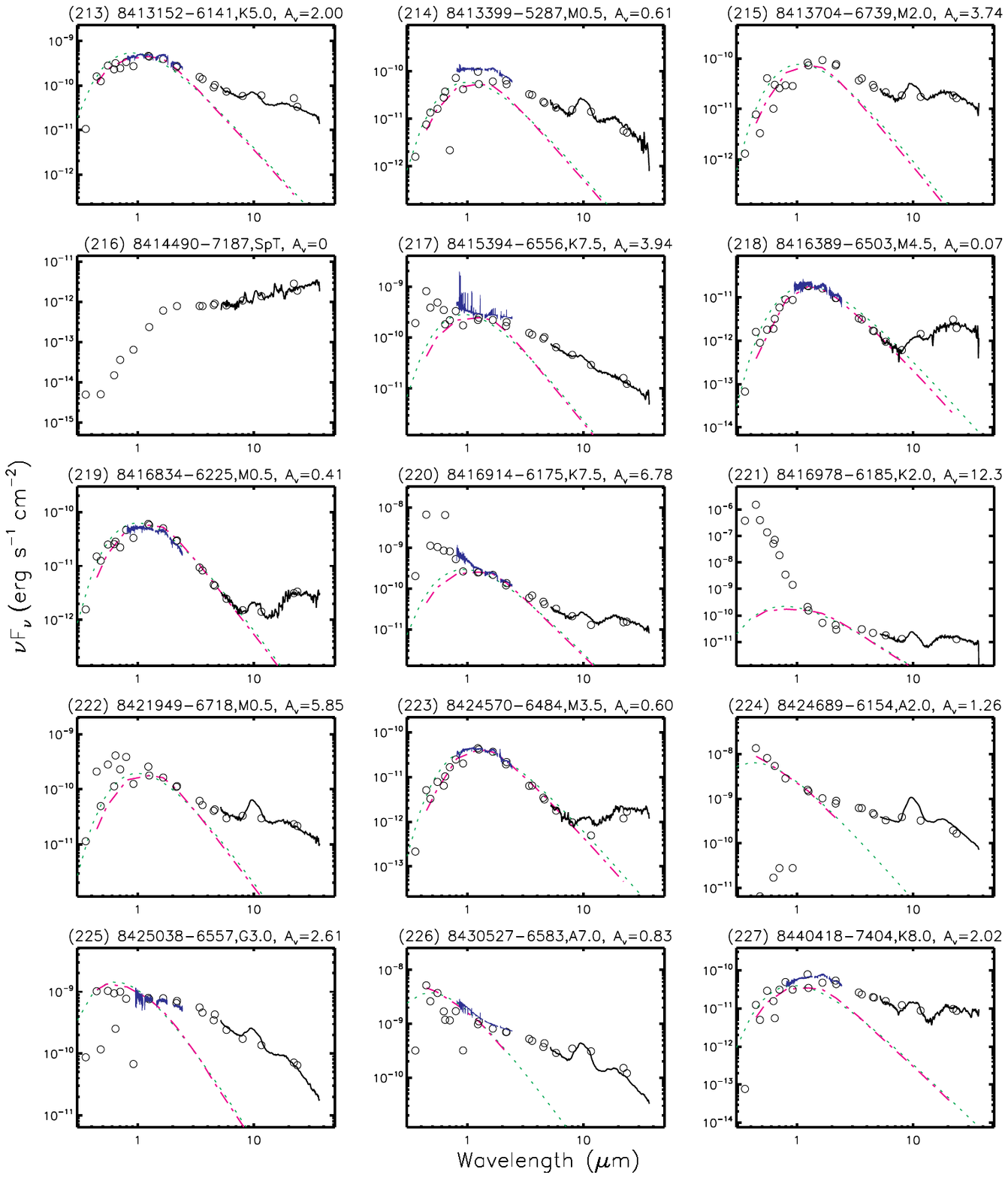}
\caption{continued. }
\end{figure}

\clearpage
\begin{figure}[!htbp]
\ContinuedFloat
\centering
\epsscale{0.9}
\plotone{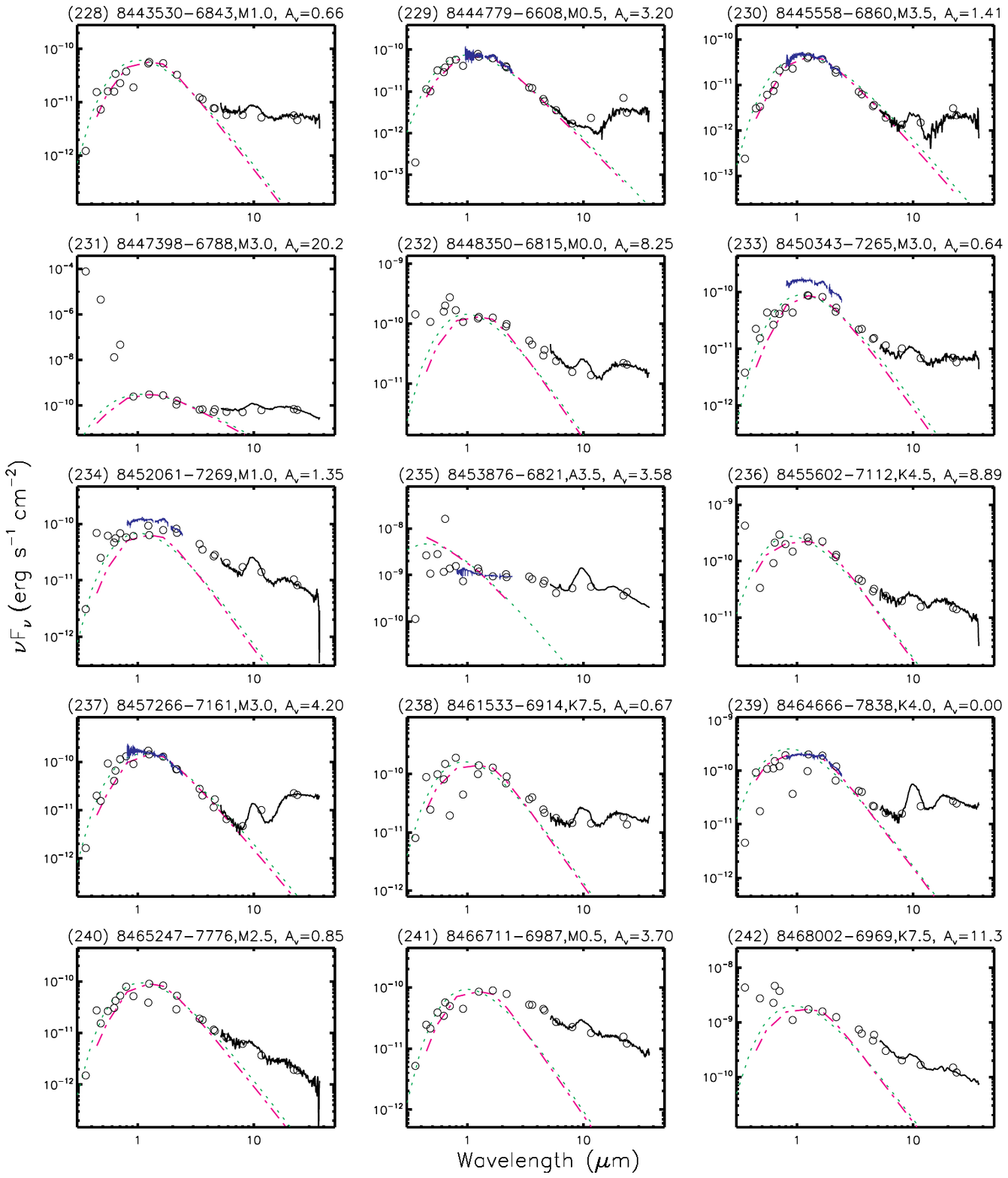}
\caption{continued. }
\end{figure}

\clearpage
\begin{figure}[!htbp]
\ContinuedFloat
\centering
\epsscale{0.9}
\plotone{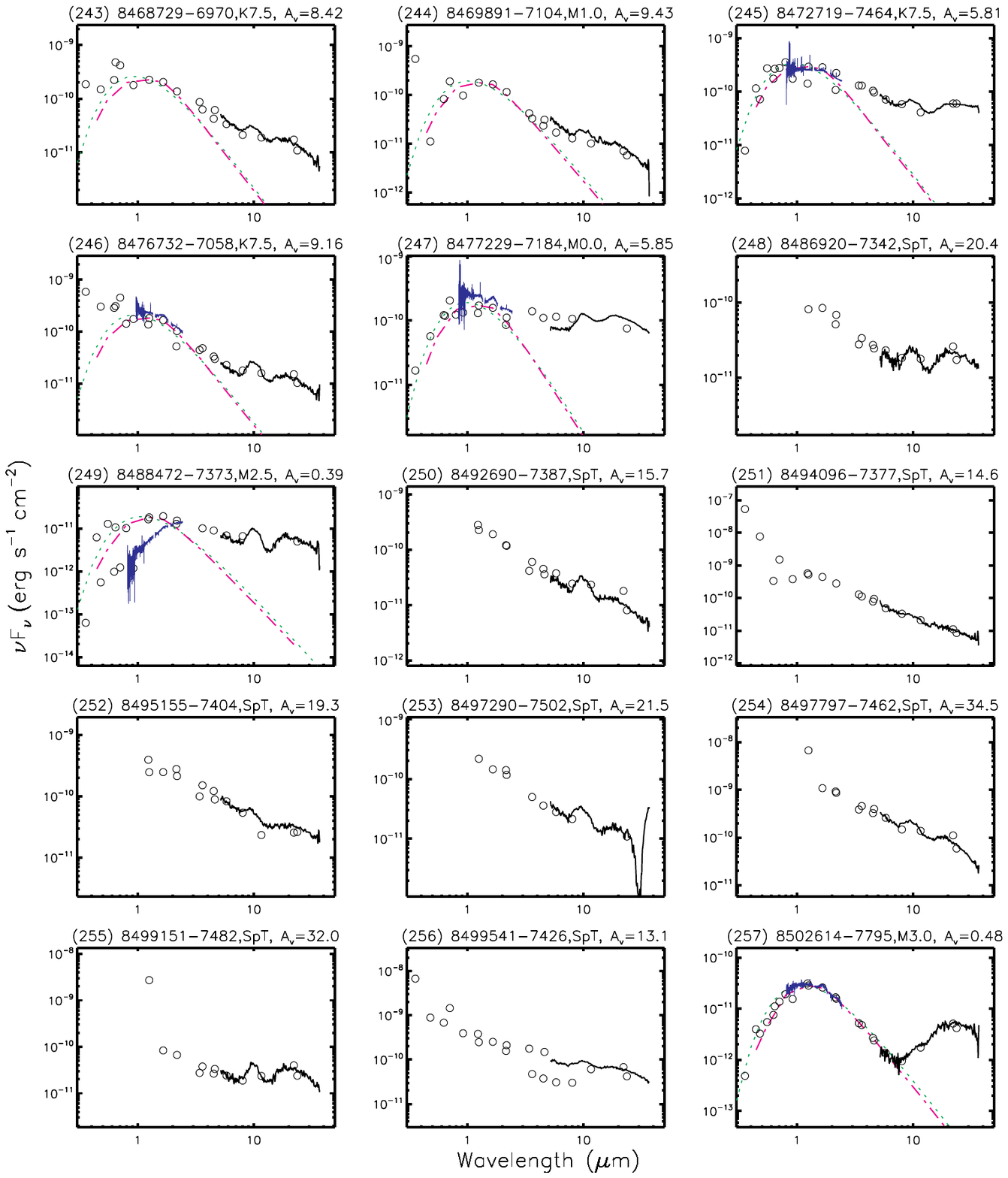}
\caption{continued. }
\end{figure}

\clearpage
\begin{figure}[!htbp]
\ContinuedFloat
\centering
\epsscale{0.9}
\plotone{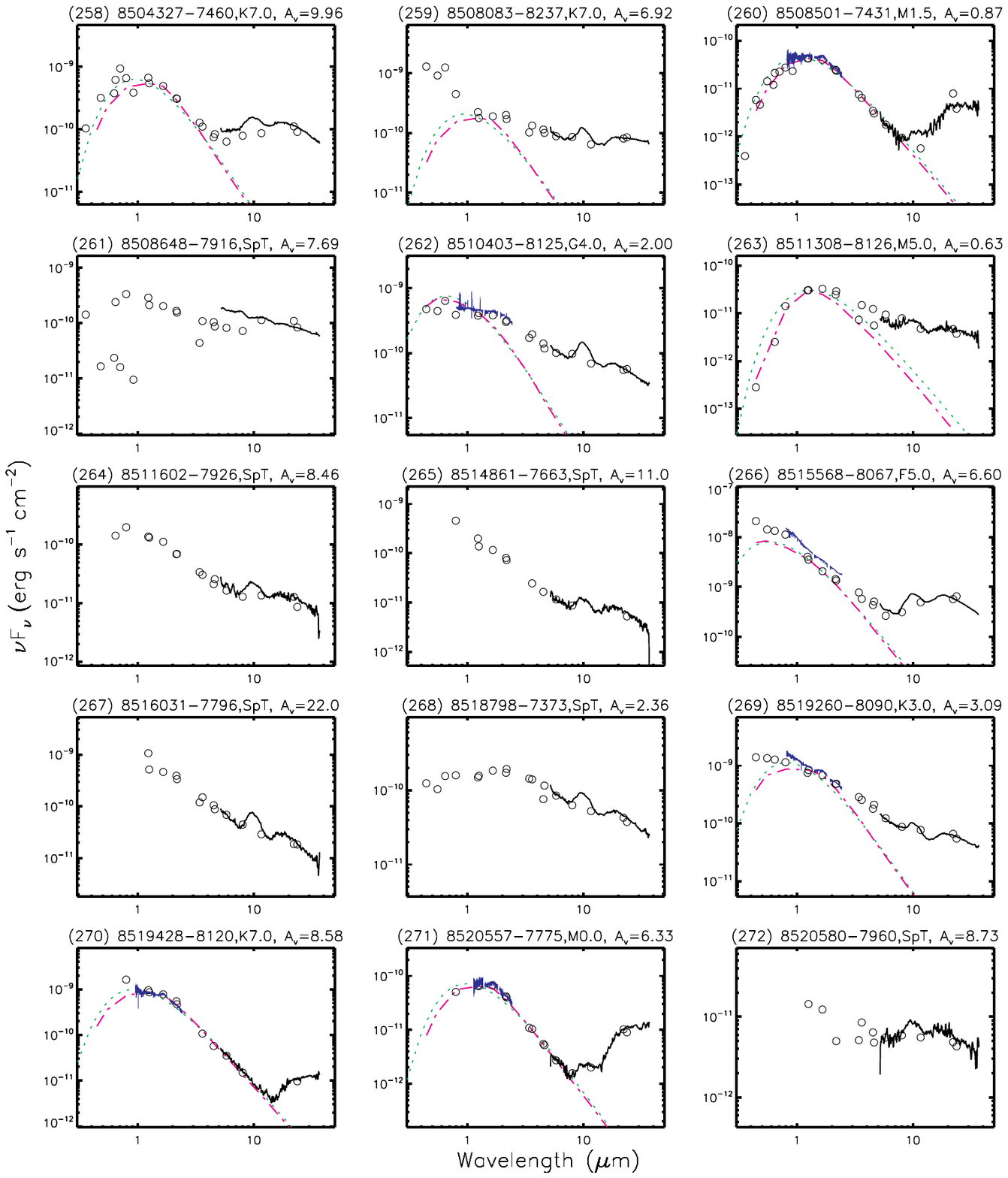}
\caption{continued. }
\end{figure}

\clearpage
\begin{figure}[!htbp]
\ContinuedFloat
\centering
\epsscale{0.9}
\plotone{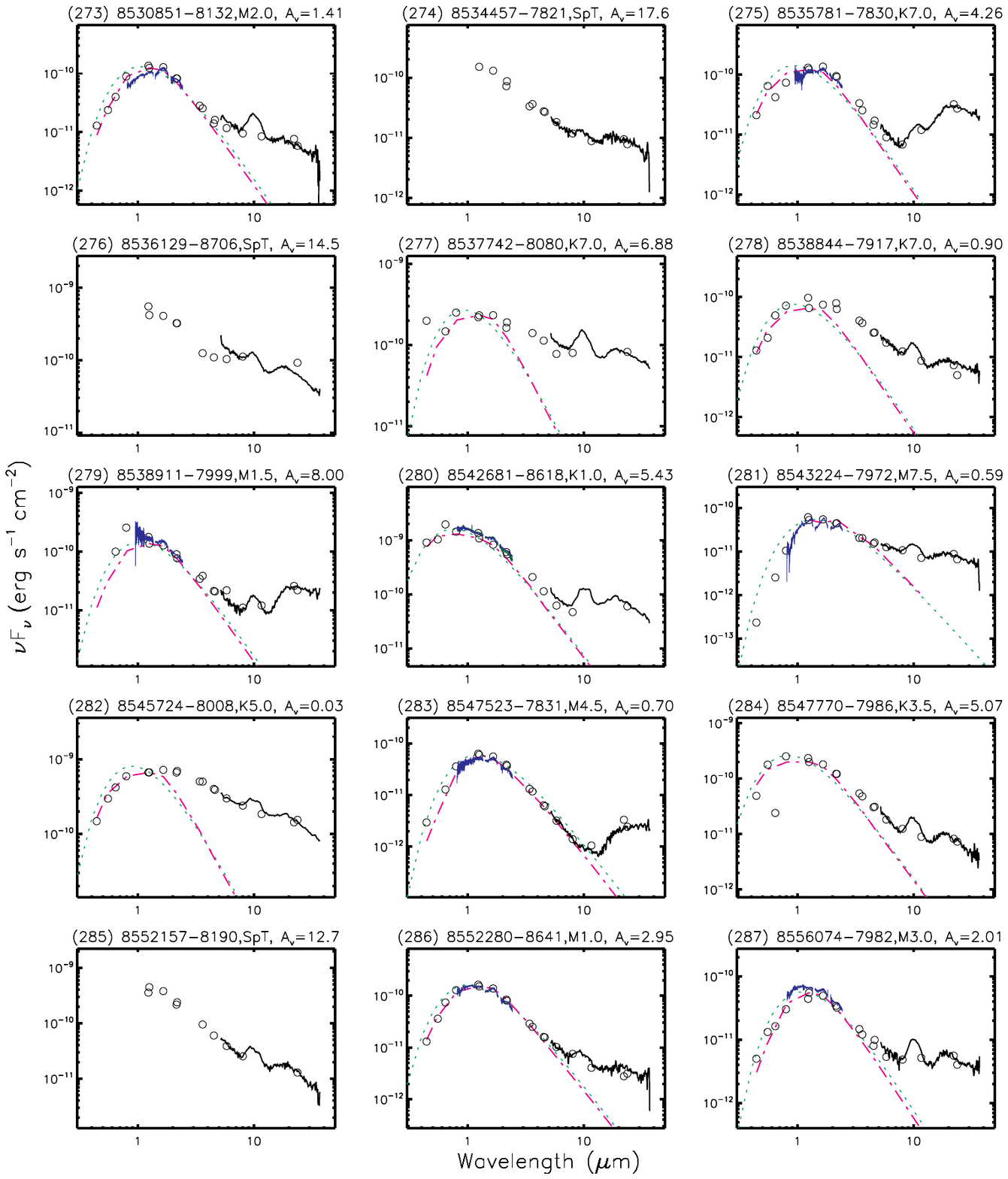}
\caption{continued. }
\end{figure}

\clearpage
\begin{figure}[!htbp]
\ContinuedFloat
\centering
\epsscale{0.9}
\plotone{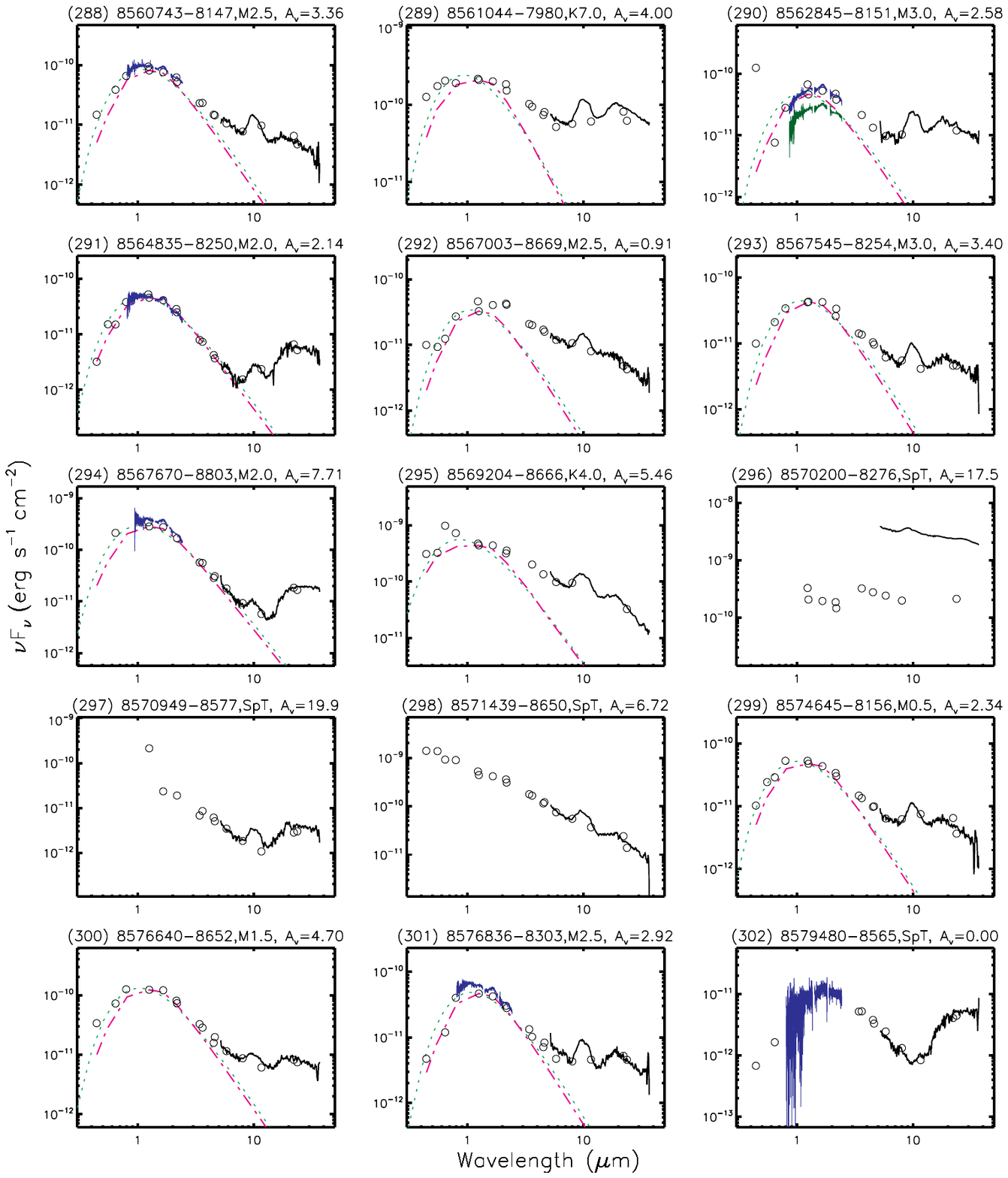}
\caption{continued. }
\end{figure}

\clearpage
\begin{figure}[!htbp]
\ContinuedFloat
\centering
\epsscale{0.9}
\plotone{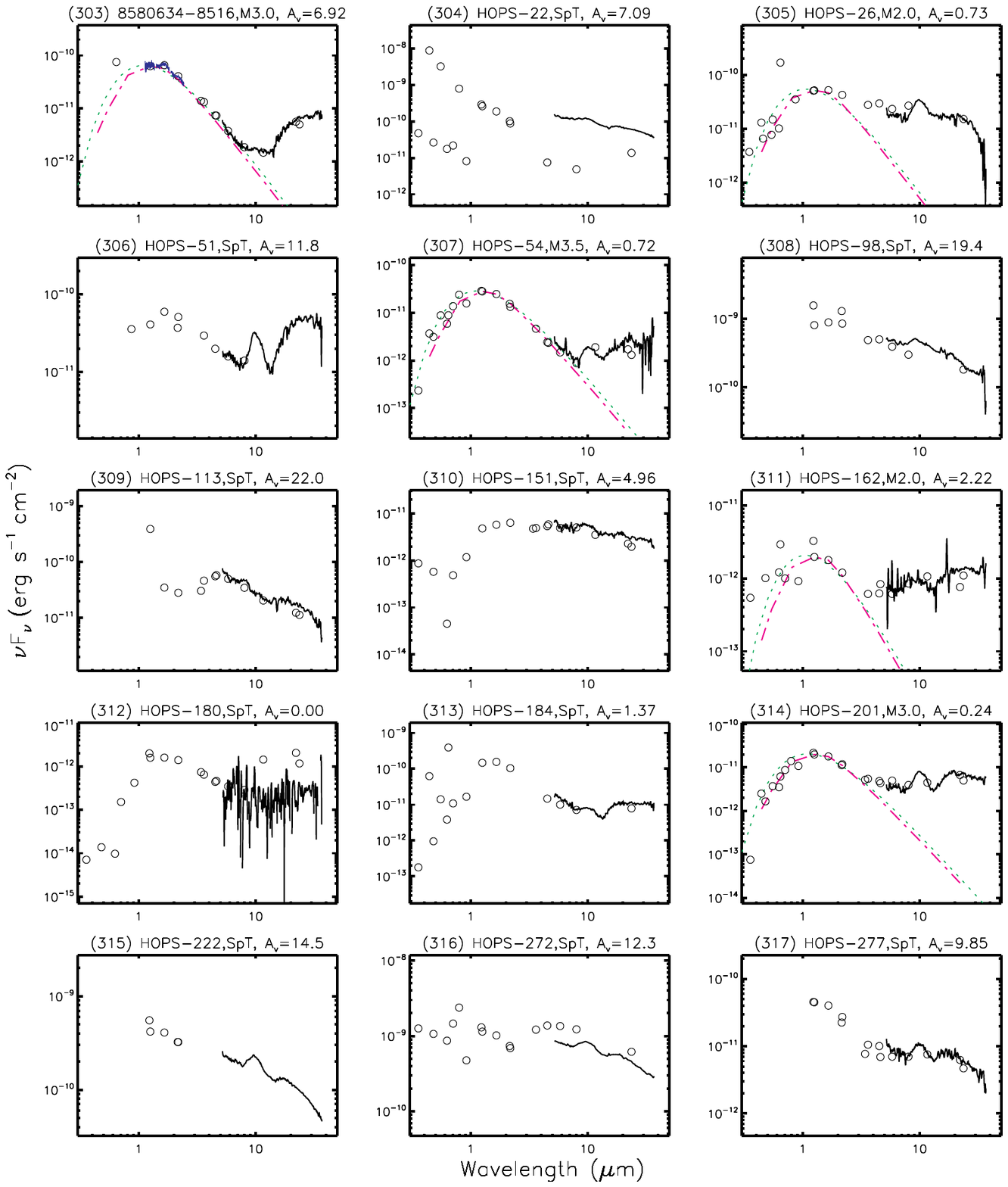}
\caption{continued. }
\end{figure}

\clearpage
\begin{figure}[!htbp]
\ContinuedFloat
\centering
\epsscale{0.9}
\plotone{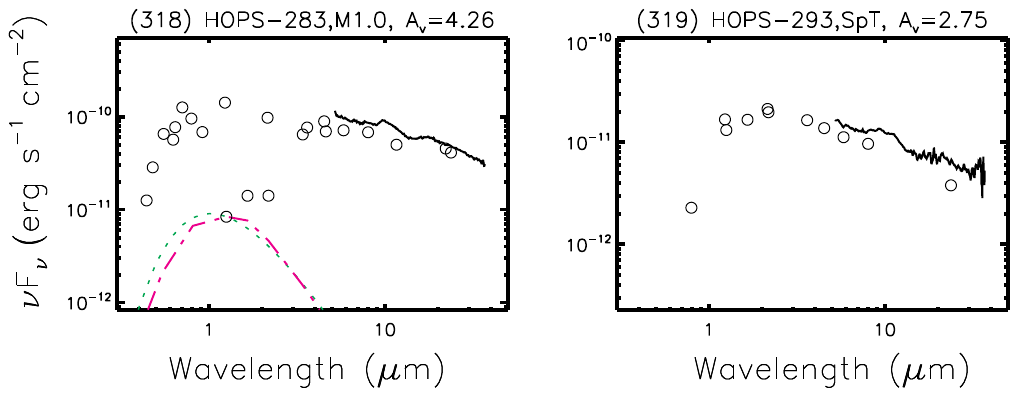}
\caption{continued. }
\end{figure}

\clearpage
\begin{figure}[!htbp]
\epsscale{0.8}
\plotone{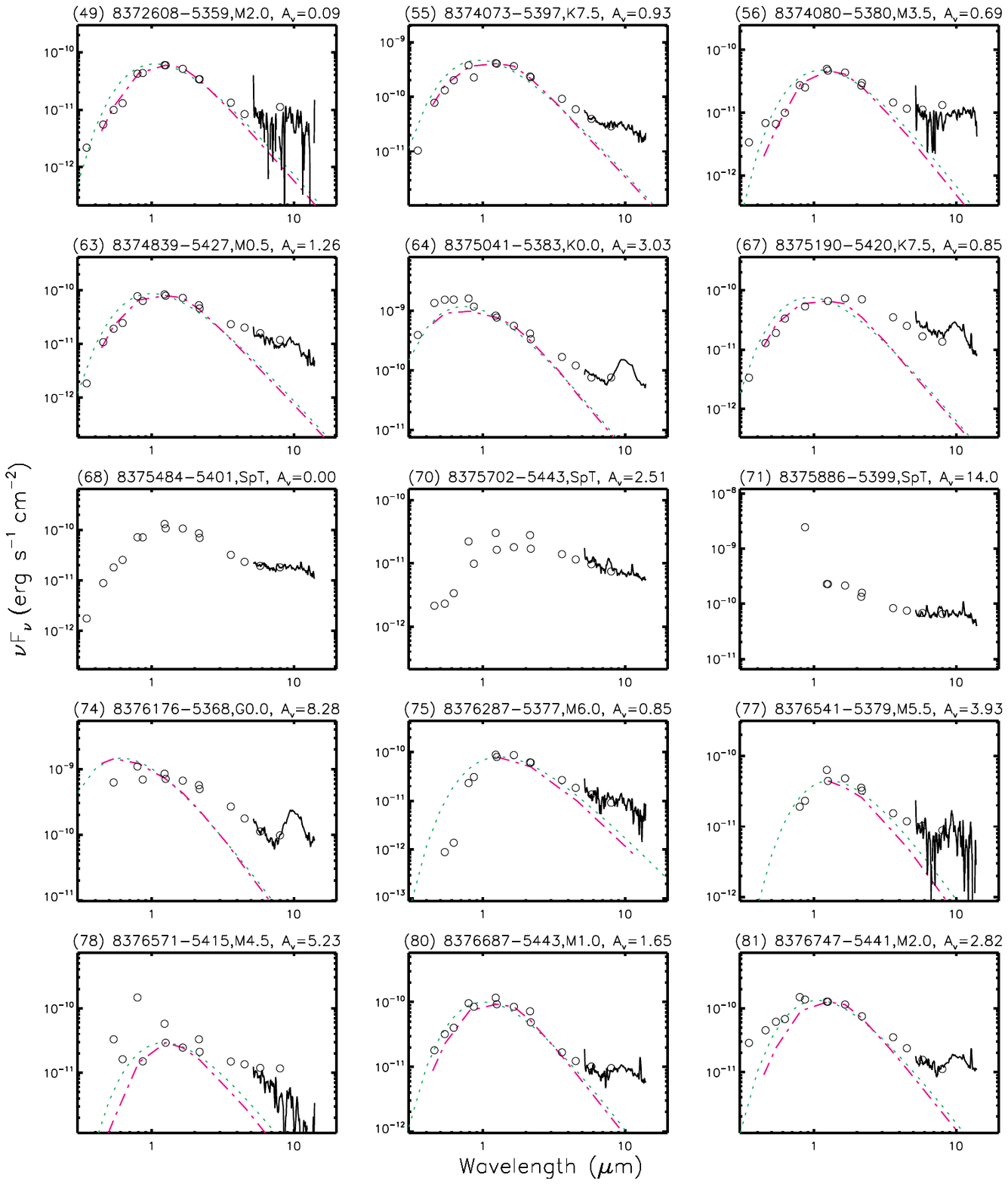}
\caption{De-reddened SEDs of the reduced samples in Trapezium, which were observed with IRS SL module only. The meanings of symbols are same as in Figure~\ref{fig-SLLL-SED}.  \label{fig-SLonly-SED}}
\end{figure}

\clearpage
\begin{figure}[!htbp]
\ContinuedFloat
\epsscale{1}
\plotone{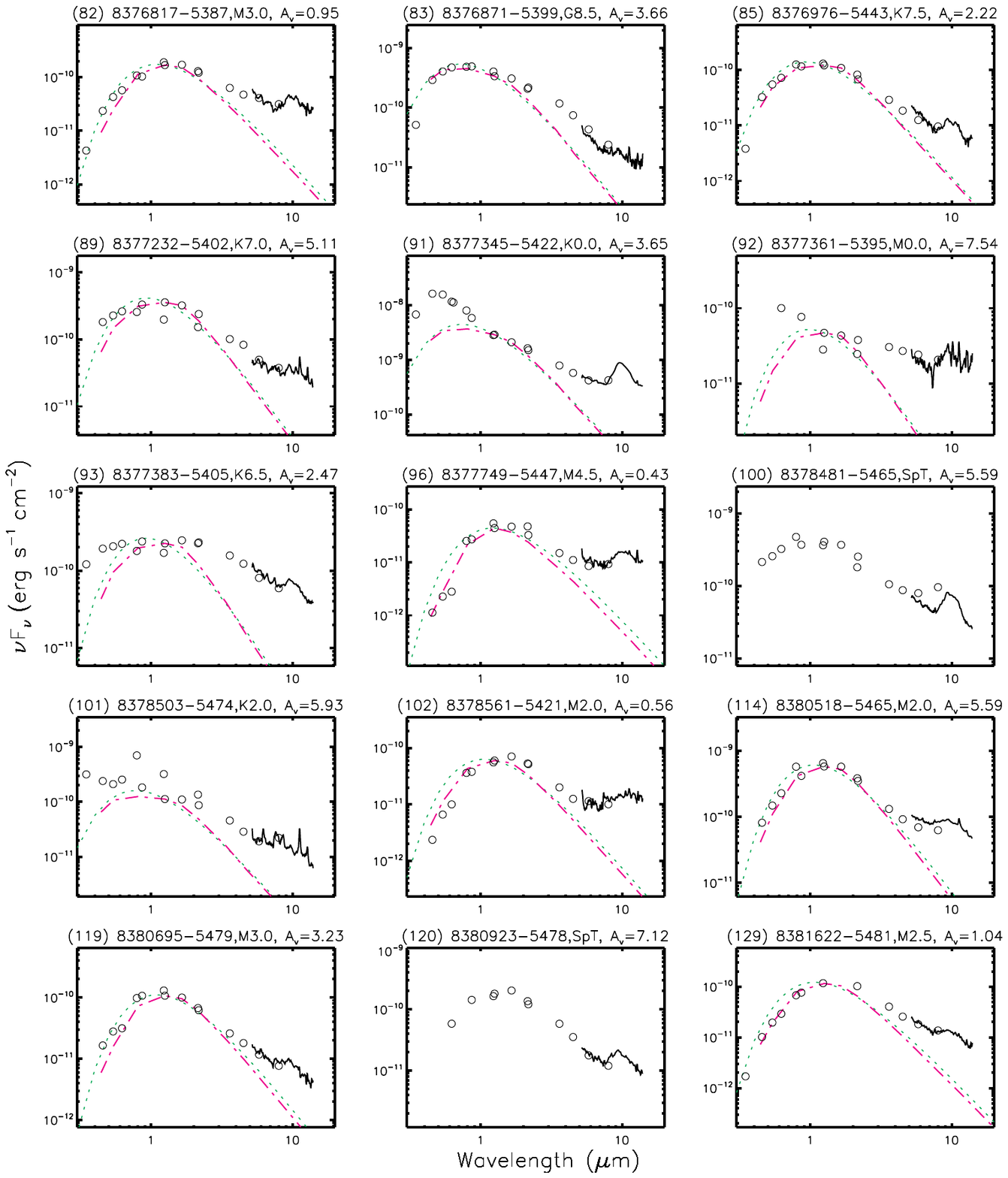}
\caption{Continued.}
\end{figure}

\clearpage
\begin{figure}[!htbp]
\ContinuedFloat
\epsscale{1}
\plotone{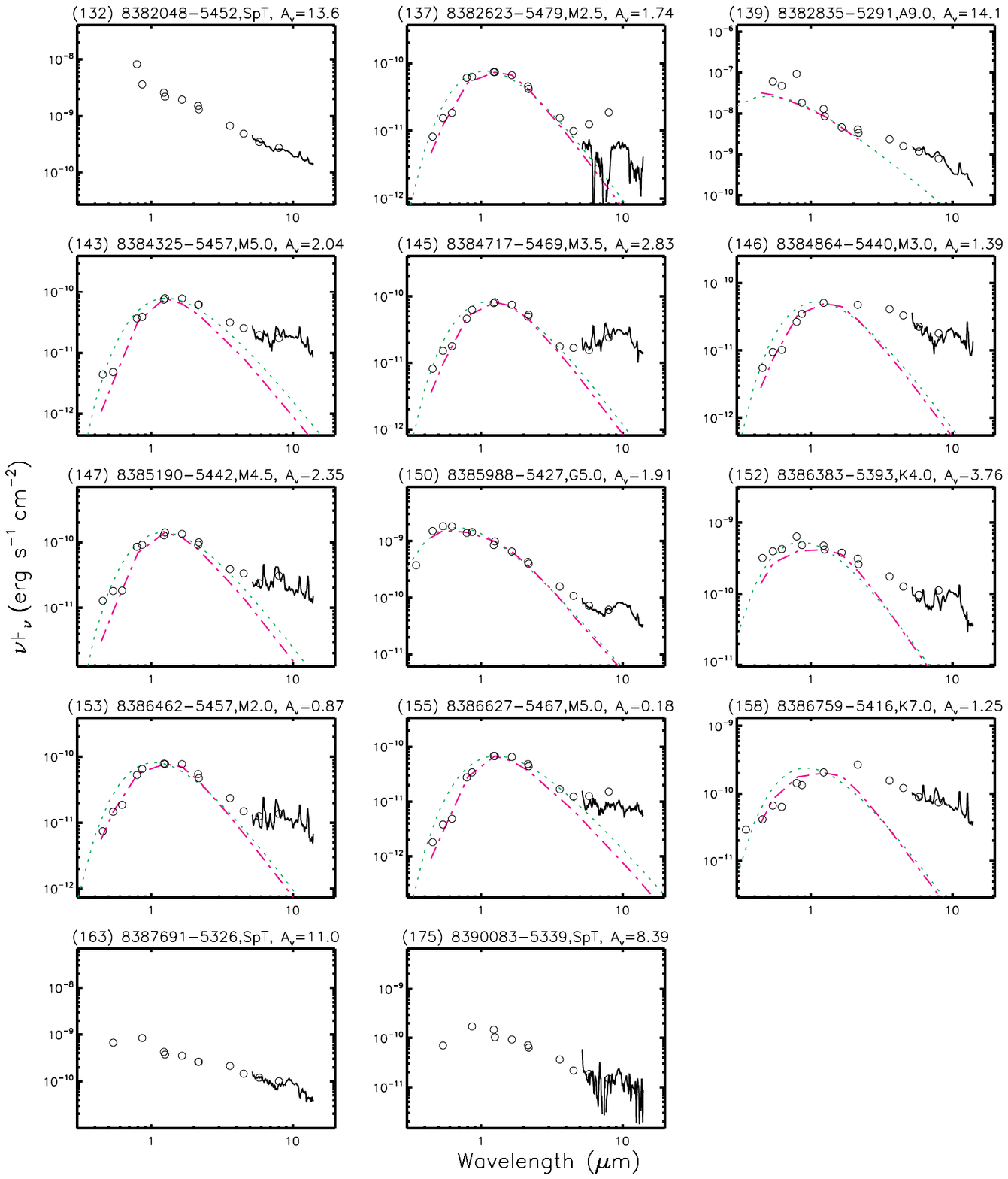}
\caption{Continued.}
\end{figure}

\begin{figure}[!htbp]
\epsscale{1}
\plotone{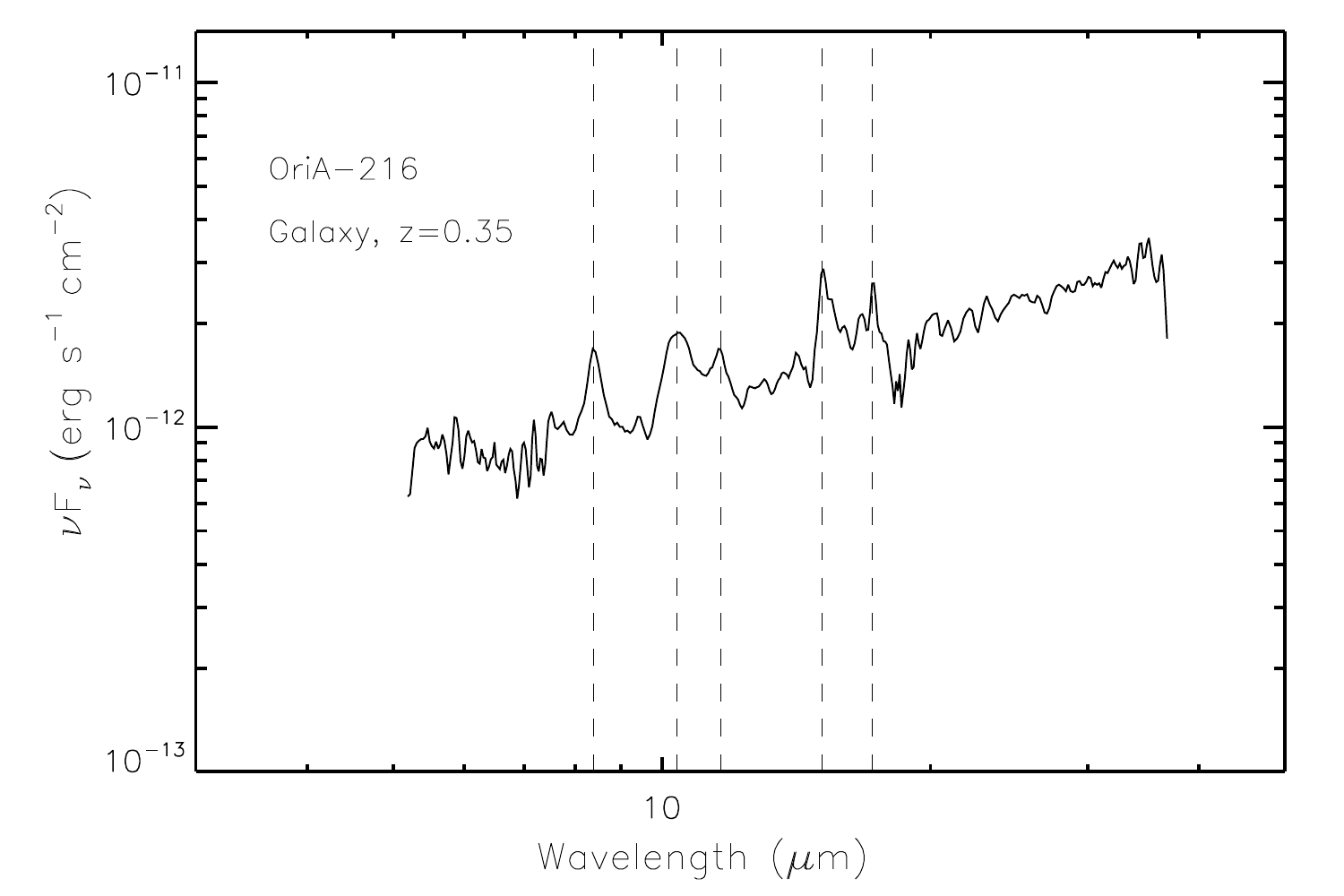}
\caption{IRS spectrum of OriA-216 which is a background galaxy with redshift, z=0.35. The dashed lines indicate the PAHs features of 6.2, 7.7, 8.6, 11.2, and 12.7 $\micron$ at the redshifted wavelengths at 8.38, 10.38, 11.63, 15.12, and 17.2 $\micron$, respectively. \label{fig-216galaxy}}
\end{figure}

\begin{figure}[!htbp]
\epsscale{0.5}
\plotone{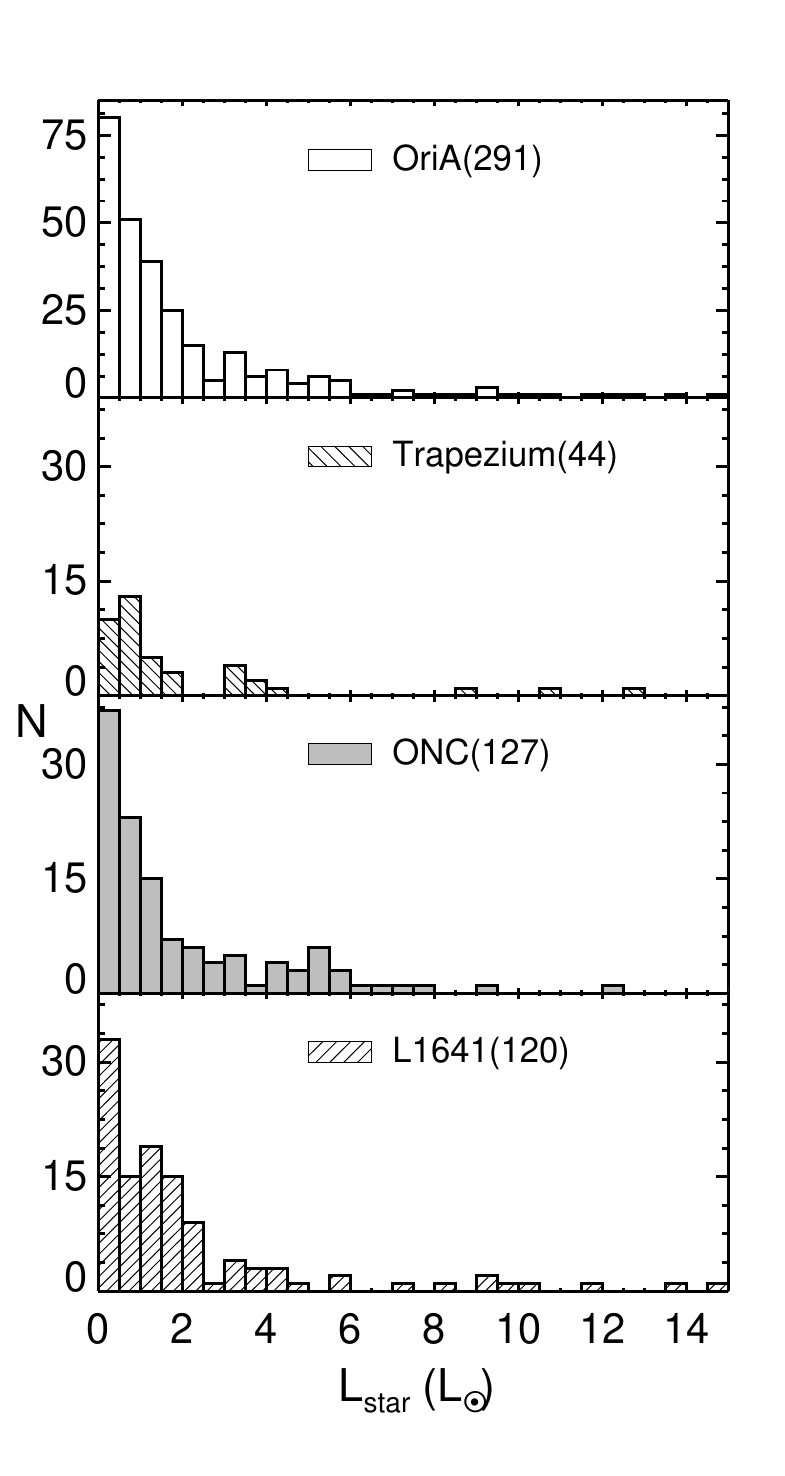}
\caption{Distributions of $L_{\star}$ of the objects in Orion A.   \label{fig-Lstar-OriA}}
\end{figure}

\begin{figure}[!htbp]
\epsscale{0.8}
\plotone{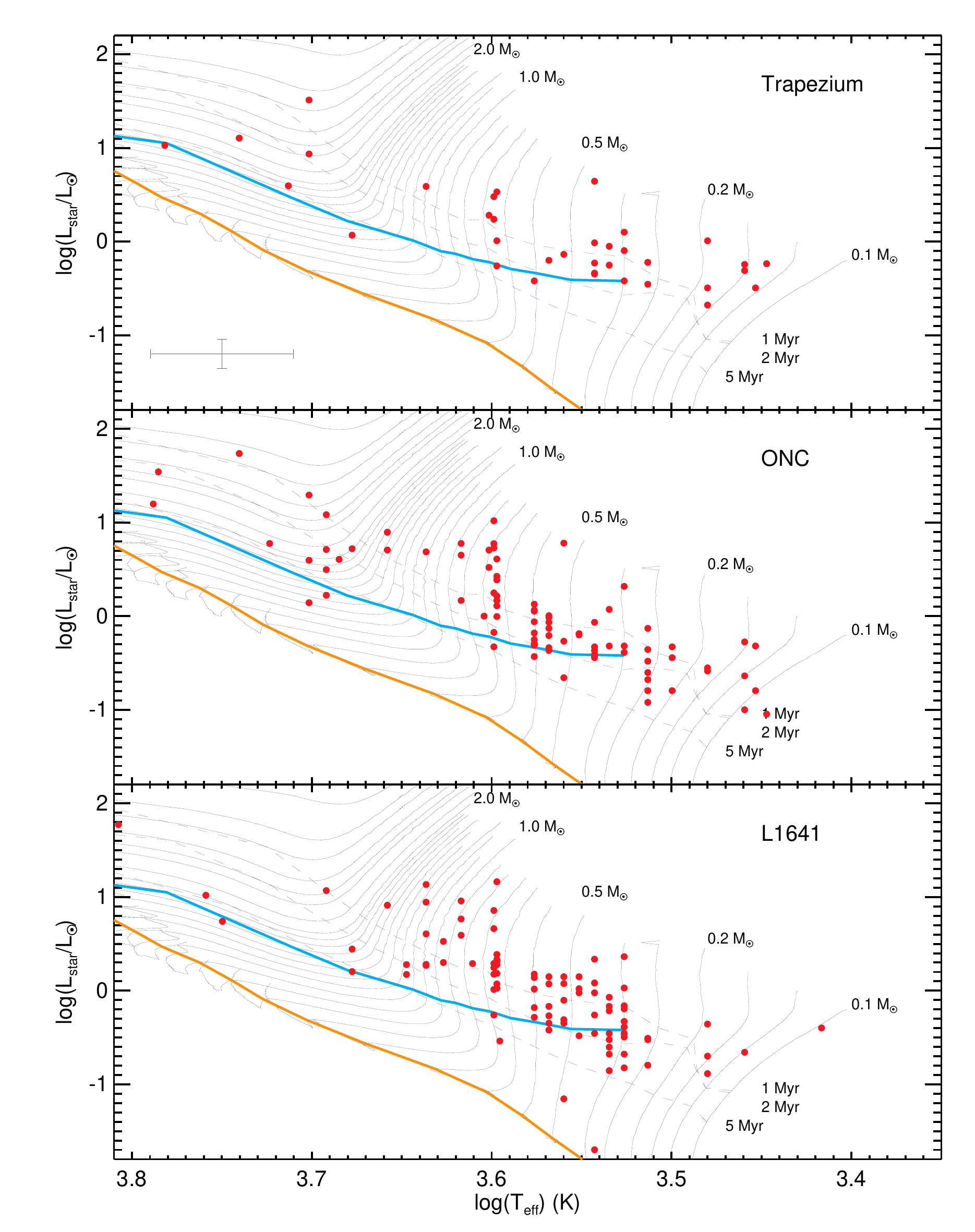}
\caption{H-R diagrams for host stars of Class II objects in Orion A. Evolutionary tracks and isochrones are from \cite{siess2000} (Z=0.02). Isochrone ages of systems of Class II objects in Orion A range from $<$ 1 Myr to $\sim$ 5 Myr. The average disk life time in Tau-Aur \citep{bertout07disktime} and the Zero Age Main Sequence (ZAMS) are also shown as cyan and orange solid lines, respectively, for reference. The cross on the left bottom of the top pannel represents a typical uncertainty in the spectral types of few ($\sim$3) subclasses. The represented uncertainty is calculated by assuming the uncertainty of $T_{eff}$ is about 345 K, which is 0.16 in $log(L_{\star}/L_{\odot})$ and 0.04 in $log(T_{eff})$.\label{fig-HRD-OriA}}
\end{figure}

\begin{figure}[!htbp]
\epsscale{0.5}
\plotone{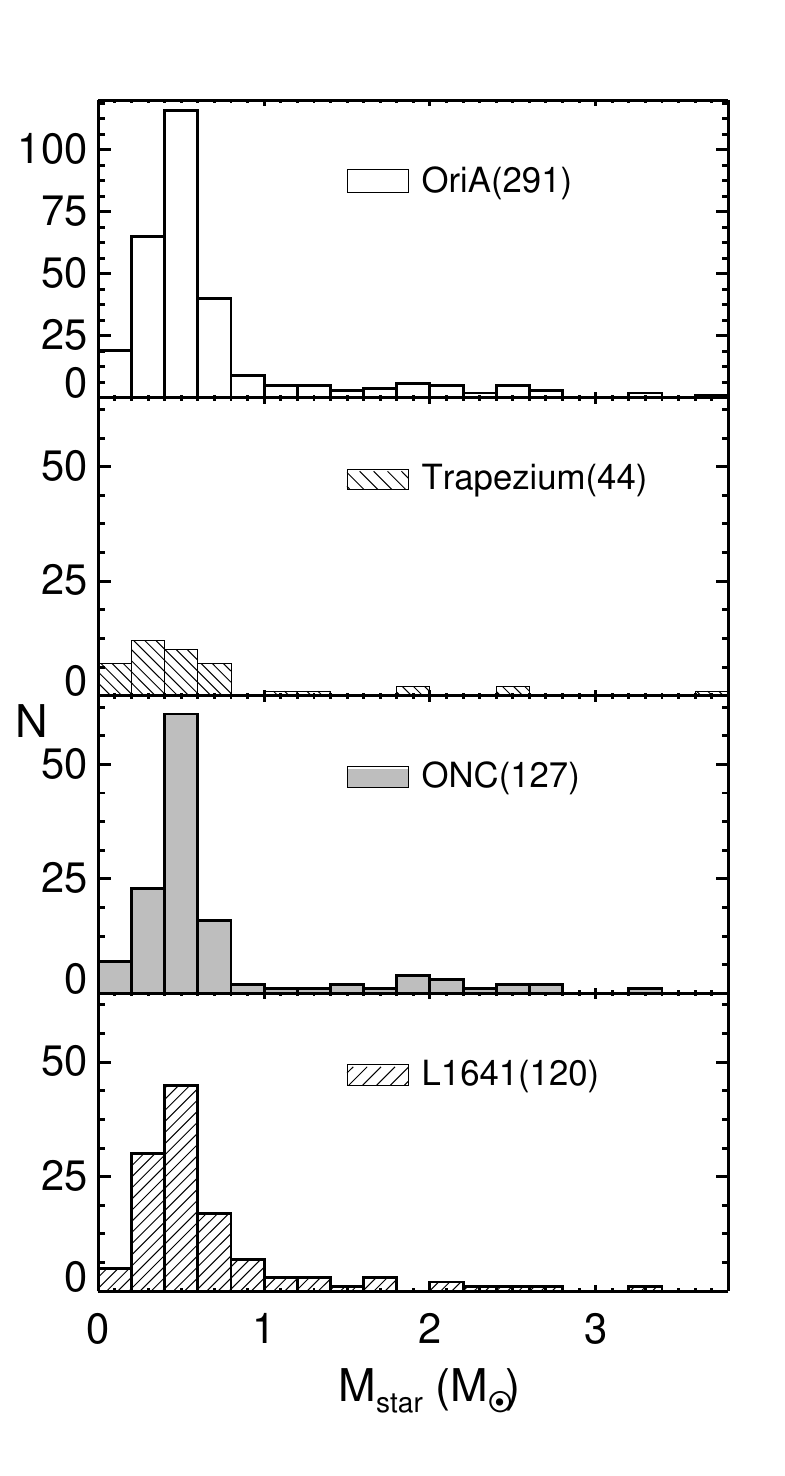}
\caption{Distributions of $M_\star$ of the objects in Orion A. \label{fig-Mstar-OriA}}
\end{figure}

\begin{figure}[!htbp]
\epsscale{0.7}
\plotone{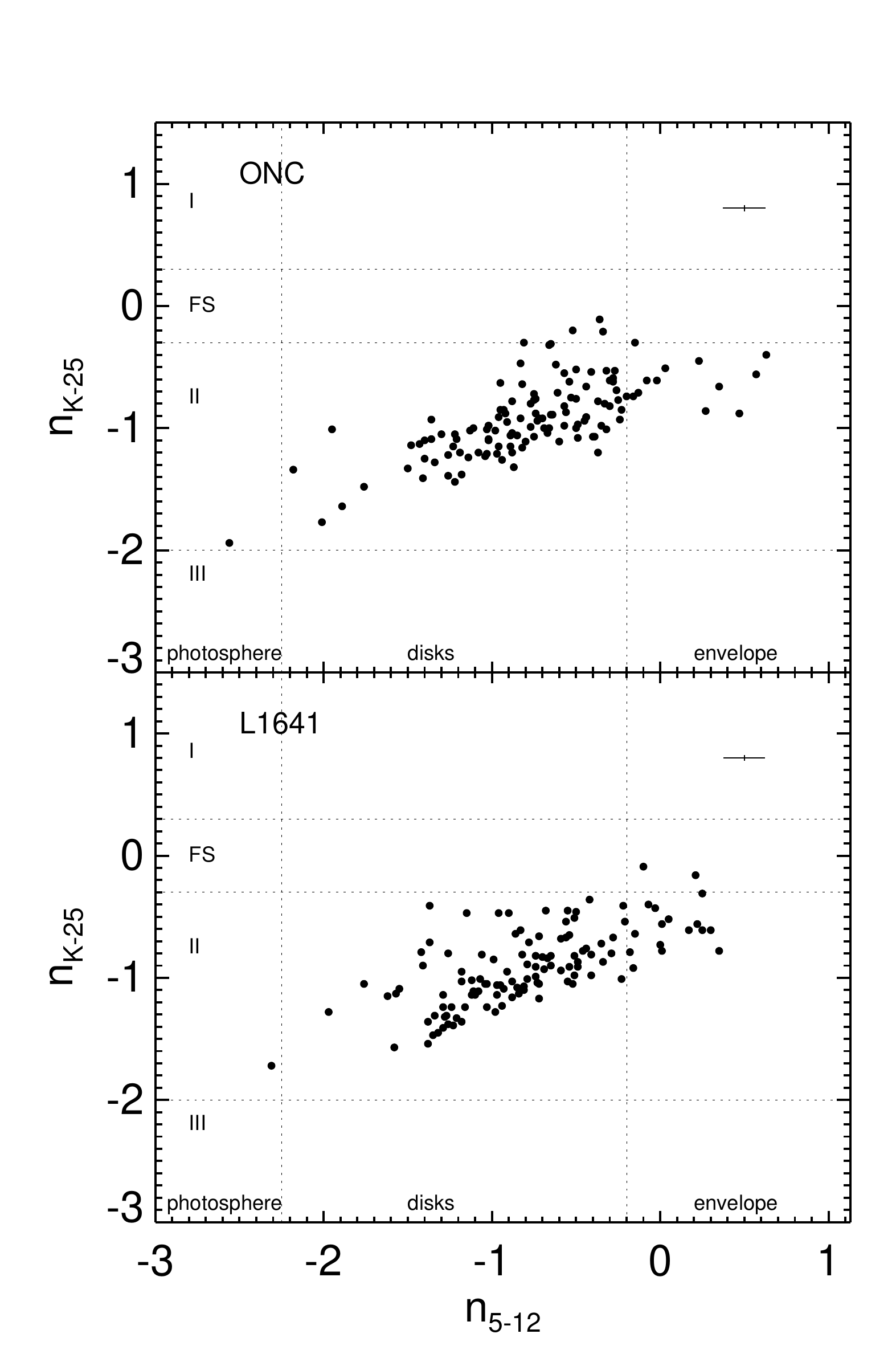}
\caption{Comparison of the observed spectral indices and disk classification by $n_{5-12}$ versus $n_{K-25}$ for ONC (top) and L1641 (bottom). The horizontal dotted lines divide the regions accupied by Class I, Flat spectrum (FS), Class II, and Class III objects by the $n_{K-25}$ criteria. The vertical dotted lines split the regions into photosphere, disks, and envelope by $n_{5-12}$. The typical errors are indicated on the top-right corner of each panel. \label{fig-n512-nk25}}
\end{figure}

\begin{figure}[!htbp]
\epsscale{0.7}
\plotone{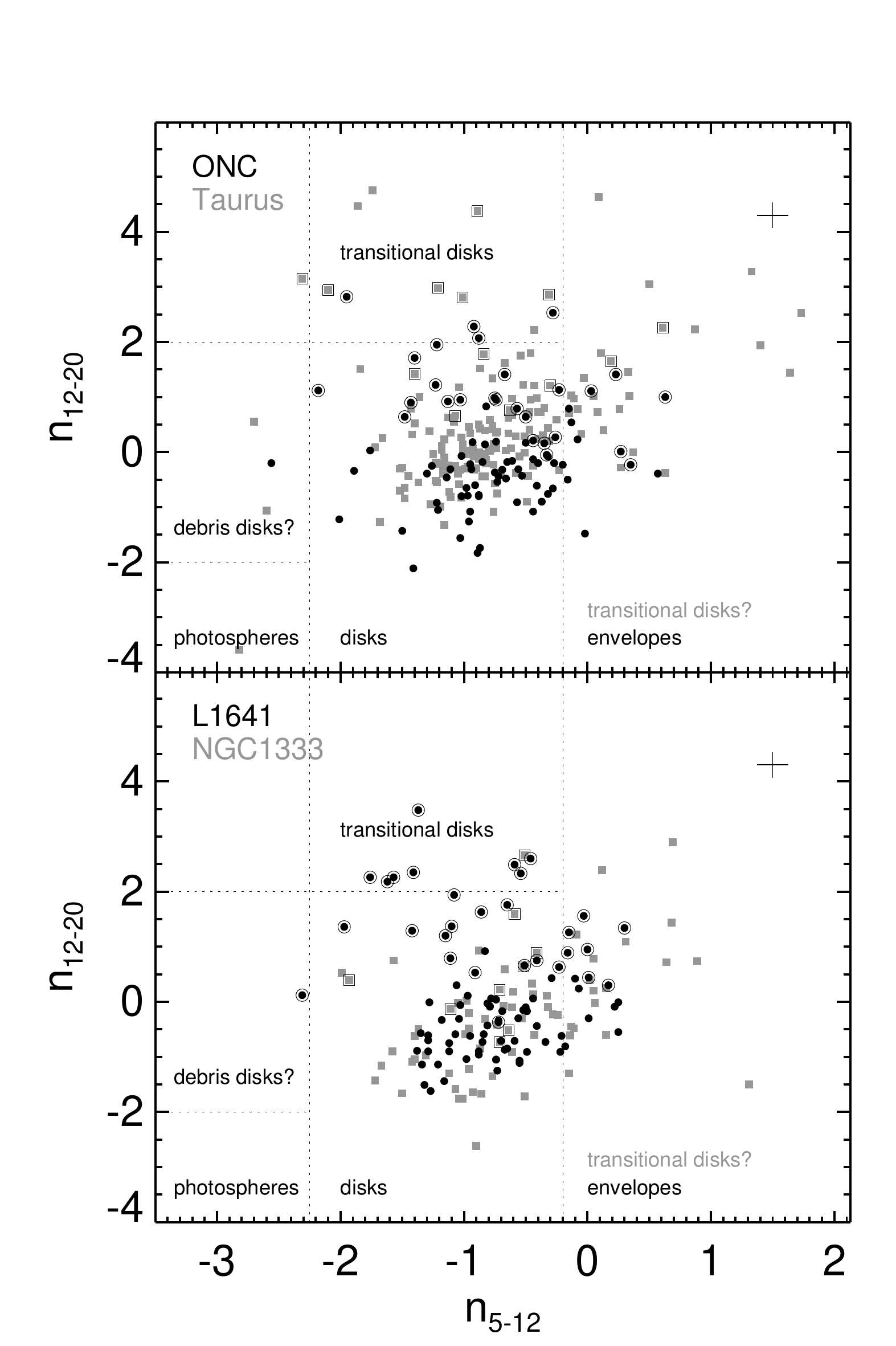}
\caption{Comparison of the observed spectral indices and disk classification by $n_{5-12}$ versus $n_{12-20}$. The verical dotted lines indicate general division of envelopes, disks, and photospheric objects. Envelopes usually lie at $n_{5-12} \geq$ -0.2, disks at -2.25 $\leq n_{5-12} <$ -0.2, and photospheres at $n_{5-12} \leq$ -0.2. Transitional disks occupy the region of disks in $n_{5-12}$, but have $n_{12-20} > 2$; a few can also be found at 0 $\leq n_{12-20} <$ 2 and $n_{5-12} \geq$ -0.2. Debris disks have $n_{5-12}$ values of photospheres, but $n_{12-20}$ in the disk range. The upper panel is for objects in ONC (black circles) and for objects in Taurus (\citet{Furlan2011Taurus}: gray suaqres) for comparison. The lower panel is for objects in L1641 (black circles) and for objects in NGC 1333 (\citet{Arnold2012}: gray squares) for comparisoin. TDs classified in \citet{khkim2013-OriATD} are indicated with larger open circles encompassing the filled circles for ONC and L1641 and larger open squares  encompassing the filled squares for Taurus and NGC 1333. The typical errors are indicated on the top-right corner of each panel. \label{fig-n512-n1220}}
\end{figure}

\begin{figure}[!htbp]
\epsscale{0.7}
\plotone{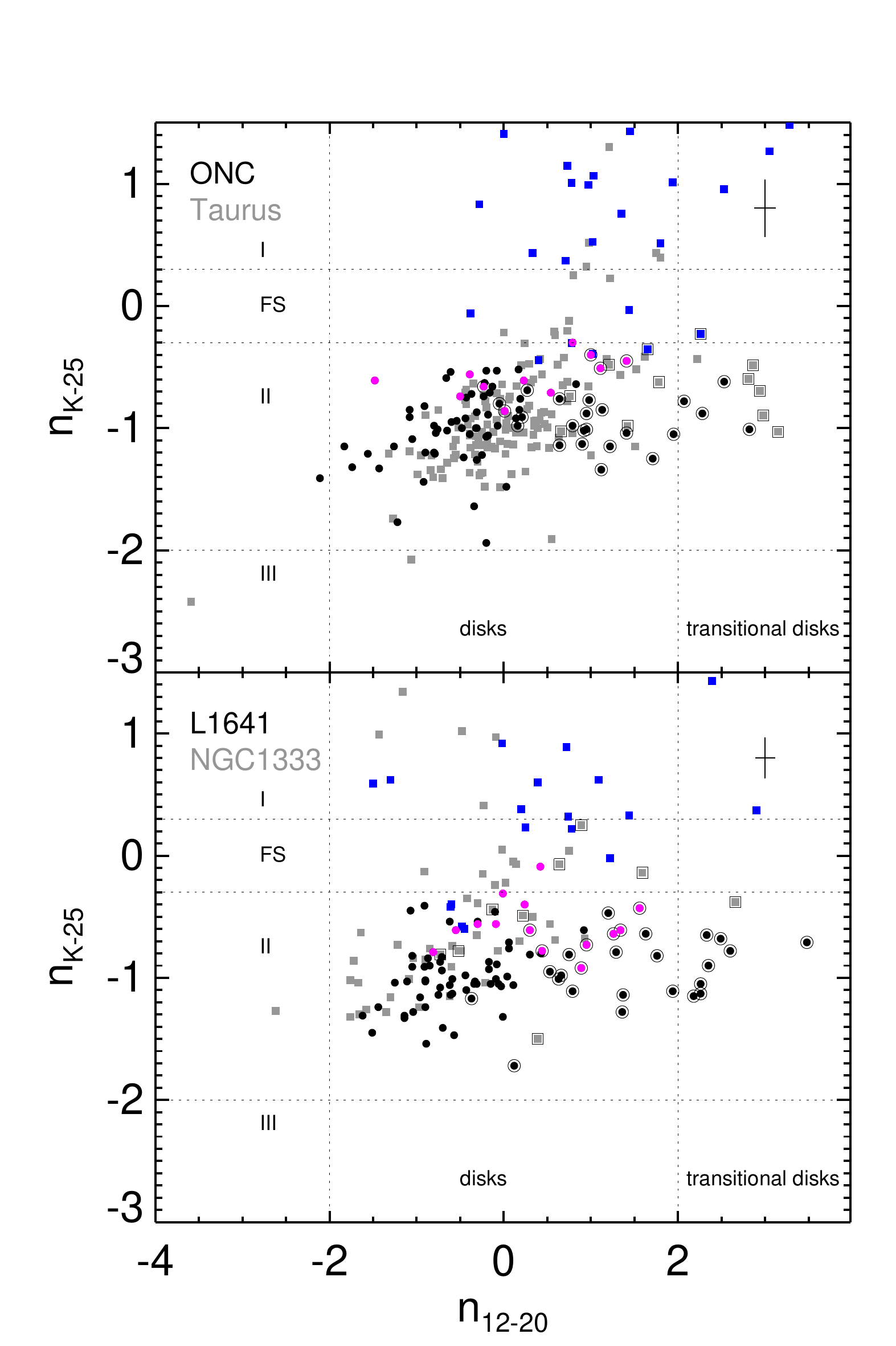}
\caption{Comparison of the observed spectral indices and disk classification by $n_{12-20}$ versus $n_{K-25}$. TDs classified in \citet{khkim2013-OriATD} are indicated with larger open circles encompassing the filled circles for ONC and L1641 and larger open squares encompassing the filled squares for Taurus and NGC 1333. The colored symbols (magenta: ONC and L1641; blue: Taurus and NGC 1333) indicate the objects with $n_{5-12} > -0.2$, which are in the ``envelopes" area in Figure~\ref{fig-n512-n1220}. The typical errors are indicated on the top-right corner of each panel.  \label{fig-n1220-nK25}}
\end{figure}

\clearpage
\begin{figure}[!htbp]
\epsscale{0.7}
\plotone{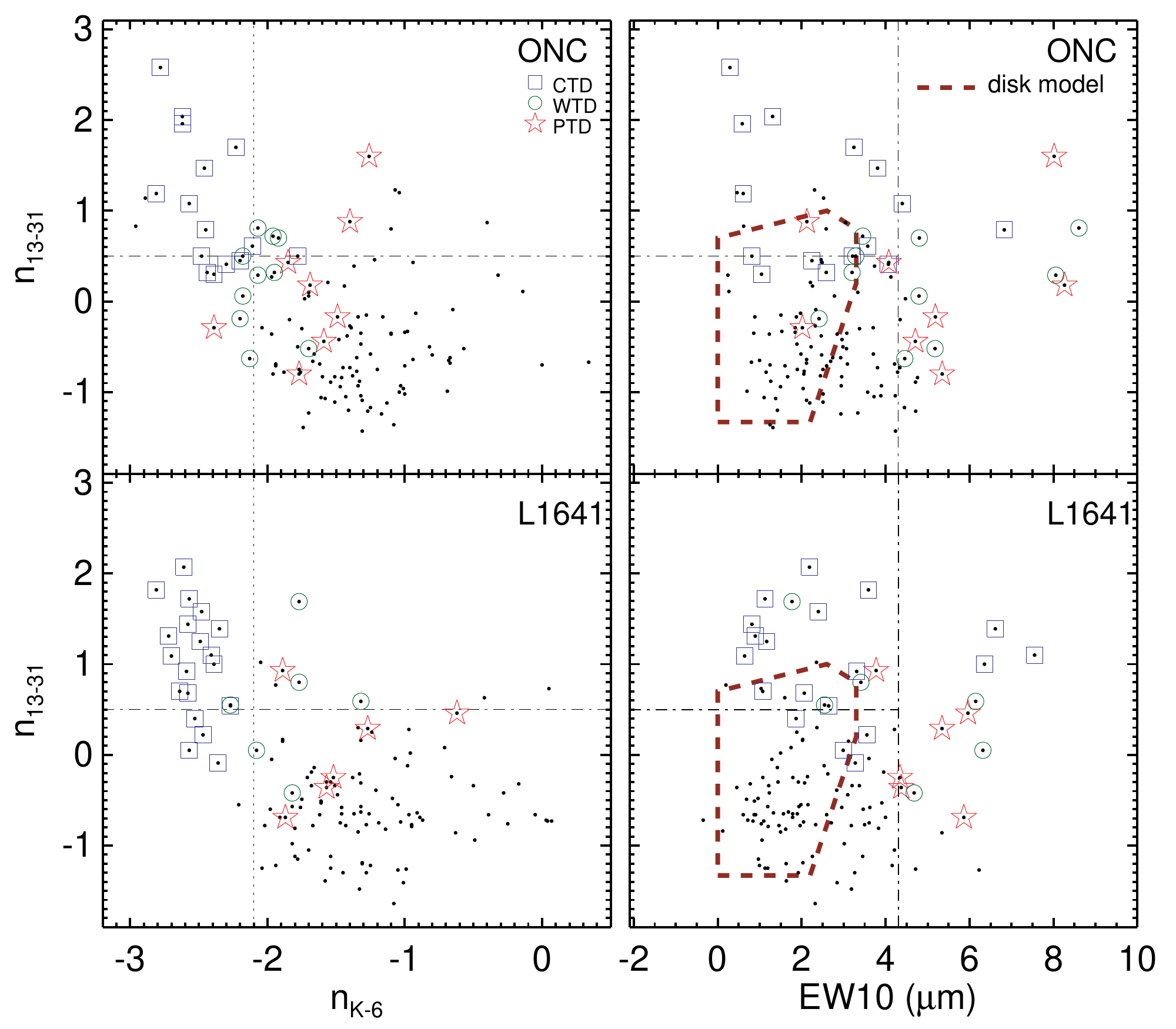}
\caption{Classification of transitional disks in Orion A, and selection by $n_{13-31}$ vs. $n_{K-6}$ (left panels) and $n_{13-31}$ vs. EW10 (right panels). In the right panels, the polygon with thick dashed brown line indicates the coverage area by a typical accretion disk model \citep{dalessio06}. The upper panels are for the TD selection in ONC, and the lower panels are for the TD selection in L1641. The dash-dotted lines indicate the upper octile; the dotted lines indicate the lower octile. The blue squares indicate the candiates of CTDs. The green circles are for WTDs, and the red stars indicate PTDs. The definition of CTD, WTD, and PTD are in the section 4.3.2. \label{fig-td-nk6-ew10-n1331}}
\end{figure}

\begin{figure}[!htbp]
\epsscale{0.9}
\plotone{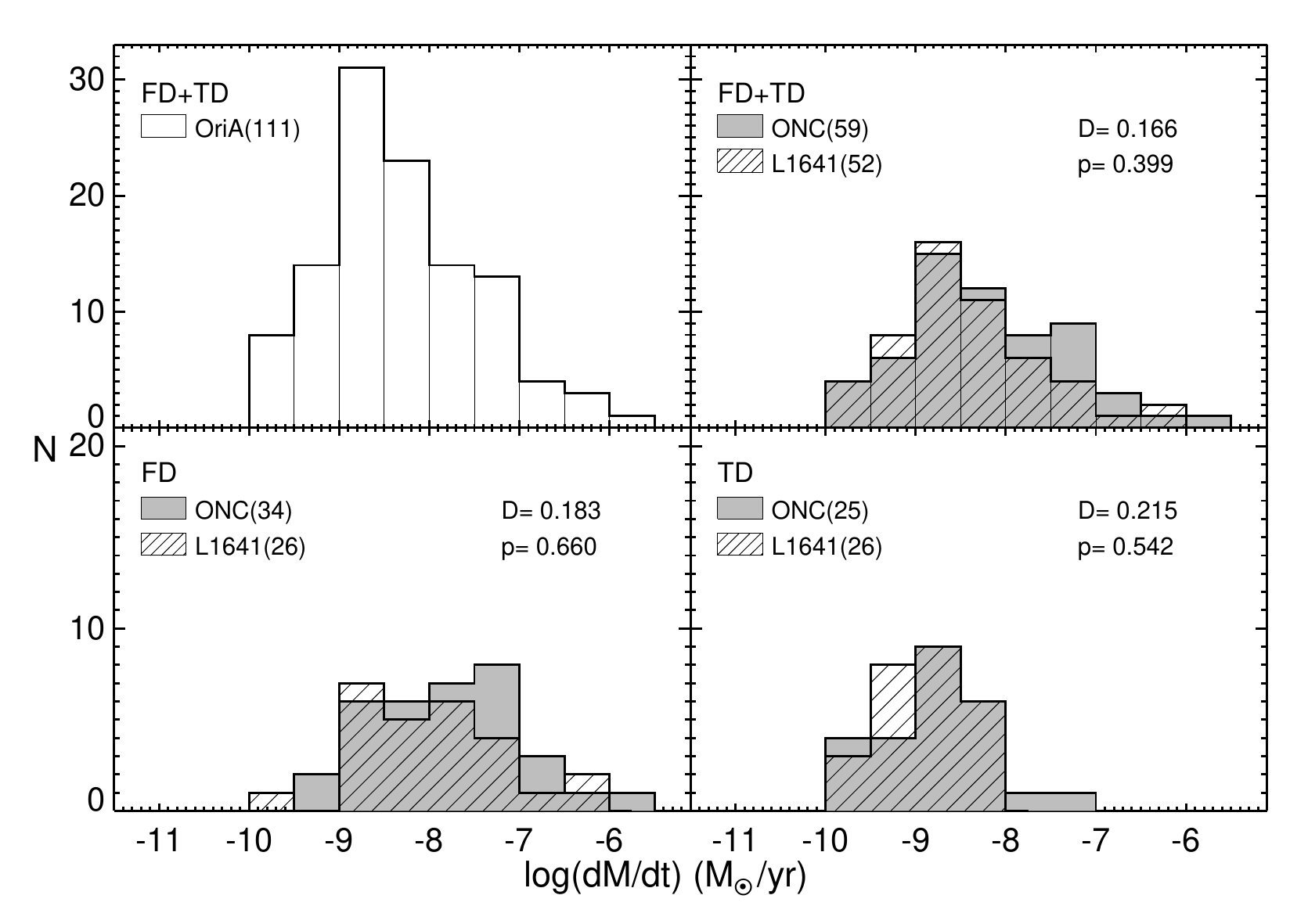}
\caption{Distribution of $\dot{M}$. All of objects with available $\dot{M}$ in Orion A is in the upper left panel. The upper right panel shows the distributions of $\dot{M}$ of objects in ONC and L1641. The lower left panel is for the comparison of $\dot{M}$ distribution of FDs in the ONC and L1641. The $\dot{M}$ distribution of TDs in ONC and L1641 is in the lower right panel. The distribution of $\dot{M}$ is slightly skewed toward to the higher value for ONC, but there is no statistically significant difference in $\dot{M}$ between ONC and L1641 when the same disk group in two regions is compared. The K-S test results (D and p) are marked in each panel. \label{fig-Mdot-OriA}}
\end{figure}

\begin{figure}[!htbp]
\epsscale{0.9}
\plotone{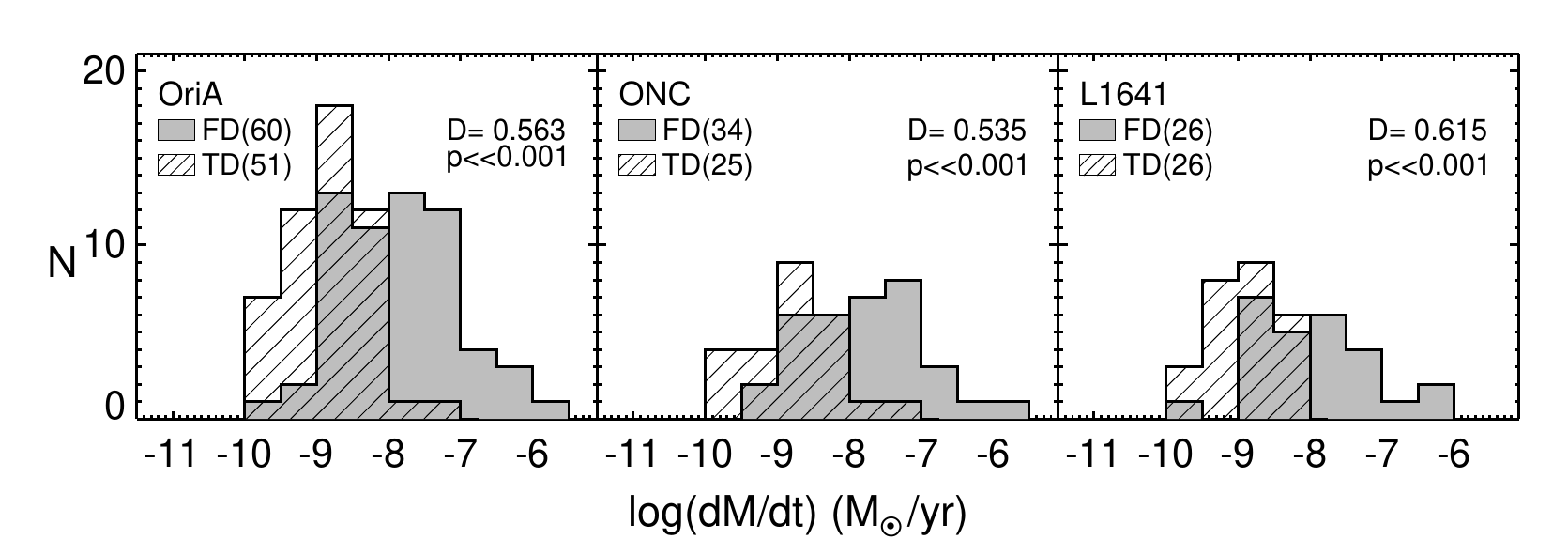}
\caption{$\dot{M}$ distributions between two groups, FD and TD, are compared in different regions: Orion A (left), ONC (middle), and L1641 (right). Distributions of $\dot{M}$ are strongly significantly different between FD and TD in all three comparisions cases. The K-S test result of each comparison is shown in each panel. \label{fig-Mdot-OriA-TDvsFD}}
\end{figure}

\clearpage
\begin{figure}[!htbp]
  \centering
  \begin{tabular}{cc}
    \includegraphics[width=65mm]{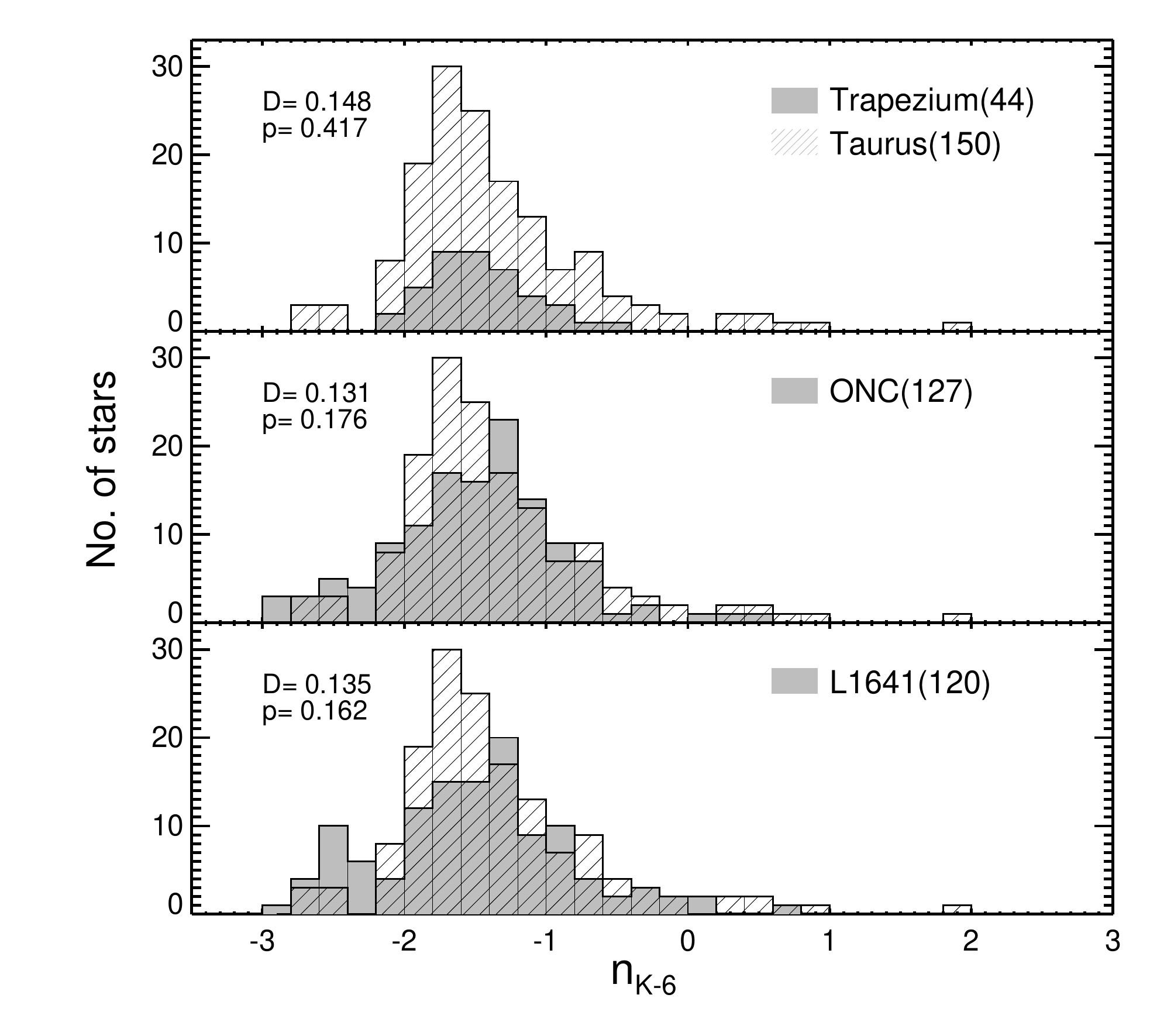}&
    \includegraphics[width=65mm]{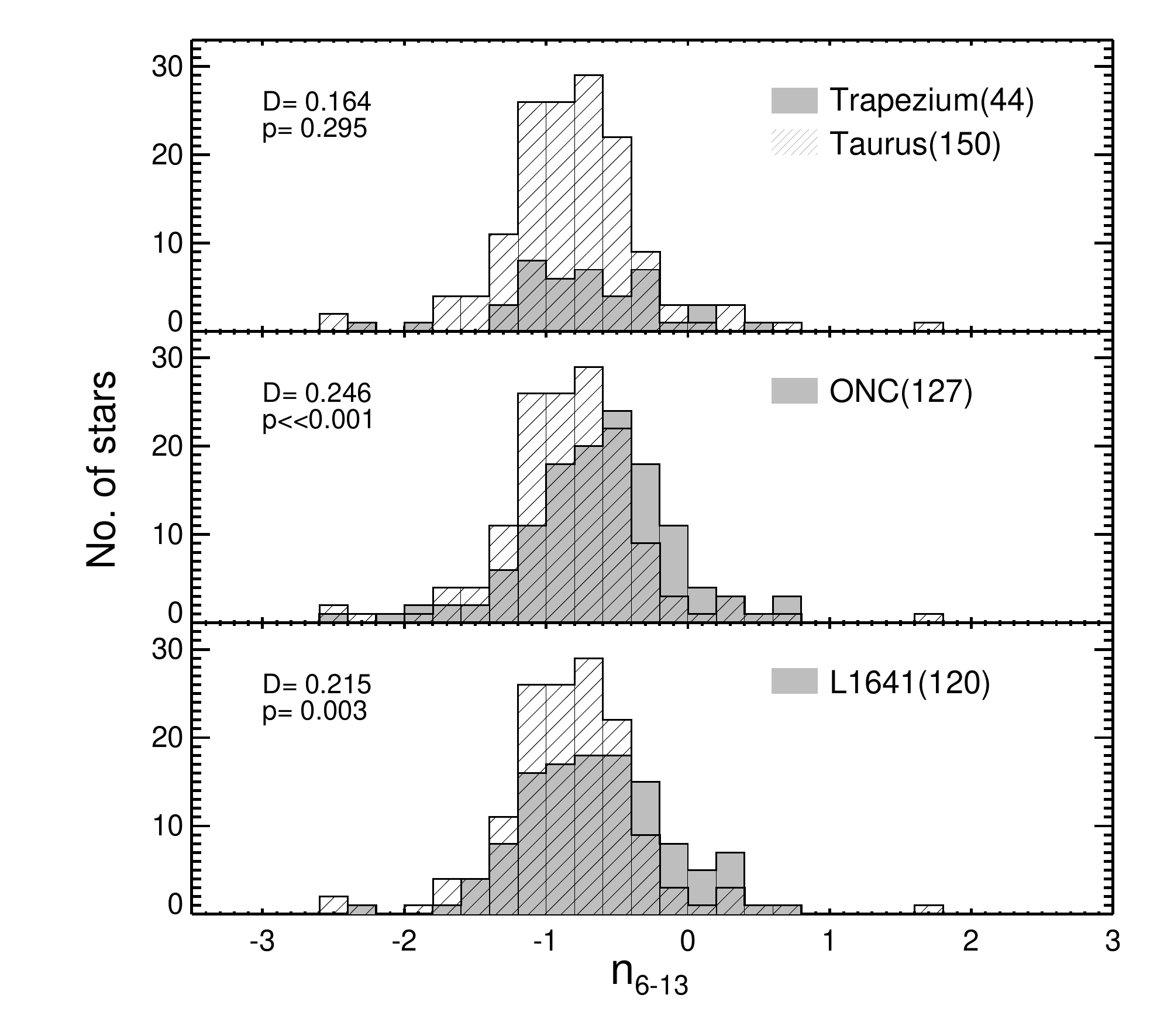}\\
    \includegraphics[width=65mm]{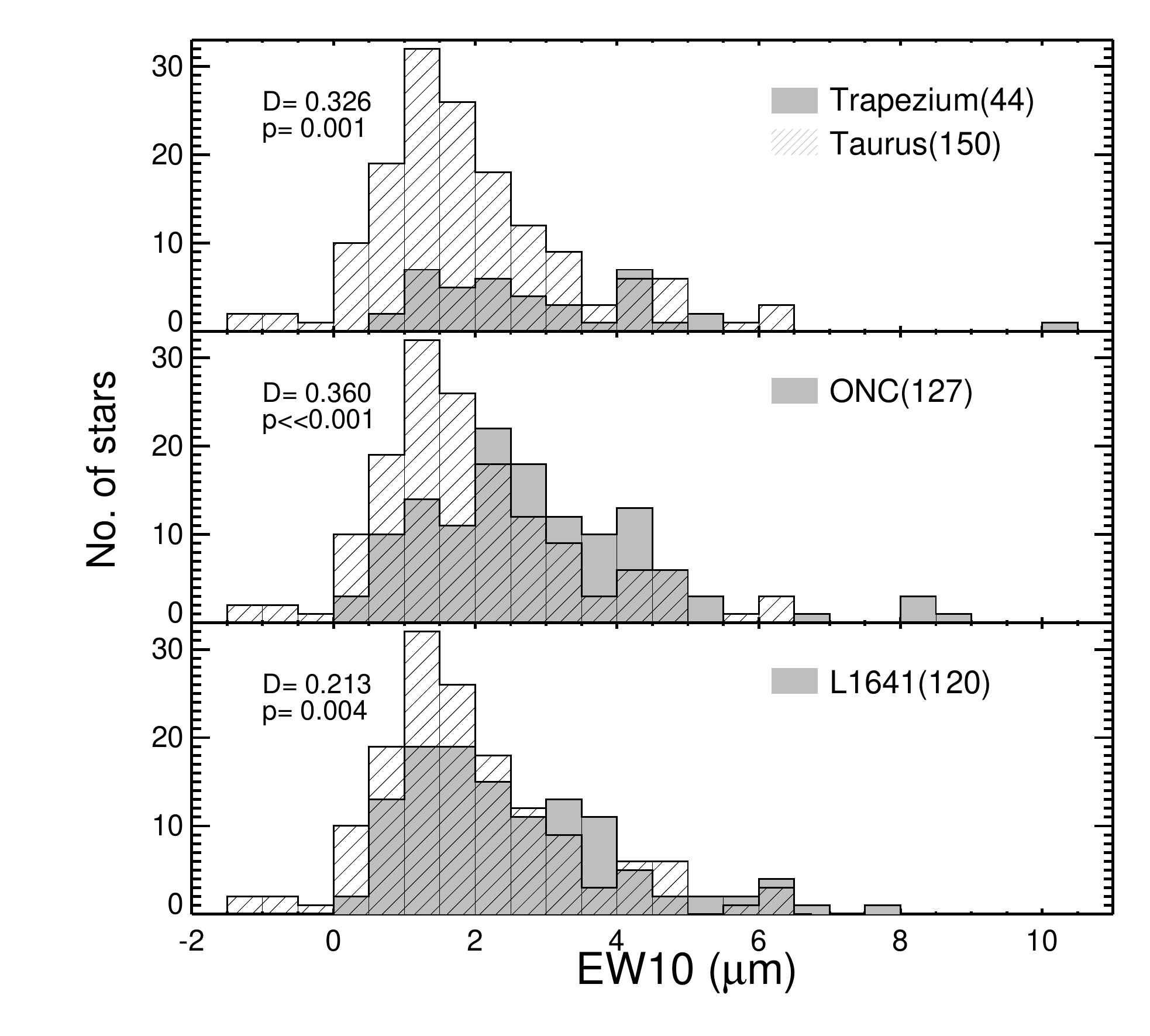}&
    \includegraphics[width=65mm]{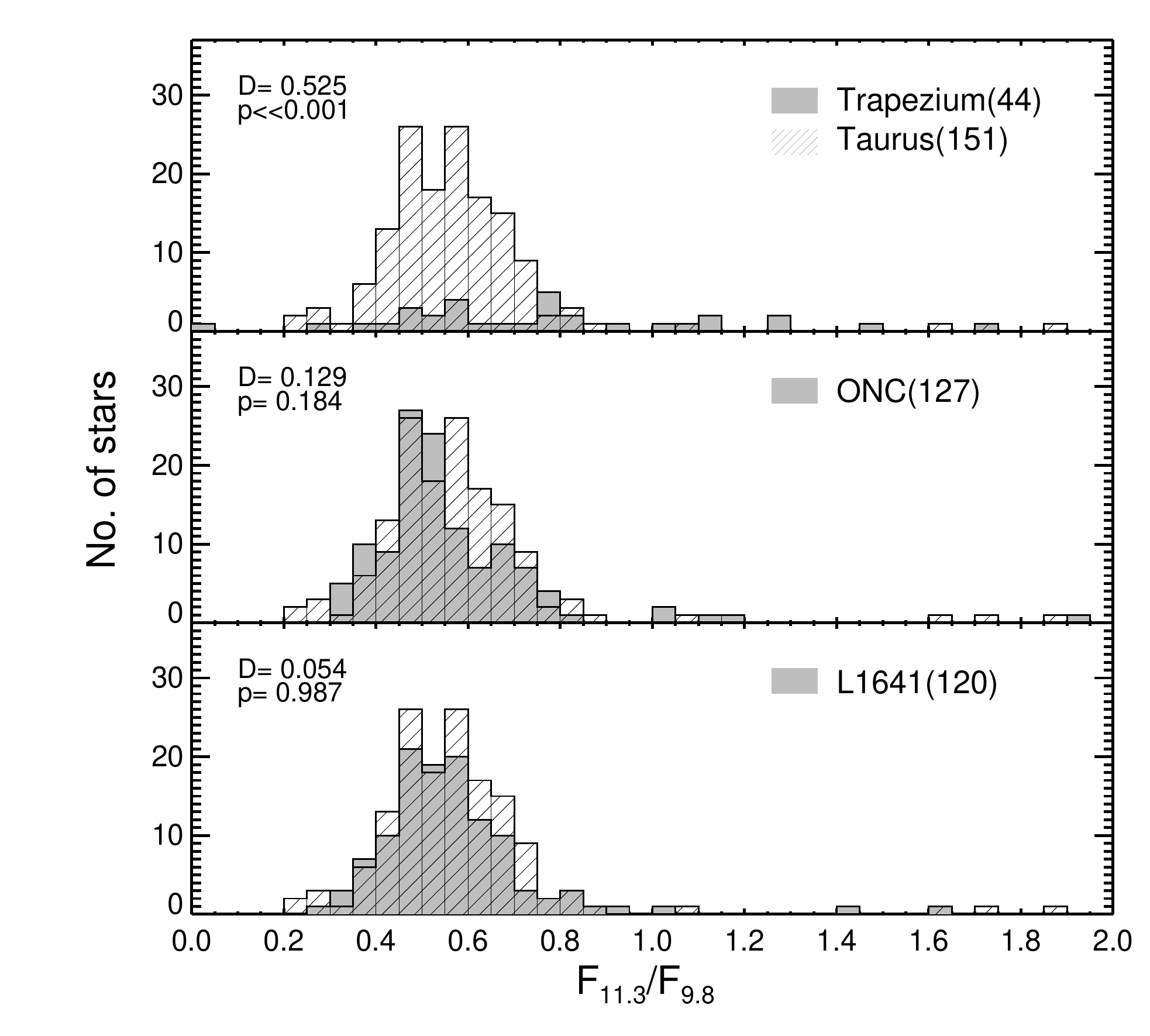}\\
  \end{tabular}
\caption{Histogram comparision of disk and dust properties measured from SL spectra. We include disks having $n_{K-6}$ (upper left), $n_{6-13}$ (upper right), EW10 (lower left), and $F_{11.3}/F_{9.8}$ (lower right) without separating the samples by their radial structures, FDs or TDs. We compare the distribution of each properties in this figure in three sub-regions of Orion A (Trapezium, ONC, and L1641) to that of Taurus.  The K-S test result of each comparison is shown in each panel.\label{fig-sl-histogram-comp}}
\end{figure}

\clearpage
\begin{figure}[!htbp]
\epsscale{0.7}
\plotone{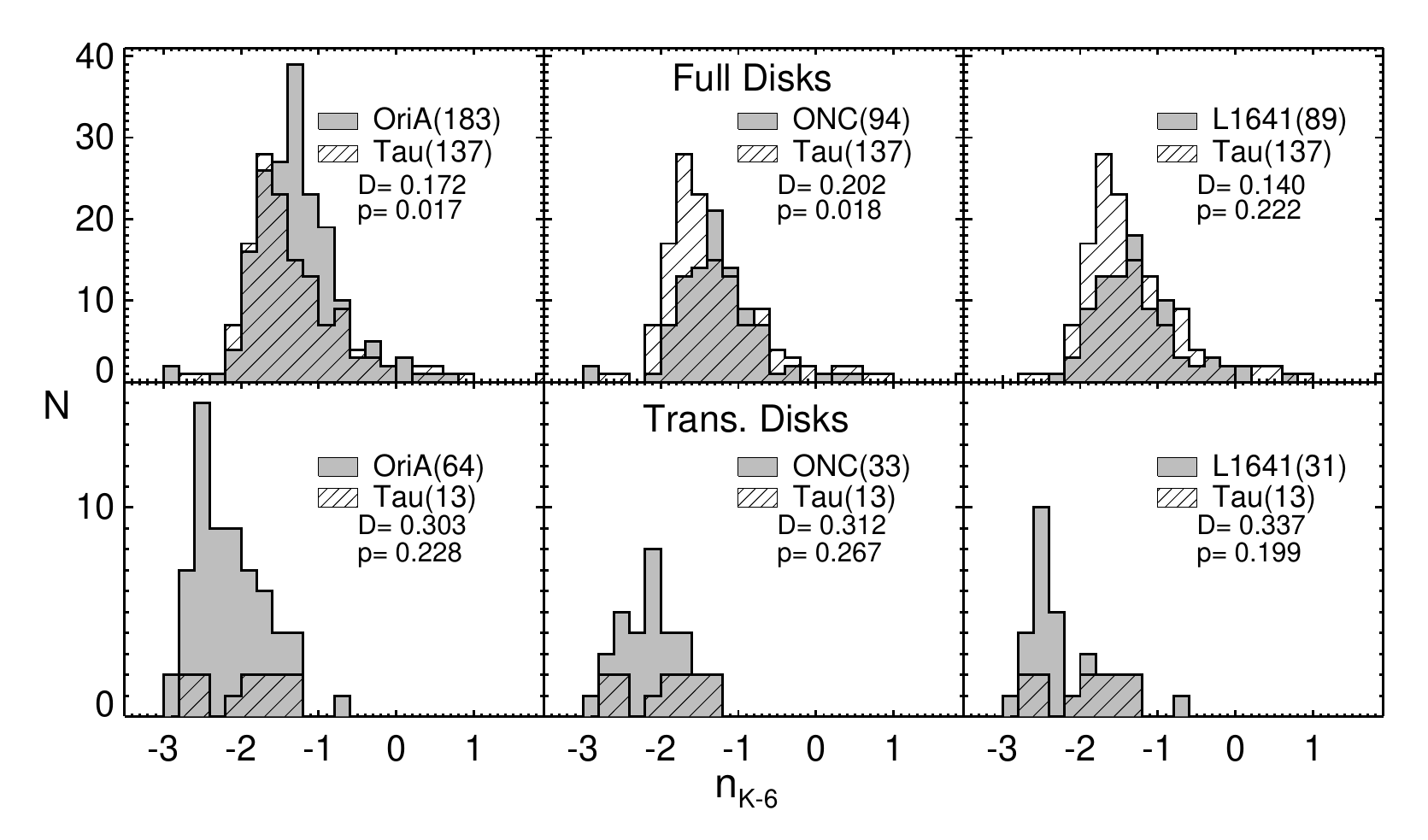}
\caption{Comparison of $n_{K-6}$ between Orion A (solid) and Taurus (hatched). The objects in the sample are separated by their radial strucuture, FDs or TDs (upper panels or lower panels), and by the sub-regions, OriA, ONC, or L1641 (left, middle, or right panels). The K-S test result of each comparison is shown in each panel. \label{fig-tdfd-nk6-OirATau}}
\end{figure}

\clearpage
\begin{figure}[!htbp]
\epsscale{0.7}
\plotone{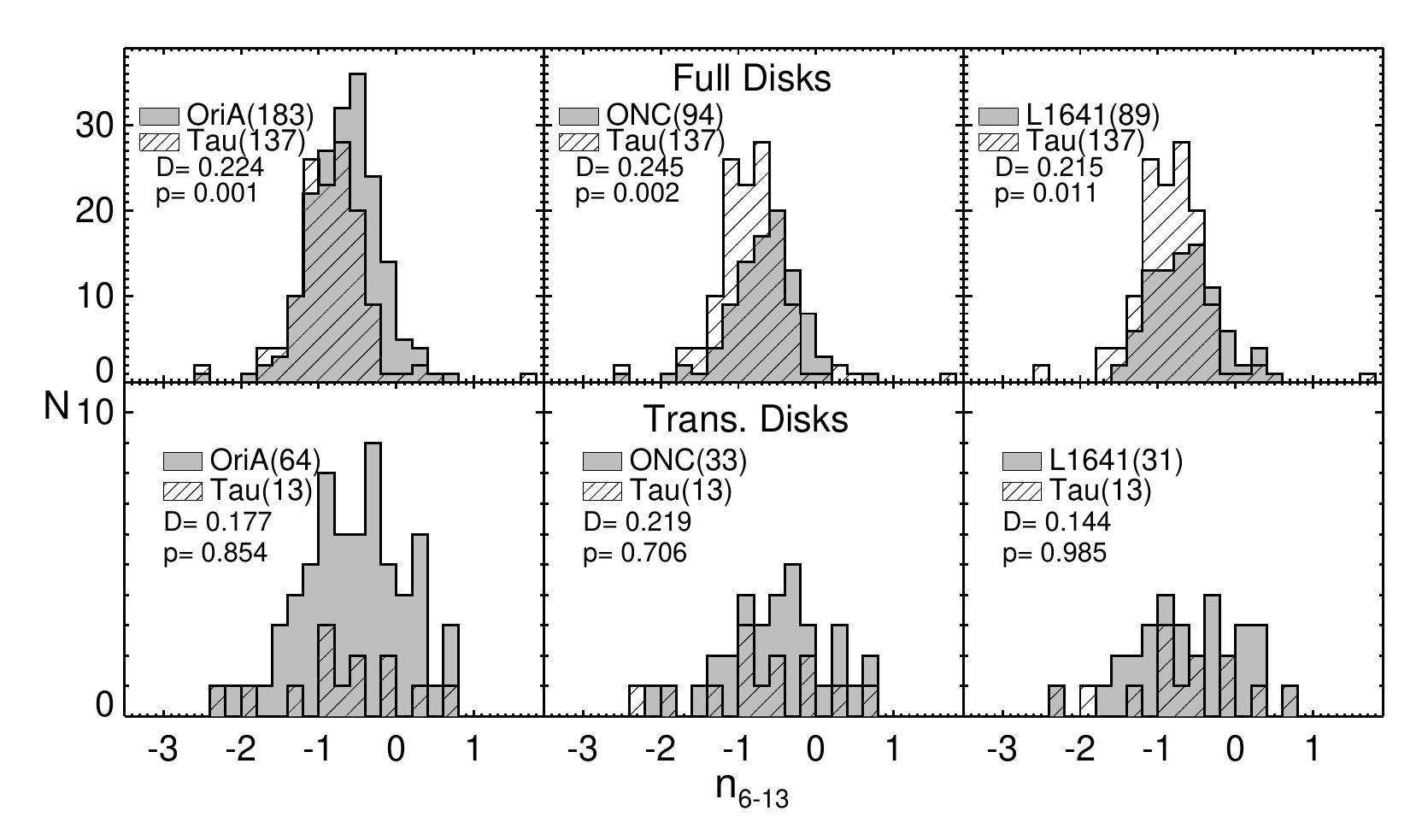}
\caption{Comparison of $n_{6-13}$ between Orion A (solid) and Taurus (hatched). The objects in the sample are separated by their radial strucuture, FDs or TDs (upper panels or lower panels), and by the sub-regions, OriA, ONC, or L1641 (left, middle, or right panels). The K-S test result of each comparison is shown in each panel. \label{fig-tdfd-n613-OirATau}}
\end{figure}

\clearpage
\begin{figure}[!htbp]
\epsscale{0.7}
\plotone{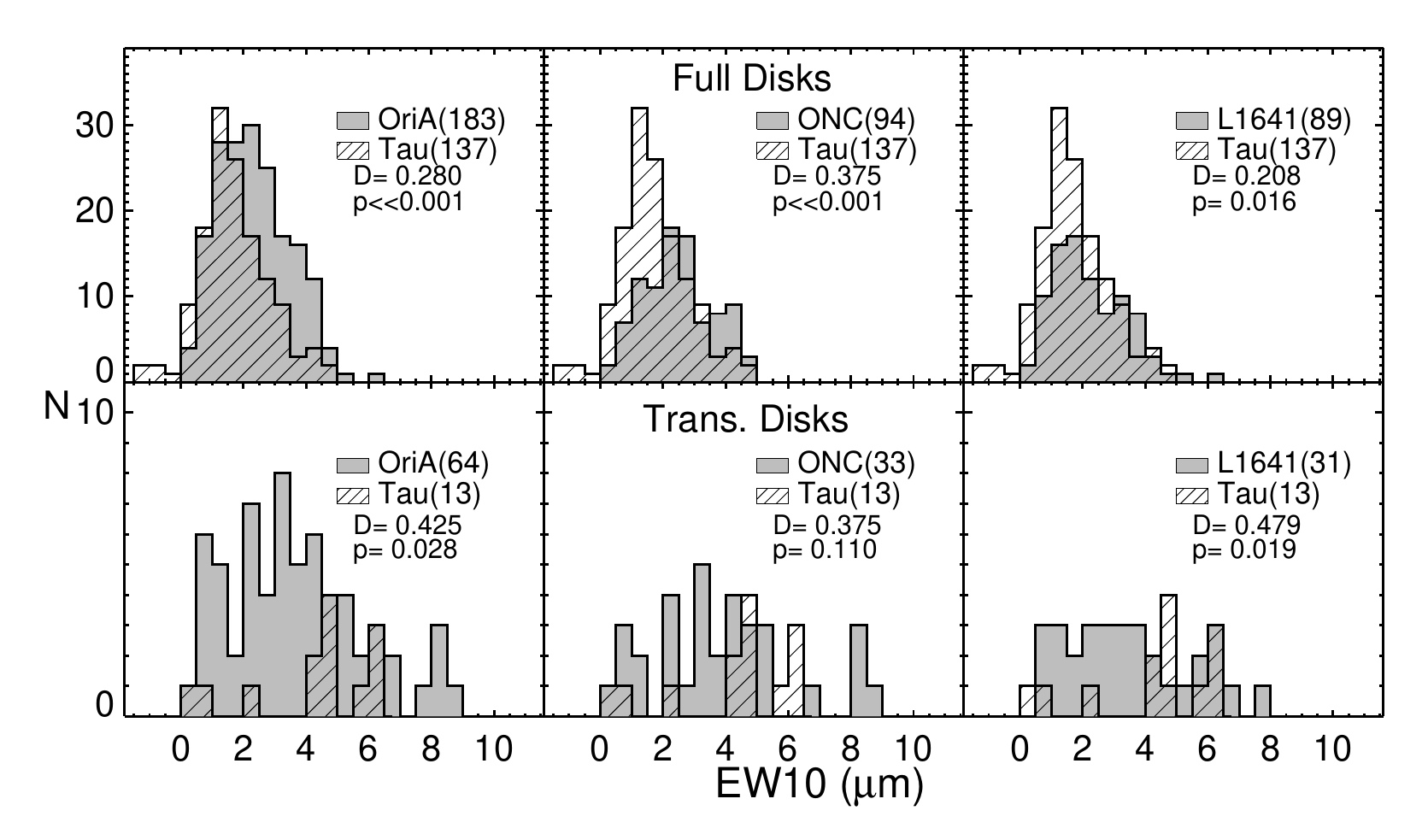}
\caption{Comparison of EW10 between Orion A (solid) and Taurus (hatched). The objects in the sample are separated by their radial strucuture, FDs or TDs (upper panels or lower panels), and by the sub-regions, OriA, ONC, or L1641 (left, middle, or right panels). The K-S test result of each comparison is shown in each panel. \label{fig-tdfd-ew10-OirATau}}
\end{figure}

\clearpage
\begin{figure}[!htbp]
\epsscale{0.7}
\plotone{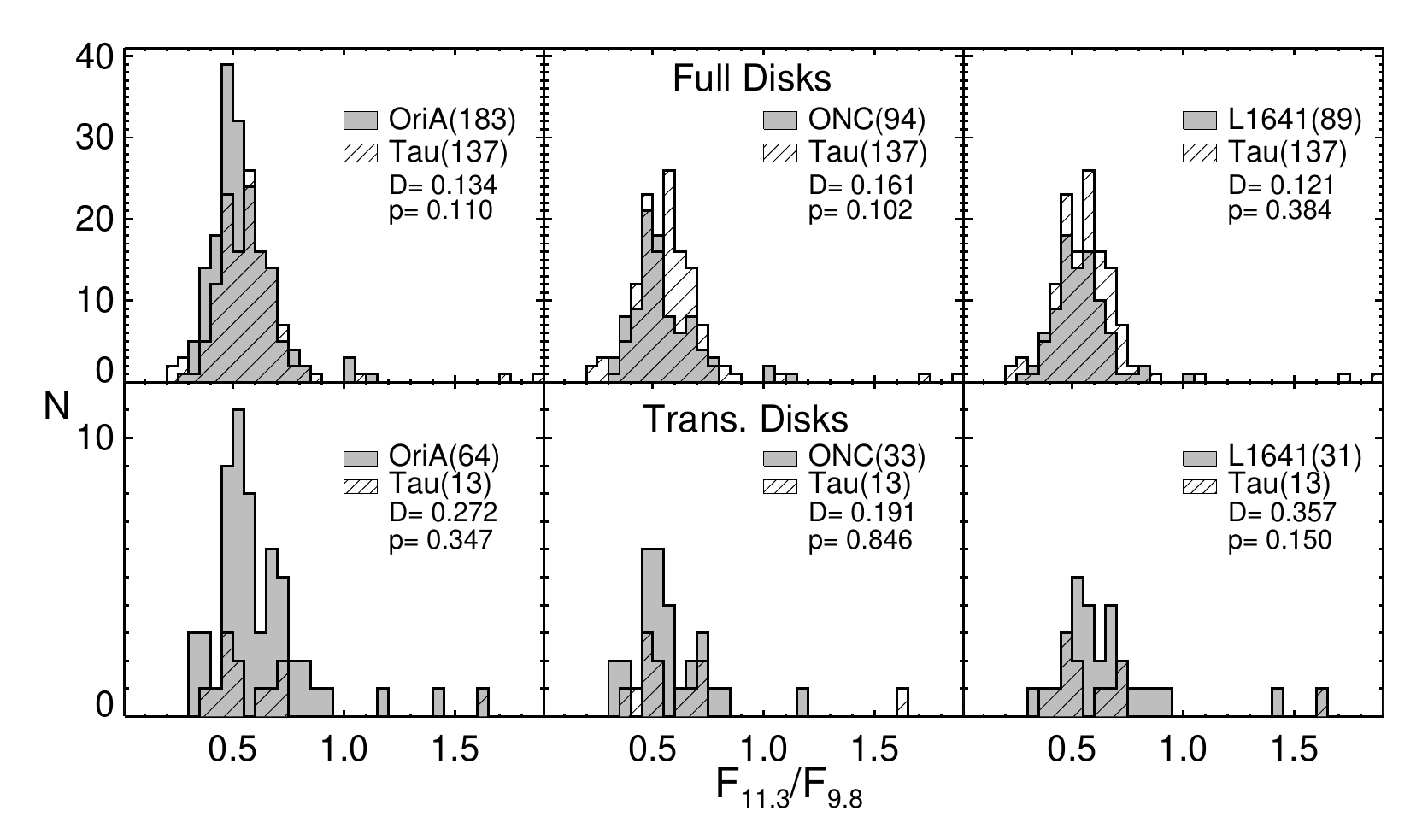}
\caption{Comparison of $F_{11.3}/F_{9.8}$ between Orion A (solid) and Taurus (hatched). The objects in the sample are separated by their radial strucuture, FDs or TDs (upper panels or lower panels), and by the sub-regions, OriA, ONC, or L1641 (left, middle, or right panels). The K-S test result of each comparison is shown in each panel. \label{fig-tdfd-f113f98-OirATau}}
\end{figure}

\clearpage
\begin{figure}[!htbp]
\epsscale{0.7}
\plotone{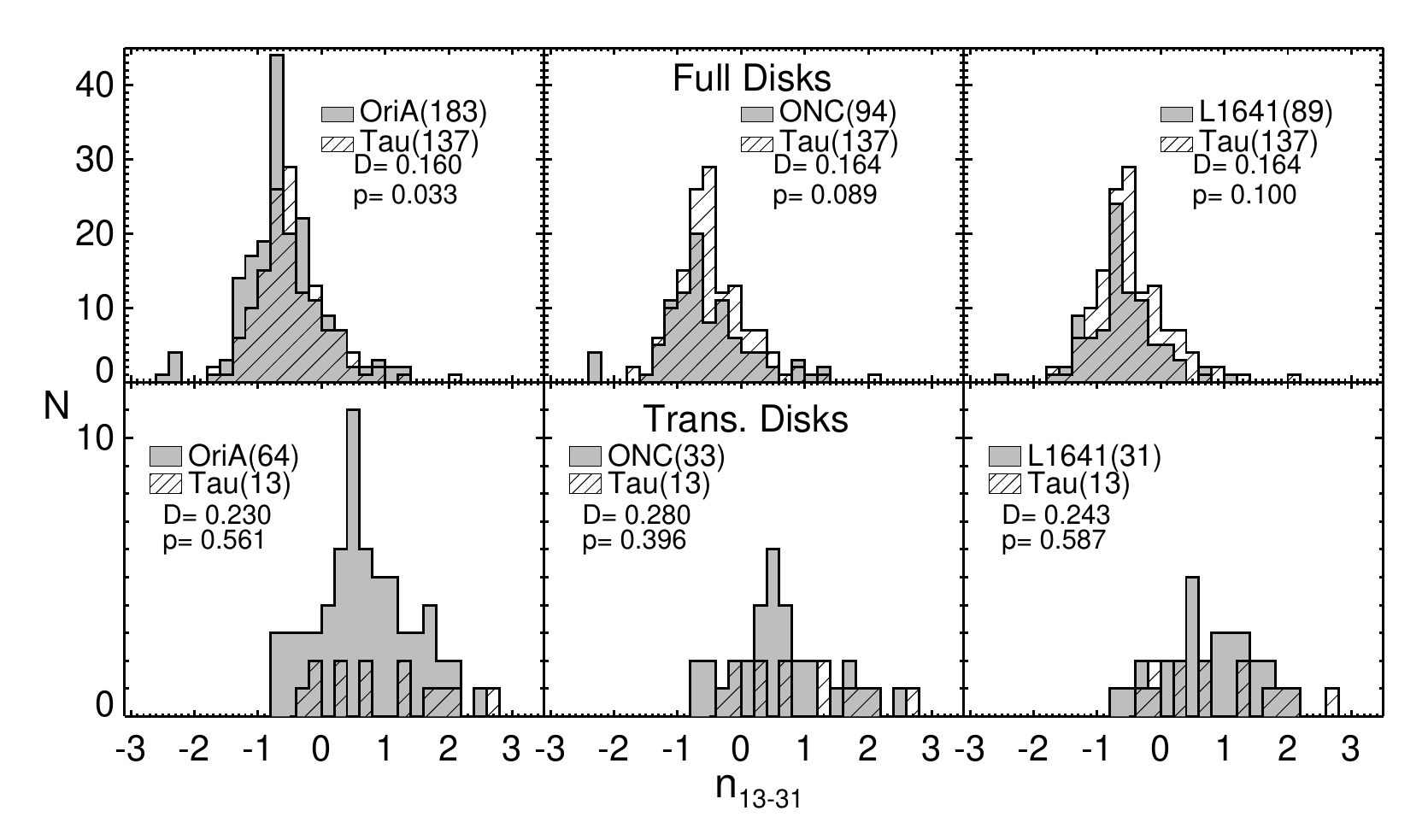}
\caption{Comparison of $n_{13-31}$ between Orion A (solid) and Taurus (hatched). The objects in the sample are separated by their radial strucuture, FDs or TDs (upper panels or lower panels), and by the sub-regions, OriA, ONC, or L1641 (left, middle, or right panels). The K-S test result of each comparison is shown in each panel. \label{fig-tdfd-n1331-OirATau}}
\end{figure}

\clearpage
\begin{figure}[!htbp]
\epsscale{1}
\plotone{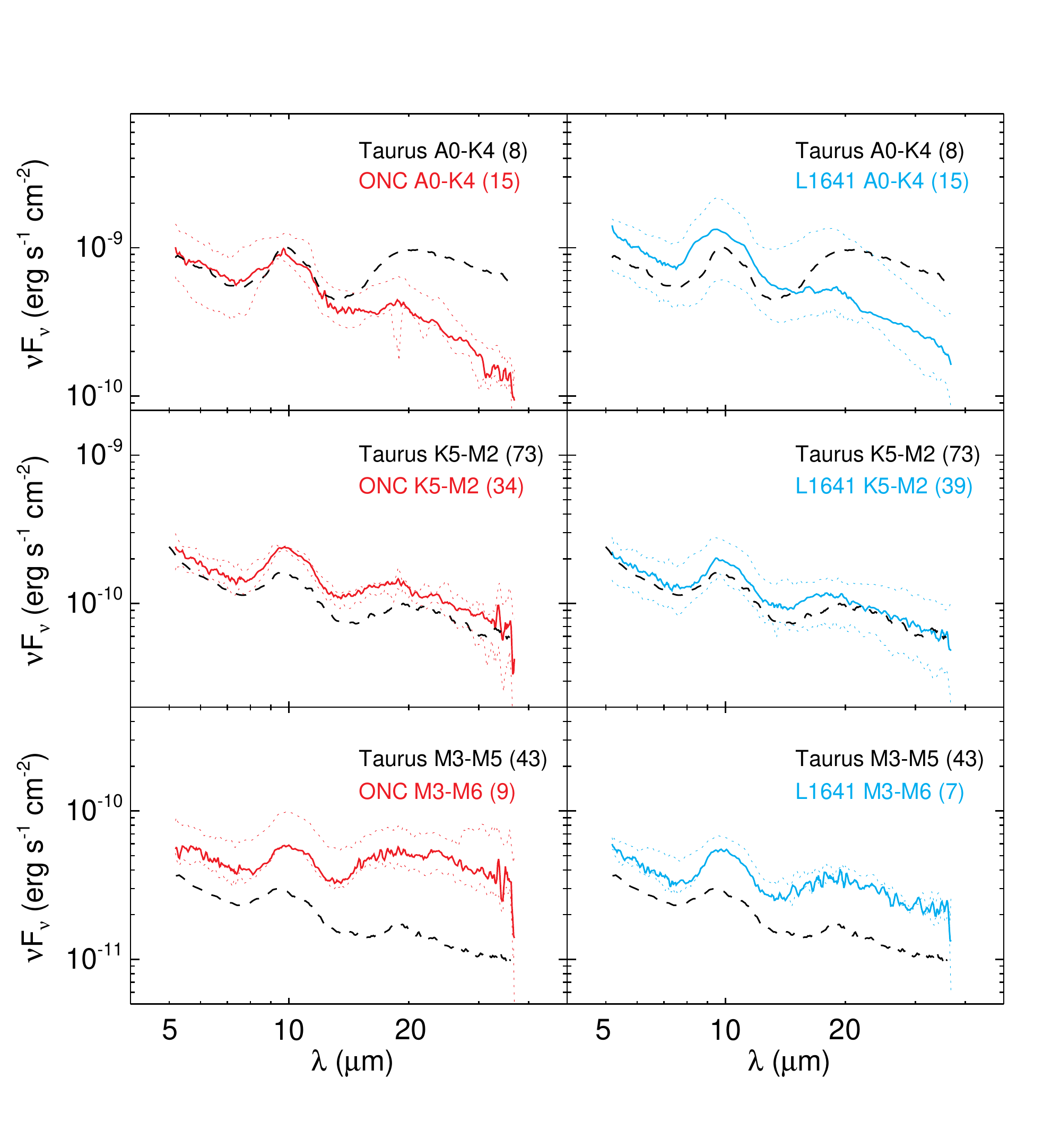}
\caption{The median spectra of FDs in Orion A, separated by sub-regions and spectral type  ranges. They are compared to Taurus median spectra. The number of objects used to generate a median spectrum is indicated next to the sign of spectral type range in each panel. The solid lines are the ONC median (left panels) and the L1641 median (right panels), and the dotted lines in each panel indicate upper and lower quartiles. The dashed lines are for the Taurus median indicated in each panel. The ONC median spectra and L1641 median spectra are normalized to the 2MASS H band median flux of the Taurus meidan spectra in each spectral range. \label{fig-median-metric-2regions}}
\end{figure}
\clearpage

\clearpage
\begin{figure}[!htbp]
\epsscale{1}
\plotone{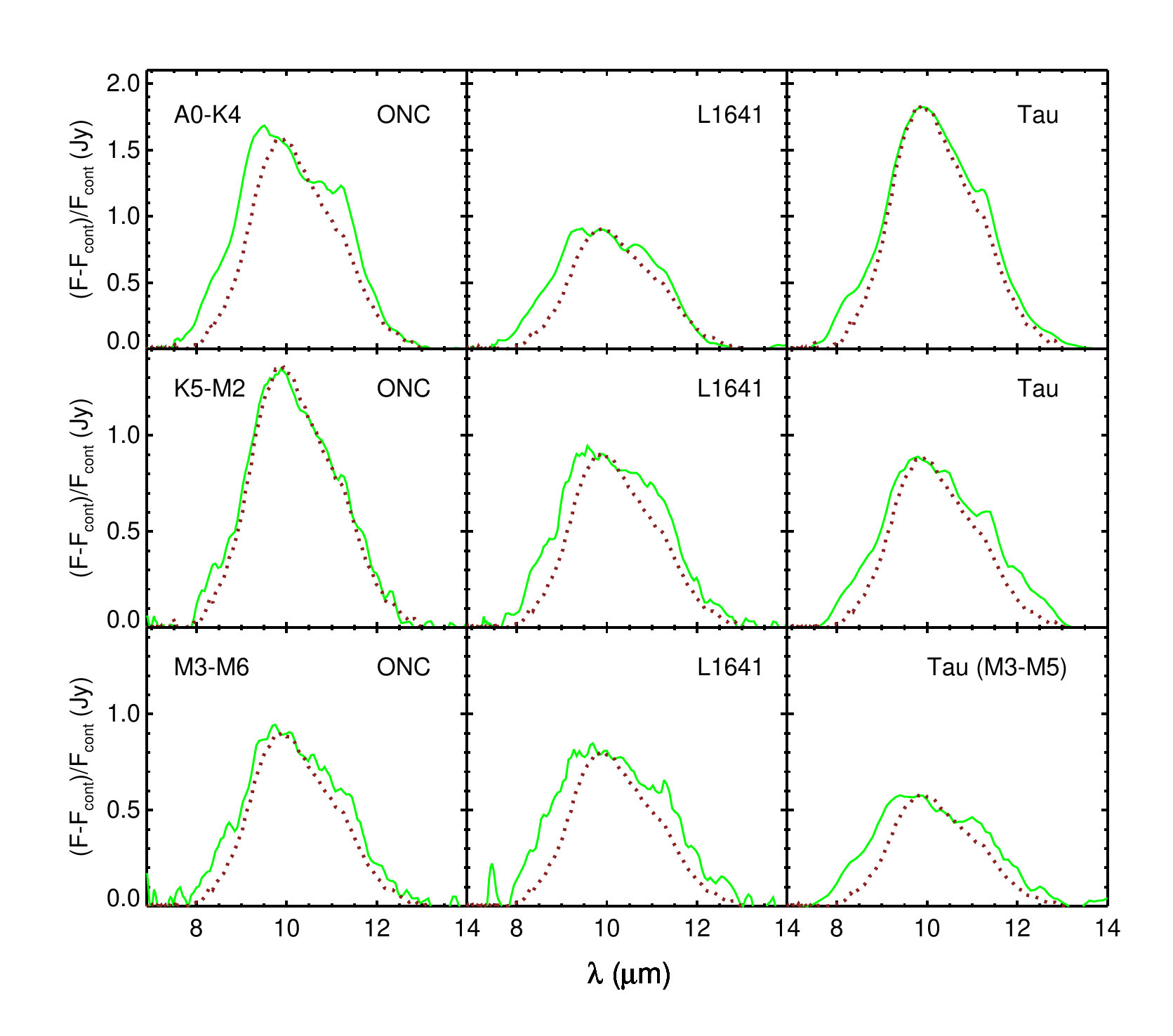}
\caption{Comparison of continuum-subtracted and -normalized 10 $\micron$ silicate feature of median spectra (solid lines) to the pristine silicate feature (dotted lines). The pristine silicate profile is derived by averaging ISM-like reference spectra of LkCa 15 and GM Aur~\citep{dmw_2009, Sargent+2009-GrainGrowth} after subtracting continuum and dividing it with continuum. The pristine silicate profile are scaled to match to each median silicate profile at 9.8 $\mu$m.  \label{fig-normflux-pristine}}
\end{figure}
\clearpage

\clearpage
\begin{figure}[!htbp]
\epsscale{1}
\plotone{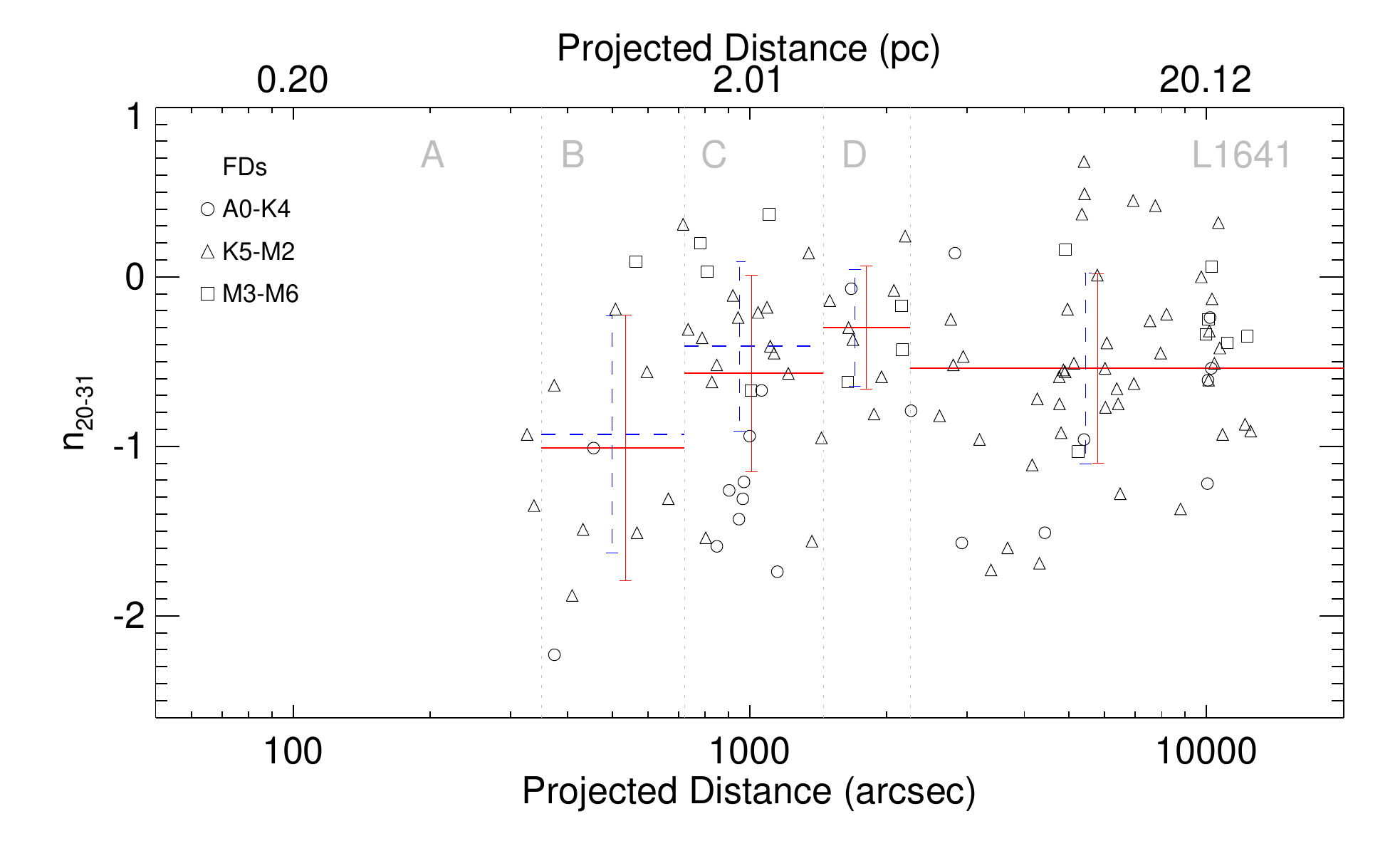}
\caption{Variation of $n_{20-31}$ with projected distance from $\theta^1$ Ori C. The empty symbols indicate the spectral type ranges to which a object belongs: circle for A0--K4; triangle for K5--M2; square for M3--M6. The dashed horizontal lines indicate the $n_{20-31}$ median of objects in K5--M2 group in each sub-section, which is indicated in Figure~\ref{fig-ONC-big} and Figure~\ref{fig-L1641-big}. The solid horizontal lines indicate the median of all objects from A to M type in each sub-section. The error bars on the medians are taken from the standard deviation of $n_{20-31}$ in each sub-section.
\label{fig-enveff-n2031-dis}}
\end{figure}

\clearpage
\begin{figure}[!htbp]
\epsscale{1}
\plotone{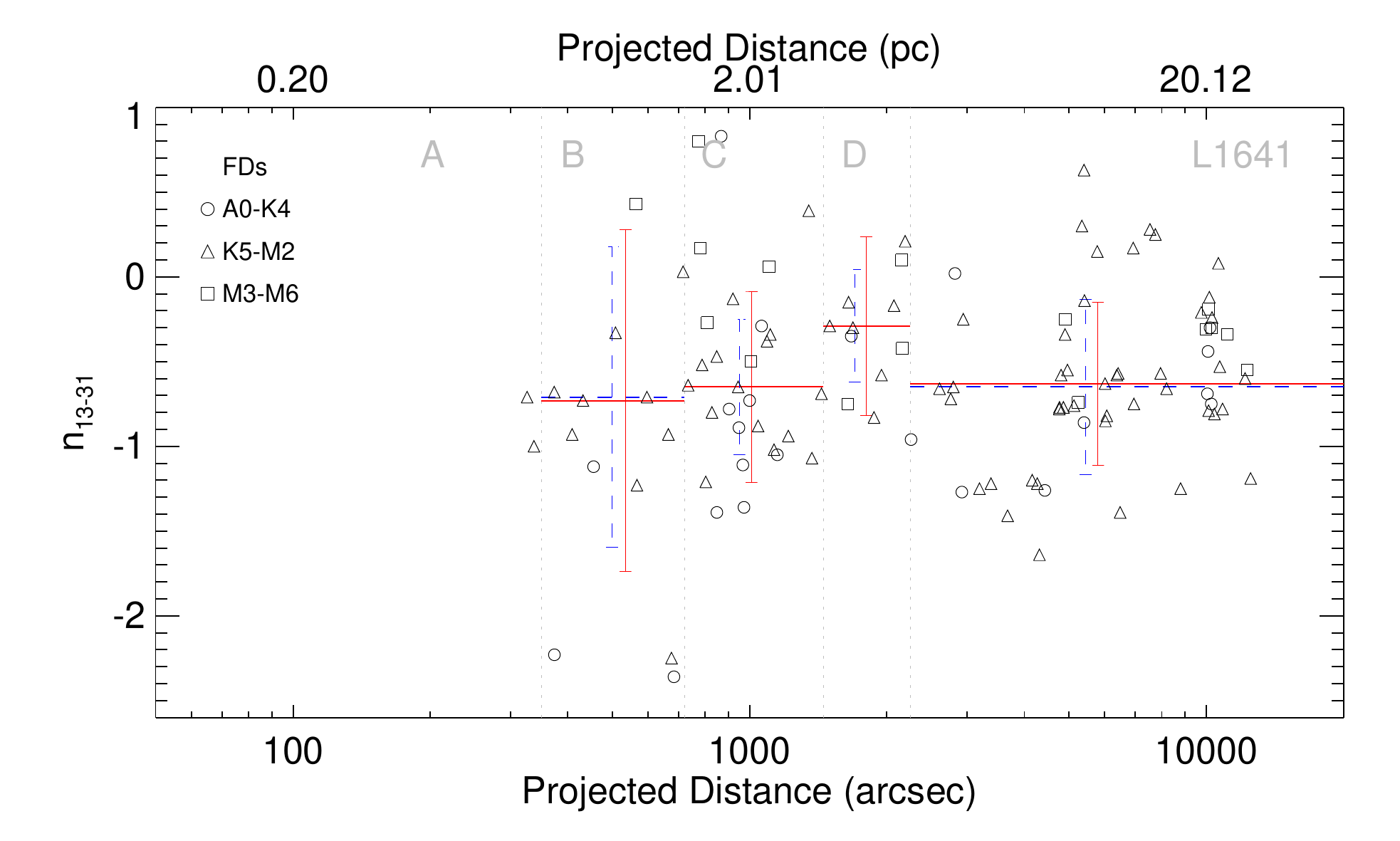}
\caption{Variation of $n_{13-31}$ with projected distance from $\theta^1$ Ori C. The meanings of symbols are same as in Figure~\ref{fig-enveff-n2031-dis}. The dashed horizontal lines indicate the $n_{13-31}$ median of objects in K5--M2 group in each sub-section, which is indicated in Figure~\ref{fig-ONC-big} and Figure~\ref{fig-L1641-big}. The solid horizontal lines indicate the median of all objects from A to M type in each sub-section. The error bars on the medians are taken from the standard deviation of $n_{13-31}$ in each sub-section.
\label{fig-enveff-n1331-dis}}
\end{figure}

\clearpage
\begin{deluxetable}{cccccccccc}
\rotate
\tablecaption{Observation and Reduction Log of IRS Spectra \label{table:IRSlog}}
\tablehead{
\colhead{Num.}& \colhead{IRS name} & \colhead{RA (J2000)} & \colhead{DEC (J2000)} & \colhead{AORID} & \colhead{IRS Camp.} & \colhead{Date observed} & \colhead{modules} & \colhead{region} & \colhead{reduction} \\
\colhead{(1)}& \colhead{(2)} & \colhead{(3)} & \colhead{(4)} & \colhead{(5)} & \colhead{(6)} & \colhead{(7)} & \colhead{(8)} & \colhead{(9)} & \colhead{(10)}
}
\tablecolumns{10}
\tablewidth{0pt}
\tabletypesize{\scriptsize}
\startdata
1	&	8336884-5290	&	5	33	28.52	&	-5	17	26.16	&	18804224	&	39	&	2007 Mar 12	&	SLLL	&	ONC	&	AdOpt		\\
2	&	8339347-5238	&	5	33	34.43	&	-5	14	17.70	&	18843904	&	39	&	2007 Mar 12	&	SLLL	&	ONC	&	AdOpt		\\
3	&	8343858-5513	&	5	33	45.26	&	-5	30	49.75	&	18825472	&	39	&	2007 Mar 12	&	SLLL	&	ONC	&	AdOpt		\\
4	&	8343917-6073	&	5	33	45.40	&	-6	04	25.86	&	18848000	&	39	&	2007 Mar 09	&	SLLL	&	L1641	&	man		\\
5	&	8343944-5609	&	5	33	45.47	&	-5	36	32.54	&	18833152	&	39	&	2007 Mar 12	&	SLLL	&	ONC	&	opse/AdOpt	\tablenotemark{a}	\\
6	&	8344455-5390	&	5	33	46.69	&	-5	23	25.44	&	18802944	&	39	&	2007 Mar 24	&	SLLL	&	ONC	&	opse		\\
7	&	8344987-5460	&	5	33	47.97	&	-5	27	38.56	&	18780672	&	44	&	2007 Oct 05	&	SLLL	&	ONC	&	man		\\
8	&	8346141-5010	&	5	33	50.74	&	-5	00	39.46	&	18781952	&	39	&	2007 Mar 22	&	SLLL	&	ONC	&	AdOpt		\\
\enddata
\tablecomments{\footnotesize Column (1) Num.: Numbers in this column are from the number sequence of 303 Class II objects observed in the IRS program ID 30706 and additional 16 Class II objects candidates identified from \textit{Herschel} Orion Protostar Survey. We use these numbers to identify objects conveniently. This number sequence will be used consistently in other future papers dealing with the objects in this work.\\
Column (2) IRS name: IRS name indicated as in the column are used to identify objects often in other works. The IRS names come from the position of each object.\\
Column (10) reduction: It indicates the methods of source extraction to get the SEDs in Figure~\ref{fig-SLLL-SED} and Figure~\ref{fig-SLonly-SED}. ECO: an automated tapered column extraction in SMART with off-nod or off-order sky subtraction; man: a manual tapered column extraction in SMART with a constant or linear sky subtraction; AdOpt: an optimal source extraction using an empirical point response function (PRF) in SMART; opse: an optimal source extraction using an analytical PRF.}
\tablenotetext{a}{\footnotesize	SL $\&$ LL1: opse; LL2: AdOpt}
\tablenotetext{b}{\footnotesize	SL: man; LL: AdOpt}
\tablenotetext{c}{\footnotesize	SL: AdOpt; LL:opse}
\tablenotetext{d}{\footnotesize	SL $\&$ LL1: AdOpt; LL2:man}
\tablenotetext{e}{\footnotesize	SL $\&$ LL2: man; LL1: opse}
\tablenotetext{f}{\footnotesize	SL: opse; LL:AdOpt}
\tablenotetext{g}{\footnotesize	SL $\&$ LL2: AdOpt; LL1:opse}
\tablenotetext{h}{\footnotesize	SL $\&$ LL2: AdOpt; LL1: ECO}
\tablenotetext{i}{\footnotesize	SL: AdOpt; LL:ECO}
\tablenotetext{j}{\footnotesize	SL: opse; LL2: AdOpt; LL1: man}
\tablenotetext{k}{\footnotesize	SL: opse; LL:man}
\tablenotetext{l}{\footnotesize	SL $\&$ LL2: AdOpt; LL1:man}
\tablenotetext{m}{\footnotesize	SL2: AdOpt; SL1:man}
\tablenotetext{n}{\footnotesize The IRS program ID 30859}
\tablenotetext{o}{\footnotesize The IRS program ID 50374}
\end{deluxetable}

\begin{deluxetable}{cccl}
\rotate
\tablecaption{Identifications of the Objects \label{table:id}}
\tablehead{
\colhead{Num.}& \colhead{IRS name} & \colhead{2MASS name} & \colhead{other names selected}
}
\tablecolumns{4}
\tablewidth{0pt}
\tabletypesize{\scriptsize}
\startdata
1	&	8336884-5290	&	05332852-0517262	&	V719 Ori,  [FHM2008] F22-ap40, Parenago 1190	\\
2	&	8339347-5238	&	05333443-0514177	&	[FHM2008] S1-ap78	\\
3	&	8343858-5513	&	05334525-0530498	&	VZ Ori, [FHM2008] S3-ap68, HBC 106, Parenago 1260	\\
4	&	8343917-6073	&	05334537-0604253	&	V1006 Ori, HBC 108, Parenago 1267	\\
5	&	8343944-5609	&	05334545-0536323	&	V726 Ori, [FHM2008] S1-ap42	\\
6	&	8344455-5390	&	05334668-0523256	&	V1897 Ori, [FHM2008] S2-ap3 	\\
7	&	8344987-5460	&	05334797-0527385	&	[FHM2008] S2-ap6	\\
8	&	8346141-5010	&	05335074-0500394	&	V1902 Ori, [FHM2008] F11-ap39	\\
\enddata
\tablecomments{\footnotesize The sources of selected objects names: \citet{Parenago} for Parenago XXX; \citet{FHM2008} for [FHM2008]XXX-XXXX; V* indicate the General Catalogue of Variable Stars (GCVS) \citep{GCVS1971}; HBC XXX indicate Herbig+Bell Catalog \citep{HBC1988Orion}; HOPS-XXX indicate the identification of these sources in \textit{Herschel} Orion Protostar Survey \citep[HOPS,][]{Megeath_HOPS_2010AAS}.}
\tablenotetext{a}{\footnotesize   The coordinate of HBC 497 (DL Ori/G3) does not exactly matched to the IRS coordinate of OriA-270.}
\tablenotetext{b}{\footnotesize	A 2MASS source is 2-3 arcsec away from the target.}
\tablenotetext{c}{\footnotesize	JHK photometry is from \citet{Robberto_ONC_JHK_2010}}
\end{deluxetable}
\clearpage

\clearpage
\begin{deluxetable}{cccccc}
\rotate
\tablecaption{Observation log of SpeX spectra \label{table:SpeXlog}}
\tablehead{
\colhead{Num.}& \colhead{Obs. Date} & \colhead{Slit Width} & \colhead{Total Int. Time} & \colhead{PA} & \colhead{note}\\
\colhead{}& \colhead{(UT)} & \colhead{(")} & \colhead{(sec)} & \colhead{($^o$)} & \colhead{} \\
\colhead{(1)}& \colhead{(2)} & \colhead{(3)} & \colhead{(4)} & \colhead{(5)} & \colhead{(6)} 
}
\tablecolumns{6}
\tablewidth{0pt}
\tabletypesize{\scriptsize}
\startdata
1	&	2011 Nov 06	&	0.5	&	480	&	$\cdots$		&	$\cdots$		\\
2	&	2011 Feb 28	&	0.8	&	480	&	-10.73	&	$\cdots$		\\
3	&	2011 Nov 07	&	0.8	&	120	&	$\cdots$		&	$\cdots$		\\
4	&	2011 Feb 28	&	0.8	&	320	&	-30	&	b	\\
5	&	2011 Feb 28	&	0.8	&	400	&	-10.73	&		$\cdots$	\\
11     &	2010 Feb 18	&	0.8	&	480	&	$\cdots$		&	$\cdots$		\\
13     &	2010 Feb 17	&	0.8	&	480	&	$\cdots$		&	$\cdots$		\\
16	&	2010 Feb 18	&	0.8	&	240	&	$\cdots$		&	$\cdots$		\\
\enddata
\tablecomments{\footnotesize Column (1): Num. The numbers are assigned in Table~\ref{table:IRSlog}. When an object was revealed having a nearby companion and it was observed separately from the target object with SpeX, we add "a" and "b" next to the Num.  Column (5): PA. The slit's positional angle for observation of a target. Column (6): Note for observation of binary systems: "a" indicates for the system which its primary and secondary were observed separately; "b" indicates for the system which its primary and secondary were observed simultaneously and extracted their spectra separately; "c" indicates for the system which is suspected to have a secondary, but the spectrum of each component were not resolved, and the spectrum is the composite spectrum of a primary and a secondary.}
\end{deluxetable}
\clearpage

\clearpage
\begin{deluxetable}{cccccccc}
\rotate
\tablecaption{Spectral Type and Extinction \label{table:SpTypeAv}}
\tablehead{
\colhead{Num.} & \colhead{IRS name} & \colhead{Spectral types} & \colhead{Spectral type} & \colhead{EW($H_\alpha$)} & \colhead{EW($Li$)} & \colhead{$A_{V}$} & \colhead{$A_{V}$ method} \\
\colhead{(1)} & \colhead{(2)} & \colhead{(3)} & \colhead{adopted (4)} &  \colhead{(5)} & \colhead{(6)} & \colhead{(7)} & \colhead{(8)}
}
\tablecolumns{8}
\tablewidth{0pt}
\tabletypesize{\scriptsize}
\startdata
1	&	8336884-5290	&	M0.0(8)	&	M0.0	&	-22.5	&	0.5	&	$0.77_{-0.39	}^{+1.02}$ &	J-H	\\
2	&	8339347-5238	&	M3.0(8)	&	M3.0	&	-12.5	&	0.4	&	$1.15_{-1.07	}^{+0.52}	$ &	J-H	\\
3	&	8343858-5513	&	K1(3)	&	K1.0	&	$\cdots$	&	$\cdots$	&	$	4.37	_{-	3.05	}^{+	2.50	}	$	&	J-H	\\
4	&	8343917-6073	&	M1.5(1); M1.0(24)	&	M1.0	&	-101.0	&	0.2	&	$	0.09	_{-	0.00	}^{+	2.70	}	$	&	CTTS J-H	\\
5	&	8343944-5609	&	M5.5(3); M5.5(8)	&	M5.5	&	-33.3	&	0.3	&	$	0.50	_{-	0.29	}^{+	1.24	}	$	&	I-J	\\
6	&	8344455-5390	&	$\cdots$	&	$\cdots$	&	$\cdots$	&	$\cdots$	&	$	3.21	_{-	1.79	}^{+	9.26	}	$	&	CTTS J-H	\\
7	&	8344987-5460	&	M6(3)	&	M6.0	&	$\cdots$	&	$\cdots$	&	$	2.42	_{-	1.09	}^{+	1.07	}	$	&	J-H	\\
8	&	8346141-5010	&	M2(3); M3.5(8)	&	M3.5	&	-8.6	&	0.3	&	$	0.23	_{-	0.00	}^{+	1.47	}	$	&	I-J	\\
\enddata
\tablecomments{\footnotesize The methods or literature reference for spectral type determination is indicated with parentheses, and the meaning of the numbers in parentheses is listed as follows:
(1) \citet{Allen_SpT_pc}, private communication; spectral types are measured from HECTOSPEC spectra
(2) \citet{Hillenbrand97_ONCpopulation_age}
(3) \citet{RHS2000}
(4) \citet{Allen_thesis_L1641}
(5) \citet{Wolff+2004}
(6) \citet{HC2000}
(7) \citet{Hernandez_SpT_MDM_pc}, private communication; spectral types measured from MDM spectroscopic data
(8) \citet{Hernandez_SpT_hecfast_pc}, private communication; spectral types are measured from HECTOSPEC spectra or FAST spectra
(9) \citet{Dario2009_ubvri}
(10) \citet{Fang2009}
(11) \citet{Parihar2009_ONC_variables_SpT4V1498Ori}
(12) \citet{Manoj2006ApJ}
(13) \citet{Parihar+2009MNRAS}
(14) \citet{HBC1988Orion}
(15) \citet{Gleboki+2005}
(16) \citet{Daemgen+2012}
(17) \citet{Grankin+2007}
(18) \citet{Riaz+2006}
(19) \citet{Rebull2001AJ}
(20) \citet{Houk+1999}
(21) \citet{SAOcatalogue}
(22) \citet{Wouterlook+1992}
(23) \citet{CarattiOGaratti2012}
(24) \citet{Hsu_L1641SpT_2012ApJ...752...59H}
(25) \citet{Fang_L1641_2013ApJS..207....5F}
(SpeX) Spectral typing with SpeX spectra in this work
}
\tablecomments{\footnotesize The unit of column (5) and (6) is $\AA$. EW($H_\alpha$) and EW($Li$) values are from \citet{Hernandez_SpT_MDM_pc, Hernandez_SpT_hecfast_pc, Hsu_L1641SpT_2012ApJ...752...59H}.  Column (7) is the adopted $A_{V}$ and the range of $A_{V}$ values calculated by several methods described in the text. Column (8) $A_{V}$ method: the method used to derive the adopted $A_{V}$ in the column (7). }
\end{deluxetable}
\clearpage

\begin{table}[!htbp]
\begin{center}
\begin{threeparttable}
\caption{K-S test for the visual extinction distribution }
\label{table:KStest-Av}
\begin{tabular}{c|cc|cc}
\space
\hline\hline
\multirow{2}{*}{Regions compared} & \multicolumn{2}{c|}{All objects} & \multicolumn{2}{|c}{available spectral types}\\
\cline{2-5}
 & $D$ & $p$ & $D$ & $p$\\
\hline
Trapezium. vs. ONC	&	0.30	&	$\ll$0.001	&	0.25	&	0.03	\\
Trapezium. vs. L1641	&	0.30	&	$\ll$0.001	&	0.25	&	0.03	\\
ONC vs. L1641	&	0.19	&	0.02	&	0.21	&	0.04	\\
\hline
\end{tabular}
   \begin{tablenotes}
     \item {\footnotesize{$D$ is the maximum deviation between the cumulative distribution of two groups. $p$ indicates the probability that there is no significant difference between the distributions}}
    \end{tablenotes}
\end{threeparttable}
\end{center}
\end{table}

\begin{deluxetable}{cccccc}
\tablecaption{Basic Stellar Properties and Mass Accretion Rate\label{table:basicprop}}
\tablehead{
\colhead{Num.} & \colhead{IRS name} & \colhead{$T_{eff}$} & \colhead{$L_{\star}$} & \colhead{$M_{\star}$} & \colhead{$\dot{M}$ ($10^{-9}$)}\\
\colhead{} & \colhead{} & \colhead{(K)} & \colhead{($L_{\odot}$)} & \colhead{($M_{\odot}$)} & \colhead{($M_{\odot}/yr$)}  
}
\tablecolumns{6}
\tablewidth{0pt}
\tabletypesize{\scriptsize}
\startdata
1	&	8336884-5290	&	3770	&	0.66	&	0.49	$\pm$	0.07	&		0.63	$\pm$	0.11	\\
2	&	8339347-5238	&	3360	&	0.41	&	0.29	$\pm$	0.06	&		1.19	$\pm$	0.96	\\
3	&	8343858-5513	&	4920	&	12.12	&	2.67	$\pm$	0.30	&		235.57	$\pm$	133.24	\\
4	&	8343917-6073	&	3630	&	0.79	&	0.40	$\pm$	0.06	&		2.91	$\pm$	0.45	\\
5	&	8343944-5609	&	2840	&	0.48	&	0.15	$\pm$	0.03	&	$<$	9.79			\\
6	&	8344455-5390	&	3770	&	0.25	&	0.50	$\pm$	0.09	&		$\cdots$			\\
7	&	8344987-5460	&	2800	&	0.09	&	0.09	$\pm$	0.03	&		$\cdots$			\\
8	&	8346141-5010	&	3260	&	0.16	&	0.24	$\pm$	0.06	&		$\cdots$			\\
\enddata
\end{deluxetable}
\clearpage

\clearpage
\begin{deluxetable}{cccccccccccccc}
\rotate
\tablecaption{Spectral Indices\label{table:sptidx}}
\tablehead{
\colhead{Num.} & \colhead{module} & \colhead{$n_{K-6}$} & \colhead{$\sigma_{n_{K-6}}$} & \colhead{$n_{6-13}$} & \colhead{$\sigma_{n_{6-13}}$} &\colhead{$n_{13-31}$} & \colhead{$\sigma_{n_{13-31}}$} &\colhead{$n_{5-12}$} & \colhead{$\sigma_{n_{5-12}}$} &\colhead{$n_{12-20}$} & \colhead{$\sigma_{n_{12-20}}$} &\colhead{$n_{K-25}$} & \colhead{$\sigma_{n_{K-25}}$}
}
\tablecolumns{14}
\tablewidth{0pt}
\tabletypesize{\scriptsize}
\startdata
1	&	SLLL	&	-2.07	&	0.10	&	-0.89	&	0.13	&	0.29	&	0.23	&	-1.03	&	0.10	&	0.95	&	0.15	&	-1.01	&	0.04	\\
2	&	SLLL	&	-1.70	&	0.08	&	-0.53	&	0.11	&	0.06	&	0.16	&	-0.65	&	0.13	&	-0.18	&	0.18	&	-0.89	&	0.03	\\
3	&	SLLL	&	-1.51	&	0.03	&	-0.27	&	0.03	&	-1.11	&	0.06	&	-0.32	&	0.04	&	-0.76	&	0.04	&	-1.01	&	0.01	\\
4	&	SLLL	&	-1.62	&	0.11	&	-0.13	&	0.13	&	-0.66	&	0.06	&	-0.49	&	0.07	&	-0.17	&	0.24	&	-0.87	&	0.02	\\
5	&	SLLL	&	-1.77	&	0.16	&	0.71	&	0.20	&	-0.80	&	0.11	&	0.35	&	0.10	&	-0.23	&	0.09	&	-0.66	&	0.04	\\
6	&	SLLL	&	0.00	&	0.05	&	-0.68	&	0.07	&	-0.70	&	0.11	&	-0.83	&	0.06	&	-0.80	&	0.07	&	-0.47	&	0.02	\\
7	&	SLLL	&	$\cdots$	&	$\cdots$	&	$\cdots$	&	$\cdots$	&	$\cdots$	&	$\cdots$	&	$\cdots$	&	$\cdots$		$\cdots$	&	$\cdots$	&	$\cdots$	&	$\cdots$	\\
8	&	SLLL	&	-2.39	&	0.07	&	-1.15	&	0.11	&	0.30	&	0.16	&	-1.40	&	0.08	&	1.71	&	0.37	&	-1.25	&	0.04	\\
\enddata
\tablecomments{\footnotesize The wavelength interval for each anchor is as follows: The K band is at 2.17 $\micron$; 5 represent for wavelengths of 5.2-5.54 $\micron$; 6 for 5.4--6.0 $\micron$; 12 for 12.7--13.1 $\micron$; 13 for 12.8--14.0 $\micron$; 20 for 19.7--19.95 $\micron$; 25 for 24.5--25.5 $\micron$; 31 for 30.3--31.0 $\micron$. }
\end{deluxetable}
\clearpage

\clearpage
\begin{deluxetable}{ccccccccccccc}
\rotate
\tablecaption{Grain Properties\label{table:ew1020}}
\tablehead{
\colhead{Num.} & \colhead{mod} & \colhead{EW10} & \colhead{$\sigma_{EW10}$} & \colhead{EW20} & \colhead{$\sigma_{EW20}$} &  \colhead{F10} & \colhead{$\sigma_{F10}$} &\colhead{F20} & \colhead{$\sigma_{F20}$} &\colhead{$F_{11.3}/F_{9.8}$} & \colhead{$\sigma_{F_{11.3}/F_{9.8}}$} & \colhead{TD/FD}\\
\colhead{} & \colhead{} & \colhead{($\mu$m)} & \colhead{($\mu$m)} & \colhead{($\mu$m)} & \colhead{($\mu$m)} &  \colhead{(10$^{-12}$erg/s/cm$^2$)} & \colhead{(10$^{-12}$erg/s/cm$^2$)} &\colhead{(10$^{-12}$erg/s/cm$^2$)} & \colhead{(10$^{-12}$erg/s/cm$^2$)} &\colhead{ } & \colhead{ } & \colhead{ }
}
\tablecolumns{13}
\tablewidth{0pt}
\tabletypesize{\tiny}
\startdata
1	&	SLLL	&	8.054	&	0.157	&	3.243	&	0.326	&	2.698	&	0.022	&	0.759	&	0.068	&	0.45	&	0.01	&	TD	\\
2	&	SLLL	&	2.781	&	0.088	&	2.392	&	0.305	&	1.127	&	0.023	&	0.553	&	0.054	&	0.53	&	0.02	&	FD	\\
3	&	SLLL	&	2.509	&	0.085	&	2.464	&	0.305	&	29.600	&	0.634	&	9.833	&	0.893	&	0.41	&	0.02	&	FD	\\
4	&	SLLL	&	2.595	&	0.086	&	2.533	&	0.309	&	3.219	&	0.075	&	1.470	&	0.145	&	0.44	&	0.02	&	FD	\\
5	&	SLLL	&	5.348	&	0.12	&	2.515	&	0.308	&	5.381	&	0.061	&	1.491	&	0.147	&	0.51	&	0.02	&	TD	\\
6	&	SLLL	&	1.247	&	0.069	&	1.485	&	0.288	&	3.608	&	0.161	&	1.449	&	0.247	&	0.49	&	0.04	&	FD	\\
7	&	SLLL	&	$\cdots$	&	$\cdots$	&	$\cdots$	&	$\cdots$	&	$\cdots$	&	$\cdots$	&	$\cdots$	&	$\cdots$	&	$\cdots$	&	$\cdots$	&	$\cdots$	\\
8	&	SLLL	&	1.045	&	0.067	&	3.142	&	0.323	&	0.049	&	0.003	&	0.119	&	0.012	&	0.54	&	0.05	&	TD	\\
\enddata
\end{deluxetable}
\clearpage

\clearpage
\begin{table}[!htbp]
\scriptsize
\begin{center}
\begin{threeparttable}
\caption{Median Mass Accretion Rates: FD vs. TD}
\label{table:prop-median-Mdot-TDFD}
\begin{tabular}{c|c|c|c|c|c}
\space
\hline\hline 
		\multicolumn{2}{c|}{}			&	\multicolumn{4}{c}{Subgroups by spectral ranges}	\\
\cline{3-6}
	\multicolumn{2}{c|}{}		&	K--M	&	K &	M	&	 M3 and later	\\
\hline																	
FD	&	N	&	56	 &	23   & 33	&		11\\
	&	$\dot{M}$	&	10.3 	&	45.4 &  3.51  & 1.93 \\	
\hline							
TD	&	N	&	50 & 13 & 37 & 18 	\\
	&	$\dot{M}$	&	1.76 & 4.77 & 1.50 & 1.34 \\
\hline
\end{tabular}
  \begin{tablenotes}
      \item N is the number of objects. The unit of $\dot{M}$ is $10^{-9}M_{\odot}$/yr.
    \end{tablenotes}
\end{threeparttable}
\end{center}
\end{table}
\clearpage

\clearpage
\begin{table}[!htbp]
\scriptsize
\begin{center}
\begin{threeparttable}
\caption{Distribution of Properties Comparison to Taurus:FD and TD}
\label{table:prop-median-KS-TDFD-OriATau}
\begin{tabular}{c|c|c|ccc|ccc|ccc}
\space
\hline\hline 
	&		&		&	\multicolumn{9}{c}{K-S test}	\\
\cline{4-12}
	&		&	Taurus	&	\multicolumn{3}{c|}{Ori A}		&	\multicolumn{3}{c|}{ONC}		&	 \multicolumn{3}{c}{L1641}	\\
\cline{4-12}
	&	parameter	&	Median	&	Median & $D$	&	$p$	&	Median  & $D$	&	$p$	&	 Median & $D$	&	$p$	\\
\hline																	
FD	&	$n_{K-6}$	&	-1.49	&	-1.33  & 0.13	&	0.14	&	-1.33 & 0.16	&	0.11	&	-1.33 & 0.13	&	0.36	\\
	&	$n_{6-13}$	&	-0.87	&	-0.64  & 0.20	&	$<$0.01	& -0.61 & 	0.23	&	$<$0.01	&	-0.64 & 0.18	&	0.06	\\
	&	$n_{13-31}$	&	-0.54	&	 -0.66 & 0.15	&	0.05	&	-0.67 & 0.16	&	0.12	&	-0.66 & 0.17	&	0.10	\\
	&	EW$(10 \micron) $	&	1.56	&	 2.21 & 0.32	&	$<$0.01	&	2.47 & 0.38	&	$<$0.01	&	1.98 & 0.26	&	$<$0.01	 \\
	&	$F_{11.3}/F_{9.8}$	&	0.55	&	 0.52 & 0.93	&	$<$0.01	&0.52 & 	0.91	&	$<$0.01	&	0.53 & 0.97	&	$<$0.01	 \\
\hline																	
TD	&	$n_{K-6}$	&	-1.89	&	 -2.23 & 0.34	&	0.14	&	-2.18 & 0.36	&	0.15	&	-2.39 & 0.36	&	0.15	\\
	&	$n_{6-13}$	&	-0.75	&	 -0.52 & 0.17	&	0.90	&	-0.47 & 0.21	&	0.77	&	-0.62 & 0.13	&	1.00	\\
	&	$n_{13-31}$	&	0.71	&	 0.59 & 0.23	&	0.55	&	0.50 & 0.30	&	0.32	&	0.70 & 0.24	&	0.60	\\
	&	EW$(10 \micron) $	&	4.67	&	3.45 & 0.41	&	0.04	&	3.57 & 0.38	&	0.10	&	3.31 & 0.44	&	0.04	 \\
	&	$F_{11.3}/F_{9.8}$	&	0.50	&	 0.57 & 0.69	&	$<$0.01	&	0.54 & 0.76	&	$<$0.01	&	0.60 & 0.62	&	$<$0.01	 \\
\hline
\end{tabular}
   \begin{tablenotes}
      \item $D$ is the maximum deviation between the cumulative distribution of two groups. $p$ indicates the probability that there is no significant difference between the distributions
    \end{tablenotes}
\end{threeparttable}
\end{center}
\end{table}
\clearpage

\clearpage
\begin{deluxetable}{cccccc}
\tablecaption{$n_{20-31}$ and the projected distance (d) from $\theta^1$ Ori C \label{table:table-n2031-Distance}}
\tablehead{
\colhead{Num.} & \colhead{$n_{20-31}$} & \colhead{$\sigma_{n_{20-31}}$} & \colhead{d (arcsec)} & \colhead{d (pc)} & \colhead{Section} 
}
\tablecolumns{6}
\tablewidth{0pt}
\tabletypesize{\scriptsize}
\startdata
1	&	-0.31	&	0.43	&	1075.8	&	2.2	&	DN	\\
2	&	0.37	&	0.3	&	1103.5	&	2.2	&	DN	\\
3	&	-1.31	&	0.1	&	966.8	&	1.9	&	CS	\\
4	&	-0.82	&	0.21	&	2607.3	&	5.2	&	L	\\
5	&	-1.12	&	0.17	&	1164.1	&	2.3	&	DS	\\
6	&	-0.49	&	0.18	&	844.0	&	1.7	&	CS	\\
8	&	-0.85	&	0.42	&	1584.0	&	3.2	&	DN	\\
9	&	-0.61	&	0.1	&	800.6	&	1.6	&	CS	\\
\enddata
\tablecomments{\footnotesize The region for each section is indicated in Figure~\ref{fig-ONC-big}. L indicates L1641 in Figure~\ref{fig-L1641-big}.}
\end{deluxetable}
\clearpage

\end{document}